\documentstyle[11pt]{article}

\renewcommand{\theequation}{\thesection.\arabic{equation}}

\begin{document}

\thispagestyle{empty}

\vspace{2cm}

\begin{flushright}
IC/2000/14
\end{flushright}

\begin{center}

\vspace{1.cm}
{\bf 
THE GENERALIZED ABEL-PLANA FORMULA.\\
APPLICATIONS TO BESSEL FUNCTIONS\\
AND\\
CASIMIR EFFECT}\\
\vspace{1cm}
Aram A. Saharian\footnote{E-mail: saharyan@server.physdep.r.am}\\
\medskip
{\em Department of Physics, Yerevan State University, 1 Alex Manoogian St,\\
375049 Yerevan, Armenia,\footnote{Permanent address.}\\
\medskip
Institute of Applied Problems in Physics NAS RA, 25 Nersessian St,\\
375014 Yerevan, Armenia,\\
and\\
The Abdus Salam International Centre for Theoretical Physics,\\
34100 Trieste, Italy}\\

\bigskip

February 28, 2000

\end{center}

\vspace{0.5cm}

\centerline{\bf Abstract}

\bigskip
 
One of the most efficient methods to obtain the vacuum expectation values
for the physical observables in the Casimir effect 
is based on using the Abel-Plana
summation formula. This allows to derive the regularized quantities
by manifestly cutoff independent way and to present them in the form 
of strongly convergent integrals. However the application of Abel-Plana
formula in usual form is restricted by simple geometries when the 
eigenmodes have a simple dependence on quantum numbers. The author 
generalized the Abel-Plana formula which essentially enlarges its 
application range. Based on this generalization,  
formulae have been obtained for various 
types of series over the zeros of some combinations of Bessel 
functions and for integrals involving these functions. It have been
shown that these results generalize the special cases existing 
in literature. Further the derived summation formulae have been used
to summarize series arising in the mode summation approach to the 
Casimir effect for spherically and cylindrically symmetric boundaries.
This allows to extract the divergent parts  from the vacuum expectation 
values for the local physical observables in the manifestly cutoff 
independent way. Present paper reviews these results. Some new 
considerations are added as well.

\newpage

\section{Introduction}

\renewcommand{\theequation}{1.\arabic{equation}}

The Casimir effect is 
among the most interesting consequences of quantum field theory
and is essentially the only macroscopic manifestation of the 
nontrivial properties of the physical
vacuum. These  properties may be determined from the responce of 
the vacuum state to classical external fields or constraints.
The simplest case is realized by boundary conditions
on quantized fields. Such conditions modify the zero point
mode spectrum and as a result can change the energy of the vacuum.
This change is manifest as an observable Casimir energy.
Since the original work by Casimir in 1948 \cite{Casimir} 
many theoretical and 
experimental works have been done on this problem, including 
various types of boundary geometry and non-zero temperature 
effects (see, e.g., \cite{Mostepanenko, Plunien, Bordag, Lamor}
and references therein). Many different approaches have been used: mode 
summation method, Green function formalism, multiple scattering 
expansions, heat-kernel series, zeta function regularization
technique, etc.. From the general theoretical point of view
the main point here is the unique separation and subsequent
removing of the divergences.
Within the framework of the mode summation method in calculations
of the expectation values for physical observables, such 
as energy-momentum tensor, one often needs to sum over the values
of a certain function at integer points, and then subtract the 
corresponding quantity for unbounded space (usually presented 
in terms of integrals). Practically, the sum and integral, 
taken separately, diverge and some physically motivated 
procedure, to handle finite result, is needed.
For a number of geometries one of the most convenient methods 
to obtain such regularized values
of the mode sums is based on the using of the Abel-Plana formula
(APF) \cite{Hardy, Erdelyi, Evgrafov}. In \cite{Mamaev} this 
formula have been used to regularize scalar field energy 
momentum-tensor on backgrounds of various Friedmann cosmological
models. Further applications to the Casimir effect for flat 
boundary geometries with corresponding references  can be found in
\cite{Mostepanenko}. Abel-Plana formula allows 
(i) to extract by cutoff independent way the Minkowski 
vacuum part and (ii) to obtain for the regularized part 
strongly convergent integrals, useful, in particular, for numerical
calculations. However the applications of APF in usual
form is restricted by the flat boundary cases when the eigenmodes
have simple dependence on quantum numbers.

In \cite{Sah, Sah1} the APF was generalized (see also \cite{Sahdis}). 
The generalized version contains two meromorphic functions. 
Choosing one of these functions in specific form APF in usual form 
is obtained. By applying the 
generalized formula to Bessel functions in \cite{Sah1, Sahdis} 
summation formulae are obtained over the zeros 
of various combinations of these functions. In particular, formulae
for Fourier-Bessel and Dini series are derived. From these formulae
by specifying the constants and choosing the order of Bessel function
equal to 1/2 one obtains a simple generalization of APF for the 
case of a function having poles. It have been shown that 
from generalized formula interseting results can be derived
for infinite integrals involving
Bessel functions. Further the obtained summation
formulae are applied to regularize the vacuum expectation 
values for the energy-momentum tensor components of the electromagnetic
field in the Casimir effect with spherically 
\cite{Grig1, Grig2, Grig3, Sah2shert} and cylindrically symmetric 
\cite{Sah2, Sah3} boundaries. As in the case of flat boundaries
the using of generalised Abel-Plana formula allow to extract 
in manifestly cutoff independent
way the contribution of the unbounded space and to present 
regularized values in terms of exponentially converging 
integrals. 

The present paper reviews these results and is organized as follows. 
In section 2 the generalized Abel-Plana formula is derived
and as a special case usual APF is obtained. It is 
indicated how to generalize this formula for the functions having
poles. The applications of generalized formula to Bessel functions are 
considered in the next section. We derive two formulae for the 
sums over zeros of $AJ_\nu (z) +BzJ'_\nu (z)$. Specific 
examples of applications of the general formulae are considered. 
For $\nu =1/2,\, B=0$ and for analytic function $f(z)$ the 
APF is obtained. In section 4 from generalized Abel-Plana formula 
by special choice of 
function $g(z)$ summation formulae are derived for the series 
over zeros of the function
$J_\nu (z)Y_\nu (\lambda z)-J_\nu (\lambda z)Y_\nu (z)$ and 
similar combinations with Bessel functions derivatives. 
Special examples are considered. The 
applications to the integrals involving Bessel 
functions and some their combinations are discussed 
in section 5. A number of interesting results for these 
integrals are presented. Specific examples of applying
these general formulae are described in the next section.
In section 7 by using generalized Abel-Plana formula 
two theorems are proved 
for the integrals involving the function
$J_\nu (z)Y_\mu (\lambda z)-J_\mu (\lambda z)Y_\nu (z)$ 
and their applications are considered. The following 
sections are devoted to the applications of generalized formula for 
the calculations of the regularized vacuum expectation values 
of the electromagnetic energy-momentum tensor inside (section 8) and
outside (section 9) a perfectly conducting spherical
shell, and for the region between two perfectly conducting
spherical surfaces (section 10). In sections 11-13 the similar
problems for the cylindrical surfaces are considered. The section
14 concludes the main results considered in this paper.

\setcounter{equation}{0}

\section{Generalized Abel-Plana formula}

\renewcommand{\theequation}{2.\arabic{equation}}

\setcounter{equation}{0}

Let $f(z)$ and $g(z)$ be meromorphic functions for $a\leq x\leq b$
in the complex plane $z=x+iy$. Let us note by $z_{f,k}$ and $z_{g,k}$ the
poles of $f(z)$ and $g(z)$ in region $a<x<b$, respectively. Assume that
${\mathrm{Im}}z_{f,k}\neq 0$ (see however the Remark to Lemma).

\bigskip

\noindent {\bf Lemma.} {\it 
If functions $f(z)$ and $g(z)$ satisfy condition
\begin{equation}
\lim_{h\to \infty}\int_{a\pm ih}^{b\pm ih}
{\left [ g(z)\pm f(z)\right ]dz}=0,
\label{cor11}
\end{equation}
then the following formula takes place
\begin{equation}
\int_{a}^{b}{f(x)dx}=R[f(z),g(z)]-\frac{1}{2}\int_{-i\infty}^{+i\infty}
{\left [ g(u)+{\mathrm{sgn}} ({{\mathrm{Im}}} z)
f(u)\right ]_{u=a+z}^{u=b+z}dz,}
\label{cor12}
\end{equation}
where
\begin{equation}
R[f(z),g(z)]=\pi i\left [ \sum_k {\mathrm{Res}}_{z=z_{g,k}}g(z)+\sum_{k}
{\mathrm{Res}}_{{\mathrm{Im}}z_{f,k}>0}f(z)-\sum_{k}
{\mathrm{Res}}_{{\mathrm{Im}}z_{f,k}<0}f(z)\right ].
\label{cor13}
\end{equation}
}

\bigskip

\noindent {\bf Proof.} Let us consider a rectangle $C_h$ with vertices 
$a\pm ih$, $b\pm ih$  described in the positive sense. 
In accordance to the residue theorem 
\begin{equation}
\int_{C_h}{g(z)dz}=2\pi i\sum_k {\mathrm{Res}}_{z=z_{g,k}}g(z),
\label{corp1}
\end{equation}
where rhs contains sum over poles within $C_h$. 
Let $C_h^+$ and $C_h^-$ denote the upper and 
lower halfs of this contour. Then one has
\begin{equation}
\int_{C_h}{g(z)dz}=\int_{C_h^+}{[g(z)+f(z)]dz}+\int_{C_h^-}{[g(z)-f(z)]dz}
-\int_{C_h^+}{f(z)dz}+\int_{C_h^-}{f(z)dz}.
\label{corp2}
\end{equation}
By the same residue theorem 
\begin{equation}
\int_{C_h^-}{f(z)dz}-\int_{C_h^+}{f(z)dz}=2\int_{a}^{b}{f(x)dx}
+2\pi i\left[ \sum_{k}{\mathrm{Res}}_{{\mathrm{Im}}z_{f,k}<0}
f(z)-\sum_{k}{\mathrm{Res}}_{{\mathrm{Im}}z_{f,k}>0}f(z)
\right] .\label{corp3}
\end{equation}
Then 
\begin{equation}
\int_{C_h^\pm }{[g(z)\pm f(z)]dz}=\pm \int_{0}^{\pm ih}
{[g(u)\pm f(u)]_{u=a+z}^{u=b+z}dz}\mp \int_{a\pm ih}^{b\pm ih}%
{[g(z)\pm f(z)]dz}. \label{corp4}
\end{equation}
Combining these results and allowing in (\ref{corp1})
$h\to \infty$ one obtains the formula (\ref{cor12}). \rule{1.5ex}{1.5ex}

\bigskip

If the functions $f(z)$ and $g(z)$ have poles 
with ${\mathrm{Re}} z_{i,k}=a,b$ ($i=f,g$) the contour have to 
pass round these points on the right or left, correspondingly.
 
\bigskip

\noindent {\bf Remark.} The formula (\ref{cor12}) 
is valid also when the function 
 $f(z)$ has real poles $z_{f,n}^{(0)}$, 
 ${\mathrm{Im}} z_{f,n}^{(0)}=0$ in the region
 $a<{\mathrm{Re}}z<b$ if the main part of its 
 Laurent expansion near of these poles
 does not contain even powers of $z-z_{f,n}^{(0)}$. In this case on 
 the left of the formula (\ref{cor12}) the integral is meant in the 
 sense of the principal value, which exists as a consequence of the
 abovementioned condition. For brevity let us consider
 the case of a single pole $z=z_0$. One has
 \begin{eqnarray}
 & & {} \int_{C_h^-}{f(z)dz}-\int_{C_h^+}{f(z)dz} = 2\left [ 
 \int_{a}^{z_0-\rho }{f(z)dz}+\int_{z_0+\rho }^{b}{f(z)dz}\right ]
 +{}\nonumber \\
 & &{}+2\pi i\left [ \sum_{k}{\mathrm{Res}}_{{\mathrm{Im}}z_{f,k}<0}f(z)-
 \sum_{k}{\mathrm{Res}}_{{\mathrm{Im}}z_{f,k}>0}f(z) \right ] + 
 \int_{\Gamma_\rho^+}{f(z)dz}+\int_{\Gamma_\rho^-}{f(z)dz},
\label{corp5}
 \end{eqnarray}
 with contours $\Gamma_\rho^+$ and $\Gamma_\rho^-$ being the 
 upper and lower circular arcs (with center at $z=z_0$) joining
 the points $z_0-\rho $ and $z_0+\rho $. By taking into 
 account that for odd negative $l$
 \begin{equation}
  \int_{\Gamma_\rho^+}{(z-z_0)^ldz}+
  \int_{\Gamma_\rho^-}{(z-z_0)^ldz}=0, \label{corp6}
 \end{equation}
in the limit $\rho\to 0$ we obtain the required result.
 \rule{1.5ex}{1.5ex}
 
\bigskip
 
 In the following on the left of (\ref{cor12}) we will write
 ${\mathrm{p.v.}}\int_{a}^{b}{f(x)dx}$, assuming that this integral converges 
in the sense of the principal value. As a direct consequence of Lemma
one obtains \cite{Sah1}:
 
 \bigskip
 
\noindent {\bf Theorem 1.} {\it
 If in addition to the conditions of Lemma one has
 \begin{equation}
\lim_{b\to \infty}\int_{b}^{b\pm i\infty }
{\left [ g(z)\pm f(z)\right ]dz}=0,
\label{th11}
\end{equation}
then
\begin{equation}
\lim_{b\to \infty}\left\{ {\mathrm{p.v.}}\int_{a}^{b}{f(x)dx}-R[f(z),g(z)]\right\}=
\frac{1}{2}\int_{a-i\infty }^{a+ i\infty }{\left [ g(z)+
{\mathrm{sgn}}({\mathrm{Im}} z) f(z)\right ]dz},
\label{th12}
\end{equation}
where on the left $R[f(z),g(z)]$ is defined as (\ref{cor13}),
$a<{\mathrm{Re}}z_{f,k},{\mathrm{Re}}z_{g,k}<b$, 
and summation goes over poles $z_{f,k}$ and
$z_{g,k}$ arranged in order ${\mathrm{Re}}z_{i,k}
\leq {\mathrm{Re}}z_{i,k+1}$, $i=f,g$.
}

\bigskip

\noindent {\bf Proof.} To proof it is sufficient 
to insert in the general formula
(\ref{cor12}) $b\to\infty $ and to use the condition (\ref{th11}). The
order of summation in $R[f(z),g(z)]$ is determined by the choice of
the integration contour $C_h$ and by limiting transition $b\to\infty $.
\rule{1.5ex}{1.5ex}

\bigskip

We will call the formula (\ref{th12}) as {\bf Generalized Abel-Plana 
Formula} (GAPF) as for $b=n+a$, $0<a<1$, 
$g(z)=-if(z)\cot\pi z$ and analytic functions $f(z)$ from (\ref{th12})
follows the Abel-Plana formula (APF) \cite{Hardy, Erdelyi, Evgrafov}
\begin{equation}
\lim_{n\to\infty}\left [ \sum_{1}^{n}f(s)-\int_{a}^{n+a}{f(x)dx}
\right ] =
\frac{1}{2i}\int_{a}^{a-i\infty }{f(z)(\cot\pi z-i)dz}-
\frac{1}{2i}\int_{a}^{a+i\infty }{f(z)(\cot\pi z+i)dz}. 
\label{apsf1}
\end{equation}
The useful form of (\ref{apsf1}) may be obtained performing the
limit $a\to 0$. By taking into account that the point $z=0$ is
a pole for integrands and therefore have to be around by arcs of 
the small circle $C_{\rho }$ on the right and performing 
$\rho \to 0$ one obtains
\begin{equation}
\sum_{n=0}^{\infty }f(n)=\int_{0}^{\infty}{f(x)dx}+
\frac{1}{2}f(0)+
i\int_{0}^{\infty }{\frac{f(ix)-f(-ix)}{e^{2\pi x}-1}dx}.
\label{apsf2}
\end{equation}
Note that now the condition (\ref{cor11}) is satisfied if
\begin{equation}
\lim_{y\to \infty}e^{-2\pi \vert y\vert }\vert f(x+iy)\vert =0
\label{cond1apsf}
\end{equation}
uniformly in any finite interval of $x$. The (\ref{apsf2}) is the
most frequently used form of APF in its physical applications.
Another useful form (in particular for fermionic 
field calculations) to sum over the values of an analytic
function at half of an odd integer points  can be obtained from 
(\ref{apsf2}) \cite{Mostepanenko, Mamaev2}:
\begin{equation}
\sum_{n=0}^{\infty }f(n+1/2)=\int_{0}^{\infty}{f(x)dx}-
i\int_{0}^{\infty }{\frac{f(ix)-f(-ix)}{e^{2\pi x}+1}dx}
\label{apsf2half}
\end{equation}
By adding to the rhs of (\ref{apsf2}) the term
\begin{equation}
\pi i\left\{ \sum_{k}{\mathrm{Res}}_{{\mathrm{Im}}z_{f,k}>0}f(z)-
 \sum_{k}{\mathrm{Res}}_{{\mathrm{Im}}z_{f,k}<0}
 f(z)-i\sum_{k}{\mathrm{Res}}_{z=z_{f,k}}
 \left[ f(z)\cot\pi z\right] \right\}
 \label{polecase}
\end{equation}
the APF may be generalized for the case when the function 
$f(z)$ has poles $z_{f,k}$, ${\mathrm{Re}}
z_{f,k}>0$, $z_{f,k}\ne 1,2,\ldots $.

As a next consequence of (\ref{th12}) a summation formula 
can be obtained over 
the points $z_n,\, {\mathrm{Re}}z_n>0$ at which the analytic
function $s(z)$ takes integer values, $s(z_n)$ is an integer,
and $s'(z_n)\ne 0$. Taking in (\ref{th12}) 
$g(z)=-if(z)\cot\pi s(z)$ one obtains the following 
formula \cite{Sah}
\begin{equation}
\sum \frac{f(z_n)}{s'(z_n)}=w+\int_{0}^{\infty}{f(x)dx}+
\int_{0}^{\infty }{\left[ \frac{f(ix)}
{e^{-2\pi is(ix)}-1}-\frac{f(-ix)}
{e^{2\pi is(-ix)}-1}\right] dx},
\label{apsf3}
\end{equation}
where
\begin{equation}
w=\left\{ \begin{array}{ll}
0, & {\textrm{if}} \quad s(0)\ne 0,\pm 1,\pm 2,\ldots \\
f(0)/[2s'(0)], & {\textrm{if}} \quad s(0)= 0,\pm 1,\pm 2,\ldots 
\end{array} \right.
\label{wintval}
\end{equation}
For $s(z)=z$ we return to APF in usual form. An example of applications
of this formula to the Casimir effect is given in \cite{Sah}.

\section{Applications to Bessel functions}

\renewcommand{\theequation}{3.\arabic{equation}}

\setcounter{equation}{0}

 The formula (\ref{th12}) contains two meromorphic functions and is 
too general. To obtain more special consequences we have to
specify the one of them. As we have seen in previous section the one of
the possible ways leads to APF. Here we will consider another
choices of the function $g(z)$ and will obtain useful formulae for
the sums over zeros of Bessel function and their combinations, as
well as some formulae for integrals involving these functions.

First of all to simplify the formulae let us introduce the notation
\begin{equation}
\bar F(z)\equiv AF(z)+BzF'(z) \label{efnot1}
\end{equation}
for a given function $F(z)$, where the prime denotes derivative 
with respect
to the argument of function, $A$ and $B$ are constants. As a 
function $g(z)$ in GAPF let us choose
\begin{equation}
g(z)=i\frac{\bar Y_{\nu}(z)}{\bar J_{\nu}(z)}f(z),
\label{gebessel}
\end{equation}
where $J_{\nu}(z)$ and $Y_{\nu}(z)$ are Bessel functions of the
first and second (Neumann function) kind. For the sum and 
difference on the right of (\ref{th12}) one obtains
\begin{equation}
f(z)-(-1)^kg(z)=\frac{\bar H^{(k)}_{\nu}(z)}%
{\bar J_{\nu}(z)}f(z), \quad k=1,2 \label{gefsum}
\end{equation}
with $H^{(1)}_{\nu}$ and $H^{(2)}_{\nu}$ being Bessel functions
of the third kind or Hankel functions. For such a choice the
integrals (\ref{cor11}) and (\ref{th11}) can be estimated by using the
asymptotic formulae for Bessel functions for fixed $\nu$ and 
$\vert z\vert \to \infty$ (see, for example, \cite{Watson, abramowiz}). 
It can be easily seen that conditions (\ref{cor11}) and (\ref{th11})
are satisfied if the function $f(z)$ is restricted by the one of
the following constraints
\begin{equation}
\vert f(z)\vert <\varepsilon (x) e^{c\vert y\vert } \quad \textrm{ or }
\quad \vert f(z)\vert <\frac{Me^{2\vert y\vert }}{\vert z\vert ^\alpha},
\quad z=x+iy,\quad \vert z\vert \to \infty ,
\label{condf}
\end{equation}
where $c<2$, $\alpha >1$ and $\varepsilon (x)\to 0$ for $x\to \infty$.
Indeed, from the asymptotic expressions for Bessel functions
it follows that
\begin{eqnarray}
\left \vert \int_{a\pm ih}^{b\pm ih}{\left [ g(z)\pm f(z)\right ]dz}
\right \vert & = & \left \vert \int_{a}^{b}
{\frac{\bar H_{\nu}^{(1,2)}(x\pm ih)}%
{\bar J_{\nu}(x\pm ih)}f(x\pm ih)dx}\right \vert <\left\{
\begin{array}{l} M_1e^{(c-2)h}\\
M'_1/h^\alpha \end{array} \right.\\
\left\vert \int_{b}^{b\pm i\infty }{\left [ g(z)\pm f(z)\right ]dz}
\right\vert & = & \left\vert \int_{0}^{\infty }{\frac{\bar 
H_{\nu}^{(1,2)}(b\pm ix)}%
{\bar J_{\nu}(b\pm ix)}f(b\pm ix)dx}\right\vert <\left\{
\begin{array}{l} N_1\varepsilon (b)\\
N'_1/b^{\alpha -1} \end{array} \right. \label{condgf}
\end{eqnarray}
with constants $M_1,\, M'_1,\, N_1,\, N'_1$, and $H^{(1)}_\nu $ 
($H^{(2)}_\nu $) corresponds to the upper (lower) sign.

Let us denote by $\lambda _{\nu ,k}\neq 0$, $k=1,2,3\ldots $ 
the zeros of $\bar J_{\nu}(z)$ in the right half-plane, 
arranged in ascending
order of the real part, ${\mathrm{Re}}\lambda _{\nu ,k}
\leq {\mathrm{Re}}\lambda _{\nu ,k+1}$, (if some of these 
zeros lie on the imaginary axis we will take only zeros with positive 
imaginary part). All these zeros are simple. Note that for real
$\nu >-1$ the function $\bar J_{\nu}(z)$ has only real zeros,
except the case $A/B+\nu <0$ when there are two purely imaginary
zeros \cite{Erdelyi, Watson}. By using the
Wronskian $W[J_{\nu}(z), Y_{\nu}(z)]=2/\pi z$ for (\ref{cor13})
one finds
\begin{equation}
R[f(z),g(z)]=2\sum_{k}T_\nu (\lambda _{\nu ,k})
f(\lambda _{\nu ,k})+r_{1\nu}[f(z)],
\label{rbessel}
\end{equation}
where we have introduced the notations
\begin{eqnarray}
T_\nu (z) & = & \frac{z}{\left( z^2-
\nu ^2\right) J_{\nu }^2(z)+z^2
J'^2_{\nu }(z)} \label{teka} \\
r_{1\nu}[f(z)] & = & \pi i
\sum_{k}{\mathrm{Res}}_{{\mathrm{Im}}z_k>0}f(z)
\frac{\bar H^{(1)}_{\nu}(z)}{\bar J_{\nu}(z)}-
\pi i\sum_{k}{\mathrm{Res}}_{{\mathrm{Im}}z_k<0}f(z)
\frac{\bar H^{(2)}_{\nu}(z)}{\bar J_{\nu}(z)}- \nonumber\\
 & & -\pi \sum_{k}{\mathrm{Res}}_{{\mathrm{Im}}z_k=0}f(z)
\frac{\bar Y_{\nu}(z)}{\bar J_{\nu}(z)}. \label{r1}
\end{eqnarray}
Here $z_{k}$ ($\neq \lambda _{\nu ,i}$) are the 
poles for the function $f(z)$ in the region $Rez>a>0$. 
Substituting (\ref{rbessel}) into (\ref{th12}) we
obtain that for the function $f(z)$ meroporphic in the half-plane
${\mathrm{Re}}z\geq a$ and satisfying the 
condition (\ref{condf}) the following
formula takes place
\begin{eqnarray}
& & \lim_{b\to +\infty}\left\{2\sum_{k=m}^{n}T_\nu (\lambda _{\nu ,k})
 f(\lambda _{\nu ,k})
+r_{1\nu}[f(z)]-{\mathrm{p.v.}}\int_{a}^{b}{f(x)dx}\right\}= 
\nonumber\\
& & -\frac{1}{2} \int_{a}^{a+i\infty}{f(z)
 \frac{\bar H^{(1)}_{\nu}(z)}{\bar J_{\nu}(z)}dz}-
 \frac{1}{2}\int_{a}^{a-i\infty}{f(z)
 \frac{\bar H^{(2)}_{\nu}(z)}{\bar J_{\nu}(z)}dz},
 \label{formgen}
\end{eqnarray}
where ${\mathrm{Re}}\lambda _{\nu ,m-1}<a<{\mathrm{Re}}\lambda _{\nu ,m}$, 
${\mathrm{Re}}\lambda _{\nu ,n}<b
<{\mathrm{Re}}\lambda _{\nu ,n+1}$, $a<{\mathrm{Re}}z_k<b$.
We will apply this formula to the function $f(z)$ meromorphic in
the half-plane ${\mathrm{Re}}z\geq 0$ taking $a\to 0$. Let us consider 
separately two cases.

\subsection{Case (a)}

Let $f(z)$ have no poles on the imaginary axis,
except possibly at $z=0$, and
\begin{equation}
f(ze^{\pi i})=-e^{2\nu \pi i}f(z)+o(z^{\beta _{\nu}}),
\quad z\to 0 \label{case21}
\end{equation}
(this condition is trivially satisfied for the function 
$f(z)=o(z^{\beta _{\nu}})$), with
\begin{equation}
\beta _{\nu}=\left\{ \begin{array}{ll}
2\vert {\mathrm{Re}}\nu \vert -1 & \textrm{ for integer $\nu $}\\
{\mathrm{Re}}\nu+\vert {\mathrm{Re}}
\nu \vert -1 & \textrm{ for noninteger $\nu $}
\end{array}\right.
\label{betanju}
\end{equation}
Under this condition for values $\nu $ for which $\bar J_{\nu}(z)$
have no purely imaginary zeros the rhs of Eq.(\ref{formgen}) in the
limit $a\to 0$ can be presented in the form
\begin{equation}
 -\frac{1}{\pi }\int_{\rho }^{\infty}{
 \frac{\bar K_{\nu}(x)}{\bar I_{\nu}(x)}\left[ e^{-\nu \pi i}
 f(xe^{\pi i/2})+e^{\nu \pi i} f(xe^{-\pi i/2})\right]dx}+
 \int_{\gamma _{\rho}^{+}}{f(z)
 \frac{\bar H^{(1)}_{\nu}(z)}{\bar J_{\nu}(z)}dz}-
\int_{\gamma _{\rho}^{-}}{f(z)
 \frac{\bar H^{(2)}_{\nu}(z)}{\bar J_{\nu}(z)}dz},
 \label{rel1}
\end{equation}
with $\gamma _{\rho}^{+}$ and $\gamma _{\rho}^{-}$ being upper and
lower halfs of the semicircle in the right half-plane  with 
radius $\rho$ and with center at point $z=0$, 
described in the positive sense
with respect to this point. In (\ref{rel1}) we have 
introduced modified Bessel functions $I_{\nu}(z)$ and $K_{\nu}(z)$ 
\cite{abramowiz}. It follows from (\ref{case21}) that for $z\to 0$
\begin{equation}
\frac{\bar H^{(1)}_{\nu}(z)}{\bar J_{\nu}(z)}f(z)=
\frac{\bar H^{(2)}_{\nu}(ze^{-\pi i})}{\bar J_{\nu}(ze^{-\pi i})}
f(ze^{-\pi i})+o(z^{-1}). \label{rel2}
\end{equation}
From here for $\rho \to 0$ one finds
\begin{equation}
D_{\nu}\equiv \int_{\gamma _{\rho}^{+}}{f(z)
 \frac{\bar H^{(1)}_{\nu}(z)}{\bar J_{\nu}(z)}dz}-
\int_{\gamma _{\rho}^{-}}{f(z)
 \frac{\bar H^{(2)}_{\nu}(z)}{\bar J_{\nu}(z)}dz}=
 -\pi {\mathrm{Res}}_{z=0}f(z)\frac{\bar Y_{\nu}(z)}{\bar J_{\nu}(z)}
\label{rel3}
\end{equation}
Indeed,
\begin{eqnarray}
D_{\nu} & = & \int_{\gamma _{\rho}^{+}}{f(z)
 \frac{\bar H^{(1)}_{\nu}(z)}{\bar J_{\nu}(z)}dz}+
 \int_{\gamma _{1\rho}^{+}}{f(ze^{-\pi i})
 \frac{\bar H^{(2)}_{\nu}(ze^{-\pi i})}{\bar 
 J_{\nu}(ze^{-\pi i})}dz}=\int_{\gamma _{\rho}^{+}+
 \gamma _{1\rho}^{+}}{f(z)
 \frac{\bar H^{(1)}_{\nu}(z)}{\bar J_{\nu}(z)}dz}+
 {}\nonumber\\
 & & +
 \int_{\gamma _{1\rho}^{+}}{o(z^{-1})dz}=
 i\int_{\gamma _{\rho}^{+}+\gamma _{1\rho}^{+}}{f(z)
 \frac{\bar Y_{\nu}(z)}{\bar J_{\nu}(z)}dz}+
 \int_{\gamma _{1\rho}^{+}}{o(z^{-1})dz}, \label{rel4}
\end{eqnarray}
where $\gamma _{1\rho}^{+}$ ($\gamma _{1\rho}^{-}$, see below) is 
the upper (lower) half of the semicircle 
with radius $\rho $ in the left half-plane 
with center at $z=0$ (described in the positive sense). 
In the last equality we have used the condition that integral 
${\mathrm{p.v.}}\int_{0}^{b}{f(x)dx}$ converges at lower limit. By 
similar way it can be seen that
\begin{equation}
D_{\nu}=i\int_{\gamma _{\rho}^{-}+\gamma _{1\rho}^{-}}{f(z)
 \frac{\bar Y_{\nu}(z)}{\bar J_{\nu}(z)}dz}+
 \int_{\gamma _{1\rho}^{-}}{o(z^{-1})dz}. \label{rel5}
\end{equation}
Combining the last two results we obtain (\ref{rel3}) in the limit
$\rho \to 0$. By using (\ref{formgen}), (\ref{rel1}) and 
(\ref{rel3}) we have \cite{Sah1}:

\bigskip

\noindent {\bf Theorem 2.} {\it If f(z) is a single valued 
analytic function in the half-plane ${\mathrm{Re}}z\geq 0$ 
(with possible branch point at 
$z=0$) except the poles $z_k$ ($\ne \lambda _{\nu ,i}$),
${\mathrm{Re}}z_k>0$ (for the case of function $f(z)$ having 
purely imaginary poles see Remark after Theorem 3), and satisfy 
conditions (\ref{condf}) and (\ref{case21}),
then in the case of $\nu $ for which the function $\bar J_{\nu}(z)$
has no purely imaginary zeros, the following formula is valid
\begin{eqnarray}
 & & {} \lim_{b\to +\infty}\left\{2\sum_{k=1}^{n}T_\nu (\lambda _{\nu ,k})
 f(\lambda _{\nu ,k})
+r_{1\nu}[f(z)]-{\mathrm{p.v.}}\int_{0}^{b}{f(x)dx}\right\}={}\nonumber\\
 & & {} = \frac{\pi}{2} 
 {\mathrm{Res}}_{z=0}f(z)\frac{\bar Y_{\nu}(z)}{\bar J_{\nu}(z)}-
 \frac{1}{\pi }\int_{0}^{\infty}{
 \frac{\bar K_{\nu}(x)}{\bar I_{\nu}(x)}\left[ e^{-\nu \pi i}
 f(xe^{\pi i/2})+e^{\nu \pi i} f(xe^{-\pi i/2})\right]dx},
 \label{sumJ1}
\end{eqnarray}
where on the left ${\mathrm{Re}}\lambda _{\nu ,n}<b
<{\mathrm{Re}}\lambda _{\nu ,n+1}$,
$0<{\mathrm{Re}}z_k<b$, and $\, T_\nu (\lambda _{\nu ,k})$ 
and $\, r_{1\nu}[f(z)]$ are determined
by relations (\ref{teka}) and (\ref{r1}).}

\bigskip

\noindent Under the condition (\ref{case21}) the integral on the right 
converges at lower limit. Recall that we assume the existence of
the integral on the left as well (see section 2).
The formula (\ref{sumJ1}) and analog ones given below
are especially useful for numerical calculations of the 
sums over $\lambda _{\nu ,k}$ as under the first conditions in
(\ref{condf}) the integral on the right converges exponentially
fast at the upper limit.

\bigskip

\noindent {\bf Remark.} Deriving the formula (\ref{sumJ1}) 
we have assumed that the function $f(z)$ is meromorphic 
in the half-plane ${\mathrm{Re}}z\geq 0$ (except possibly at
$z=0$).
However this formula is valid also for some functions having 
branch points on the imaginary axis, for example,
\begin{equation}
f(z)=f_1(z)\prod_{l=1}^{k}\left( z^2+c^2_l\right) ^{\pm 1/2},
\label{fbranch1}
\end{equation} 
with meromorphic function $f_1(z)$. The proof for (\ref{sumJ1}) 
in this case is similar to the given above with difference that branch
points $\pm ic_l$ have to be around on the right along
contours with small radii. In view of the further applications
to the Casimir effect (see below) let us consider the case $k=1$.
By taking into account that
\begin{equation}
\left( z^2+c^2\right) ^{1/2}=\left\{ \begin{array}{lll}
\left\vert z^2+c^2\right\vert ^{1/2} & \textrm{if} 
 & \vert z\vert<c \\
\left\vert z^2+c^2\right\vert ^{1/2}e^{i\pi /2} & 
\textrm{if} & {\mathrm{Im}}z>c \\
\left\vert z^2+c^2\right\vert ^{1/2}e^{-i\pi /2} & 
\textrm{if} & {\mathrm{Im}}z<-c
\end{array}\right.
\label{rel6}
\end{equation}
from (\ref{sumJ1}) one obtains
\begin{eqnarray}
 & & {} \lim_{b\to +\infty}\left\{2\sum_{k=1}^{n}T_\nu (\lambda _{\nu ,k})
 f(\lambda _{\nu ,k})
+r_{1\nu}[f(z)]-{\mathrm{p.v.}}\int_{0}^{b}{f(x)dx}\right\}= \frac{\pi}{2} 
 {\mathrm{Res}}_{z=0}f(z)\frac{\bar Y_{\nu}(z)}{\bar J_{\nu}(z)}-{}
 \nonumber\\
 & & {}-\frac{1}{\pi }\int_{0}^{c}{
 \frac{\bar K_{\nu}(x)}{\bar I_{\nu}(x)}\left[ e^{-\nu \pi i}
 f_1(xe^{\pi i/2})+e^{\nu \pi i} f_1(xe^{-\pi i/2})\right]
 \left( c^2-x^2\right) ^{\pm 1/2}dx}\mp {}\nonumber\\
 & & {} \mp \frac{i}{\pi }\int_{c}^{\infty}{
 \frac{\bar K_{\nu}(x)}{\bar I_{\nu}(x)}\left[ e^{-\nu \pi i}
 f_1(xe^{\pi i/2})-e^{\nu \pi i} f_1(xe^{-\pi i/2})\right]
 \left( x^2-c^2\right) ^{\pm 1/2}dx},
 \label{sumJbranch}
\end{eqnarray}
where $f(z)=f_1(z)\left( z^2+c^2\right) ^{\pm 1/2},\, c>0$.
In Section 11 we apply this formula with analytic function
$f_1(z)$ to derive the expressions for the regularized values
of the energy-momentum tensor components in the region inside
the perfectly conducting cylindrical shell. \rule{1.5ex}{1.5ex}

\bigskip

For an analitic function $f(z)$ the formula (\ref{sumJ1}) yields
\begin{eqnarray}
& &  \sum_{k=1}^{\infty}\frac{2\lambda _{\nu ,k}
f(\lambda _{\nu ,k})}{\left( \lambda _{\nu ,k}^2-
\nu ^2\right)J_{\nu }^2(\lambda _{\nu ,k})+\lambda _{\nu ,k}^2
J'^2_{\nu }(\lambda _{\nu ,k})}=
\int_{0}^{\infty }{f(x)dx}+\frac{\pi}{2} 
 {\mathrm{Res}}_{z=0}f(z)\frac{\bar Y_{\nu}(z)}{\bar J_{\nu}(z)}-{}
 \nonumber\\
 & & {}- \frac{1}{\pi }\int_{0}^{\infty}{
 \frac{\bar K_{\nu}(x)}{\bar I_{\nu}(x)}\left[ e^{-\nu \pi i}
 f(xe^{\pi i/2})+e^{\nu \pi i} f(xe^{-\pi i/2})\right]dx}.
 \label{sumJ1anal}
\end{eqnarray}
By taking in this formula $\nu =1/2$, $A=1, \, B=0$ (see the
notation (\ref{efnot1})) as a particular case we immediately
receive the APF in the form (\ref{apsf2}). In like manner 
substituting $\nu =1/2$, $A=1$, $B=2$ we obtain APF in the form
(\ref{apsf2half}). Consequently the 
formula (\ref{sumJ1}) is a generalization of APF for general
$\nu$ (with restrictions given above) and for functions $f(z)$
having poles in the right half-plane.

Having in mind the further applications to the Casimir effect
in Sections 8 and 11 let us choose in (\ref{sumJ1anal})
\begin{equation}
f(z)=F(z)J_{\nu +m}^2(zt), \quad t>0,\quad {\mathrm{Re}}\nu \geq 0 
\label{casapf1}
\end{equation}
with $m$ being an integer. Now the conditions 
(\ref{condf}) formulated in terms of $F(z)$ are in form
\begin{equation}
\vert F(z)\vert <\vert z\vert \varepsilon e^
{(c-2t)\vert y\vert } \quad \textrm{ or }
\quad \vert f(z)\vert <\frac{Me^{2(1-t)\vert y\vert }}
{\vert z\vert ^{\alpha -1}},
\quad z=x+iy,\quad \vert z\vert \to \infty \label{condFcas}
\end{equation}
with the same notations as in (\ref{condf}).
In like manner from the condition (\ref{case21}) for $F(z)$
one has
\begin{equation}
F(ze^{\pi i})=-F(z)+o(z^{-2m-1}), \quad z\to 0.
\label{case21F}
\end{equation}
Now as a consequence of (\ref{sumJ1anal}) we obtain that if the
conditions (\ref{condFcas}) and (\ref{case21F}) are satisfied,
then for the function $F(z)$ analytic in the right half-plane,
the following formula takes place
\begin{eqnarray}
& &  2\sum_{k=1}^{\infty}T_\nu (\lambda _{\nu ,k})
F(\lambda _{\nu ,k})J^2_{\nu +m}(\lambda _{\nu ,k}t)=
\int_{0}^{\infty }{F(x)J^2_{\nu +m}(xt)dx}-{}
 \nonumber\\
 & & {}- \frac{1}{\pi }\int_{0}^{\infty}{
 \frac{\bar K_{\nu}(x)}{\bar I_{\nu}(x)}I_{\nu +m}^2(xt)
 \left[ F(xe^{\pi i/2})+F(xe^{-\pi i/2})\right] dx}
 \label{sumJ1analcas}
\end{eqnarray}
for ${\mathrm{Re}}\nu \geq 0$ and ${\mathrm{Re}}\nu +m\geq 0$.

\subsection{Case (b)}

Let $f(z)$ be a function satisfying the
condition
\begin{equation}
f(xe^{\pi i/2})=-e^{2\nu \pi i}f(xe^{-\pi i/2})
\label{caseb}
\end{equation}
for real $x$. It is clear that if $f(z)$ have purely imaginary
poles, then they are complex conjugate: $\pm iy_k$, $y_k>0$.
By (\ref{caseb}) the rhs of (\ref{formgen}) for $a\to 0$ 
and ${\mathrm{arg}}\lambda _{\nu ,k}=\pi /2$ may
be written as
\begin{equation}
\left( \int_{\gamma _\rho ^+}+\sum_{\sigma _k=iy_k,
\lambda _{\nu ,k}}\int_{C_\rho (\sigma _k)}\right)
\frac{\bar H^{(1)}_{\nu}(z)}{\bar J_{\nu}(z)}f(z)dz-
\left( \int_{\gamma _\rho ^-}+\sum_{\sigma _k=-iy_k,
-\lambda _{\nu ,k}}\int_{C_\rho (\sigma _k)}\right)
\frac{\bar H^{(2)}_{\nu}(z)}{\bar J_{\nu}(z)}f(z)dz,
\label{rel7}
\end{equation}
where $C_\rho (\sigma _k)$ denotes the right half of the circle
with center at the point $\sigma _k$ and radius $\rho $,
described in the positive sense, and the contours $\gamma _\rho ^\pm$
are the same as in (\ref{rel1}). We have used the fact the purely
imaginary zeros of $\bar J_\nu (z)$ are complex conjugate
numbers, as $\bar J_\nu (ze^{\pi i})=e^{\nu \pi i}\bar J_\nu (z)$.
We have used also the fact that on the right of (\ref{formgen}) the
integrals (with $a=0$) along straight 
segments of the upper and lower imaginary semiaxes 
are canceled, as in accordance of (\ref{caseb})
for ${\mathrm{arg}}z=\pi /2$
\begin{equation}
\frac{\bar H^{(1)}_{\nu}(z)}{\bar J_{\nu}(z)}f(z)=
\frac{\bar H^{(2)}_{\nu}(ze^{-\pi i})}{\bar J_{\nu}
(ze^{-\pi i})}f(ze^{-\pi i}).
\label{rel8}
\end{equation}
Let us show that from (\ref{rel8}) for 
$z_0=x_0e^{\pi i/2}$ it follows that this relation is 
valid for any $z$ in a
small enough region including this point. Namely, as the
function $f(z)\bar H^{(p)}_{\nu}(z)/\bar J_{\nu}(z)$,
$p=1,2$ is meromorphic near the point $(-1)^{p+1}x_0e^{\pi i/2}$,
there exists a neighbourhood of this point where this function
is presented as a Laurent expansion
\begin{equation}
\frac{\bar H^{(p)}_{\nu}(z)}{\bar J_{\nu}(z)}f(z)=
\sum_{n=-n_0}^{\infty }\frac{a_n^{(p)}}{\left[ z-
(-1)^{p+1}x_0e^{\pi i/2}\right]^n}\, .
\label{rel9laur}
\end{equation}
From (\ref{rel8}) for $z=xe^{\pi i/2}$ one concludes
\begin{equation}
\sum_{n=-n_0}^{\infty }\frac{a_n^{(1)}e^{-n\pi i/2}}{\left( x-
x_0\right)^n}= \sum_{n=-n_0}^{\infty }
\frac{(-1)^na_n^{(2)}e^{-n\pi i/2}}{\left( x-
x_0\right)^n}, 
\label{rel10}
\end{equation}
and hence $a_n^{(1)}=(-1)^na_n^{(2)}$. Our statement follows
directly from here. By this it can be seen that
\begin{equation}
\int_{C_\rho (\sigma _k)}{\frac{\bar H^{(1)}_{\nu}(z)}{
\bar J_{\nu}(z)}f(z)dz}-\int_{C_\rho (-\sigma _k)}{
\frac{\bar H^{(2)}_{\nu}(z)}{\bar J_{\nu}(z)}f(z)dz}
=2\pi i{\mathrm{Res}}_{z=\sigma_k}\frac{\bar H^{(1)}_{\nu}(z)}{
\bar J_{\nu}(z)}f(z)
\label{rel11}
\end{equation}
where $\sigma _k=iy_k,\, \lambda _{\nu ,k}$, 
${\mathrm{arg}}\lambda _{\nu ,k}=\pi /2$. Now by taking into account
(\ref{rel3}) and letting $\rho \to 0$ we get \cite{Sah1, Sahdis}:

\bigskip

\noindent {\bf Theorem 3.} {\it Let $f(z)$ be meromorphic function
in the half-plane ${\mathrm{Re}}z\geq 0$ (except possibly at $z=0$)
with poles $z_k,\, {\mathrm{Re}}z_k>0$ and $\pm iy_k,\, y_k>0$, $k=1,2,...$
($\ne \lambda _{\nu ,p}$). If this function satisfy the conditions
(\ref{condf}) and (\ref{caseb}) then
\begin{eqnarray}
& & \lim_{b\to +\infty}\left\{2\sum_{k=1}^{n}T_\nu (\lambda _{\nu ,k})
 f(\lambda _{\nu ,k})
+r_{1\nu}[f(z)]-{\mathrm{p.v.}}\int_{0}^{b}{f(x)dx}\right\}= \nonumber\\
& & =-\frac{\pi i}{2} \sum_{\eta _k=0,iy_k}\left( 2-
 \delta_{0\eta_k}\right)
 {\mathrm{Res}}_{z=\eta _k}f(z)\frac{\bar H^{(1)}_{\nu}(z)}{\bar J_{\nu}(z)},
 \label{sumJ2}
\end{eqnarray}
where on the left $0<{\mathrm{Re}}z_k<b$, 
${\mathrm{Re}}\lambda _{\nu ,n}<b<{\mathrm{Re}}
\lambda _{\nu ,n+1}$ and $r_{1\nu }$ is defined by (\ref{r1}).}

\bigskip

\noindent Note that the residue terms in (\ref{rel11}) with 
$\sigma _k=\lambda _{\nu ,k}$, ${\mathrm{arg}}\lambda _{\nu ,k}=\pi /2$ 
are equal to $4T_\nu (\lambda _{\nu ,k})f(\lambda _{\nu ,k})$ and 
are included in the first sum on the left of (\ref{sumJ2}).

\bigskip

\noindent {\bf Remark.} Let $\pm iy_k,\, y_k>0$ and 
$\pm \lambda _{\nu ,k}$, ${\mathrm{arg}}\lambda _{\nu ,k}=\pi /2$ 
are purely imaginary poles of function $f(z)$ and purely 
imaginary zeros of $\bar J_\nu (z)$, correspondingly. Let function
$f(z)$ satisfy condition
\begin{equation}
f(z)=-e^{2\nu \pi i}f(ze^{-\pi i})+o\left( (z-\sigma _k)^{-1}
\right), \quad z\to \sigma _k,\quad \sigma _k=iy_k, 
\lambda _{\nu ,k}. \label{caseaplusb}
\end{equation}
Now in the limit $a\to 0$ the rhs of (\ref{formgen}) can be 
presented in the form (\ref{rel7}) plus integrals along the 
straight segments of the imaginary axis between the poles. Using 
the arguments similar those given above we obtain the relation
(\ref{rel11}) with additional contribution from the last term
on the right of (\ref{caseaplusb}) in the form 
$\int_{C_\rho (-\sigma _k)}{o\left( (z-\sigma _k)^{-1}\right) }dz$.
In the limit $\rho \to 0$ the latter vanishes and sum of the 
integrals along the straight segments of the imaginary axis gives 
the principal value of the integral on the right of (\ref{sumJ1}).
As a result the formula (\ref{sumJ1}) can be generalized for
functions having purely imaginary poles and satisfying 
condition (\ref{caseaplusb}) writing instead of residue 
term on the right the sum of residues from the right of 
(\ref{sumJ2}) and taking the principle value of the integral
on the right. The latter exists due to the condition 
(\ref{caseaplusb}). \rule{1.5ex}{1.5ex}

\bigskip

It follows from (\ref{sumJ2}) an interesting result. Let 
$\lambda _{\mu ,k}^{(1)}$ be zeros of the function
$A_1J_\mu (z)+B_1zJ'_\mu (z)$ with some real constants $A_1$
and $B_1$. Let $f(z)$ be an analytic function in the right 
half-plane satisfying condition (\ref{caseb}) and $f(z)=
o(z^\beta )$ for $z\to 0$, where $\beta ={\mathrm{max}}(\beta _\mu ,
\beta _\nu )$ (the definition $\beta _\nu $ see (\ref{betanju})).
For this function from (\ref{sumJ2}) we get
\begin{equation}
\sum_{k=1}^{\infty }T_{\mu }(\lambda ^{(1)}_{\mu ,k}) 
f(\lambda ^{(1)}_{\mu ,k})
=\sum_{k=1}^{\infty }T_\nu (\lambda _{\nu ,k}) f(\lambda _{\nu ,k}), 
\quad \mu =\nu +m \, .
\label{equalsumJ}
\end{equation}
 For the case of Fourier-Bessel and Dini series this result 
is given in \cite{Watson}.

Let us consider some applications of the formula (\ref{sumJ2})
to the special types of series. Firstly we choose in this formula
\begin{equation}
f(z)=F_1(z)J_\mu (zt),\quad t>0, \label{examp1}
\end{equation}
where the function $F_2(z)$ is meromorphic on the right half-plane
and satisfy conditions
\begin{equation}
\vert F_1(z)\vert <\varepsilon _1(x)e^{(c-t)\vert y\vert }
\quad \textrm{or} \quad 
\vert F_1(z)\vert <M\vert z\vert ^{-\alpha _1}e^{(2-t)\vert y\vert },
\quad \vert z\vert \to \infty,
 \label{condex1}
\end{equation}
with $c<2,\, \alpha _1>1/2$, $\varepsilon _1(x)=o(\sqrt{x})$ for
$x\to +\infty $, and condition 
\begin{equation}
F_1(xe^{\pi i/2})=-e^{(2\nu -\mu )\pi i}F_1(xe^{-\pi i/2}).
\label{condex12}
\end{equation}
From (\ref{condex1}) it follows that the integral 
${\mathrm{p.v.}}\int_{0}^{\infty }{F_1(x)J_\mu (xt)dx}$
converges at the upper limit and hence in this case the formula
(\ref{sumJ2}) may be written in the form
\begin{eqnarray}
 & & {} \sum_{k=1}^{\infty }T_\nu (\lambda _{\nu ,k})F_1(\lambda _{\nu ,k})
 J_\mu (\lambda _{\nu ,k}t)=
\frac{1}{2} {\mathrm{p.v.}}\int_{0}^{\infty }{F_1(x)J_\mu (xt)dx}-
\frac{1}{2}r_{1\nu }\left[ F_1(z)J_\mu (zt)\right]- {}
 \nonumber\\
 & & {}-\frac{\pi i}{4} \sum_{\eta _k=0,iy_k}\left( 2-
 \delta_{0\eta_k}\right)
 {\mathrm{Res}}_{z=\eta _k}F_2(z)J_{\mu }(zt)
 \frac{\bar H^{(1)}_{\nu}(z)}{\bar J_{\nu }(z)},
 \label{sumJ3}
\end{eqnarray}
For example, it follows from here that for $t<1$, 
${\mathrm{Re}}\sigma ,\, {\mathrm{Re}}\nu  >-1$
\begin{equation}
\sum_{k=1}^{\infty }\frac{T_\nu (\lambda _{\nu ,k})}
{\lambda _{\nu ,k}^\sigma }
J_{\sigma +\nu +1}(\lambda _{\nu ,k}) J_\nu (\lambda _{\nu ,k}t)=
\frac{1}{2}\int_{0}^{\infty }{J_{\sigma +\nu +1}(z) J_\nu (zt)
\frac{dz}{z^\sigma }}=\frac{\left( 1-t^2\right) ^\sigma  
t^\nu }{2^{\sigma +1}\Gamma (\sigma +1)} \label{examp1ap}
\end{equation}
(for the value of integral see, e.g., \cite{Watson}). For $B=0$
this result is given in \cite{Prudnikov}. In a similar manner
taking $\mu =\nu +m$,
\begin{equation}
F_1(z)=z^{\nu +m+1}\frac{J_\sigma (a\sqrt{z^2+z_1^2})}{\left( z^2+
z_1^2\right) ^{\sigma /2}}, \quad a>0
\label{examp2apF2}
\end{equation}
with ${\mathrm{Re}}\nu \geq 0$ and ${\mathrm{Re}}\nu +m\geq 0$,
from (\ref{sumJ3}) for $a<2-t$, ${\mathrm{Re}}\sigma >
{\mathrm{Re}}\nu +m$ one finds
\begin{eqnarray}
 & & \sum_{k=1}^{\infty }T_\nu (\lambda _{\nu ,k})
\lambda _{\nu ,k}^{\nu +m+1}J_{\nu +m}(\lambda _{\nu ,k}t)
\frac{J_\sigma (a\sqrt{\lambda _{\nu ,k}^2+z_1^2})}
{\left( \lambda _{\nu ,k}^2+z_1^2\right) ^{\sigma /2}}=\label{examp2ap}\\
 & & =\frac{1}{2}\int_{0}^{\infty }{x^{\nu +m+1}J_{\nu +m}(xt)
 \frac{J_\sigma (a\sqrt{x^2+z_1^2})}{\left( x^2+z_1^2
\right) ^{\sigma /2}}
 dx}= \frac{t^{\nu +1}}{a^\sigma }\left( -z_1\right) ^{m+1}
\frac{J_{m+1}(z_1\sqrt{a^2-t^2})}{\left( a^2-t^2\right) ^{(m+1)/2}}
 , \, a>t
\nonumber
\end{eqnarray}
and the sum is zero when $a<t$. Here we have used
the known value for Sonine integral \cite{Watson}.

If an addition to (\ref{condex1}), (\ref{condex12}) the 
function $F_1$ satisfies conditions
\begin{equation}
F_1(xe^{\pi i/2})=-e^{\mu \pi i}F_1(xe^{-\pi i/2})
\label{condex13}
\end{equation}
and
\begin{equation}
\vert F_2(z)\vert <\varepsilon _1(x)e^{c_1t\vert y\vert }
\quad \textrm{or} \quad 
\vert F_2(z)\vert <M\vert z\vert ^{-\alpha _1}e^{t\vert y\vert },
\quad \vert z\vert \to \infty,
 \label{condex1nor}
\end{equation}
when the formula (\ref{intJform42})(see below) with $B=0$ may be applied to
the integral on the right of (\ref{sumJ3}). This gives \cite{Sah1, Sahdis}

\bigskip

\noindent {\bf Corollary 1.} {\it Let $F(z)$ be meromorphic function in
the half-plane ${\mathrm{Re}}z\geq 0$ (except possibly at $z=0$) 
with poles $z_k,\, {\mathrm{Re}}z_k>0$ and 
$\pm iy_k,\, y_k>0$ ($\ne \lambda _{\nu ,i}$).
If $F(z)$ satisfy condition
\begin{equation}
F(xe^{\pi i/2})=(-1)^{m+1}e^{\nu \pi i}F(xe^{-\pi i/2})
\label{cor1cond1}
\end{equation}
with an integer $m$, and to one of inequalities
\begin{equation}
\vert F(z)\vert <\varepsilon _1(x)e^{a\vert y\vert }
\quad \textrm{or} \quad 
\vert F(z)\vert <M\vert z\vert ^{-\alpha _1}e^{a_0\vert y\vert },
\quad \vert z\vert \to \infty,
 \label{cor1cond2}
\end{equation}
with $a<{\mathrm{min}}(t,2-t)\equiv a_0$, $\varepsilon _1(x)=
o(x^{1/2}),\, x\to 
+\infty $, $\alpha _1>1/2$, the following formula is valid
\begin{eqnarray}
& &  \sum_{k=1}^{\infty }T_\nu (\lambda _{\nu ,k})F(\lambda _{\nu ,k})
 J_{\nu +m}(\lambda _{\nu ,k}t)= \nonumber\\
& & =\frac{\pi i}{4} \sum_{\eta _k=0,iy_k,z_k}\left( 2-
 \delta_{0\eta_k}\right)
 {\mathrm{Res}}_{z=\eta _k}\left\{ \left[ J_{\nu +m}(zt)\bar Y_\nu (z)-
Y_{\nu +m}(zt)\bar J_\nu (z)\right] 
 \frac{F(z)}{\bar J_{\nu }(z)}\right\} .
 \label{cor1form}
\end{eqnarray}
}

\bigskip

\noindent Recall that for the imaginary zeros $\lambda _{\nu ,k}$, 
in lhs of (\ref{cor1form}) the zeros with positive imaginary parts 
enter only. By using the formula (\ref{cor1form}) a number of 
Furier-Bessel and Dini series can be summarized (see, for instance,
below).

\bigskip

\noindent {\bf Remark.} The formula (\ref{cor1form}) may be obtained 
also by considering the integral
\begin{equation}
\frac{1}{\pi }\int_{C_h}{ \left[ 
H^{(2)}_{\nu +m}(zt)\bar H^{(1)}_\nu (z)-
H^{(1)}_{\nu +m}(zt)\bar H^{(2)}_\nu (z)\right] 
 \frac{F(z)}{\bar J_{\nu }(z)} dz},
 \label{contint}
\end{equation}
where $C_h$ is an rectangle with vertices $\pm ih,\, b\pm ih$, described 
in the positive sense (purely imaginary poles of 
$F(z)/\bar J_{\nu }(z)$ and the origin are around by 
semicircles in the right half-plane with small radii). This integral
is equal to the sum of residues over the poles within
$C_h$ (points $z_k$, $\lambda _{\nu ,k}$, ($Rez_k,\, 
{\mathrm{Re}}\lambda _{\nu ,k}>0$)). On the other hand it follows from 
(\ref{cor1cond1}) that integrals along the segments of the imaginary
axes cancel each other. The sum of integrals along the conjugate
semicircles give the sum of residues over purely imaginary poles in the
upper half plane. The integrals along the remained three segments
of $C_h$ in accordance with (\ref{cor1cond2}) approach to zero in
the limit $b,\, h \to \infty $. Equating these expressions for 
(\ref{contint}) one immediately obtains the result (\ref{cor1form}).
\rule{1.5ex}{1.5ex}

\bigskip

From (\ref{cor1form}) for $t=1,\, F(z)=J_\nu (zx)$, $m=1$ one
obtains \cite{Erdelyi, Watson}
\begin{equation}
 \sum_{k=1}^{\infty }T_\nu (\lambda _{\nu ,k}) J_{\nu }(\lambda _{\nu ,k}x)
J_{\nu +1}(\lambda _{\nu ,k}) =
 \frac{x^\nu }{2}, \quad 0\leq x<1.
\label{examp2}
\end{equation}
By similar way choosing $m=0$, $F(z)=zJ_\nu (zx)/\left( z^2-
a^2\right)$, $B=0$ we obtain the Kneser-Sommerfeld expansion
\cite{Watson}:
\begin{equation}
 \sum_{k=1}^{\infty }\frac{J_{\nu }(\lambda _{\nu ,k}t)
 J_{\nu }(\lambda _{\nu ,k}x)}{\left(\lambda _{\nu ,k}^2-a^2
 \right) J_{\nu +1}^2(\lambda _{\nu ,k})}=
\frac{\pi }{4}\frac{J_{\nu }(ax)}{J_{\nu }(a)}\left[ J_{\nu }(at)
Y_\nu (a)-Y_{\nu }(at)J_\nu (a)\right],\quad 0\leq x\leq t\leq 1.
 \label{examp3}
\end{equation}
In (\ref{cor1form}) as a function $F(z)$ one may choose, for 
example, the following functions
\begin{eqnarray}
& & \! \! \! z^{\rho -1}\prod_{l=1}^{n}
\left( z^2+z_l^2\right) ^{-\mu _{l}/2}
J_{\mu _l}(b_l\sqrt{z^2+z_l^2}),{} \label{func1}\\
& & \qquad \textrm{for}\quad {\mathrm{Re}}\nu<\sum_{l=1}^{n}
{\mathrm{Re}}\mu _l+n/2+2p+3/2-m
-\delta _{ba_0},\quad b\leq a_0,\, b=\sum_{l=1}^{n}b_l;{}
\nonumber\\
& & \! \! \! z^{\rho -2n-1}\prod_{l=1}^{n}\left[ 1-J_0(b_lz)\right] ,{}
\label{func2}\\
& & \qquad \textrm{for}\quad {\mathrm{Re}}\nu<2n+2p+3/2-m
-\delta _{ba_0};{}\nonumber\\
& & \! \! \! z^{\rho -1}\prod_{l=1}^{n}
\left( z^2+z_l^2\right) ^{\mu _{l}/2}
Y_{\mu _l}\left( b_l\sqrt{z^2+z_l^2}\right) ,\, \mu _l>0 \, 
\textrm{-half of an odd integer,}{} \label{func3}\\
& & \qquad \textrm{for}\quad {\mathrm{Re}}\nu<-\sum_{l=1}^{n}
\mu _l+n/2+2p+3/2-m
-\delta _{ba_0}; {}\nonumber\\
& & \! \! \! z^{\rho -1}\prod_{l=1}^{n}z^{\vert k_l\vert }
\left[ J_{\mu _l+k_l}(a_lz)Y_{\mu _l}(b_lz)-
Y_{\mu _l+k_l}(a_lz)J_{\mu _l}(b_lz)\right],\quad k_l \, \textrm{- 
 integer,}{} \label{func4}\\
& & \qquad \textrm{for}\quad {\mathrm{Re}}
\nu<n+2p+3/2-m-\sum \vert k_l\vert
-\delta _{\tilde{a},a_0},\quad \tilde{a}\equiv \sum_{l=1}^{n}
\vert a_l-b_l\vert \leq a_0;\nonumber 
\end{eqnarray}
with $\rho =\nu +m-2p$ ($p$ - integer), as well as any products
between these functions and with $\prod_{l} \left( z^2-
c_l^2\right) ^{-p_{l}}$, provided the condition (\ref{cor1cond2})
is satisfied. For example, the following formulae take place
\begin{eqnarray}
& & \! \! \sum_{k=1}^{\infty }j_{\nu ,k}^{\nu -2}
\frac{J_\nu (j_{\nu ,k}t)}
{J_{\nu +1}^2(j_{\nu ,k})}
\prod_{l=1}^{n}\left[ J_{\mu _l}(a_lj_{\nu ,k})Y_{\mu _l}(b_l
j_{\nu ,k})-
Y_{\mu _l}(a_lj_{\nu ,k})J_{\mu _l}(b_lj_{\nu ,k})\right]={}
\nonumber\\
& & \qquad =\frac{2^{\nu -2}}{\pi ^nt^\nu }\left( 1-t^{2\nu }\right)
\prod_{l=1}^{n}\frac{b_l^{\mu _l}}{\mu _la_l^{\mu _l}}
\left[ 1-\left( \frac{a_l}{b_l}\right)^{2\mu _l}\right] ,
\quad 0<t\leq 1,
\label{examp4}\\
& &  \qquad c\equiv \sum_{l=1}^{n}
\vert a_l-b_l\vert \leq t, \quad a_l,b_l>0,
\quad {\mathrm{Re}}\mu _l\geq 0,\quad 
{\mathrm{Re}}\nu <n+3/2-\delta_{ct}; \nonumber\\
& & \! \! \sum_{k=1}^{\infty }\frac{J_\nu (j_{\nu ,k}t)
J_{\nu +1}(\lambda j_{\nu ,k})}
{j_{\nu ,k}^{2n+3}J_{\nu +1}^2(j_{\nu ,k})}
\prod_{l=1}^{n}\left[ 1-J_0(b_lj_{\nu ,k})\right]=
\frac{\lambda ^{\nu +1}\left( 1-t^{2\nu }\right)}{4^{n+1}
\nu (\nu +1)t^\nu }\prod_{l=1}^{n}b_l^2, \label{examp5}\\
& & \qquad \lambda +\sum_{l=1}^{n}b_l\leq t\leq 1,\quad \lambda ,b_l>0;
\nonumber\\
& & \! \! \sum_{k=1}^{\infty }\frac{J_\mu (j_{\nu ,k}b)
J_{\nu +1}(\lambda j_{\nu ,k})J_\nu (j_{\nu ,k}t)}
{\left(j_{\nu ,k}^2 -a^2\right) j_{\nu ,k}^{\mu +1}
J_{\nu +1}^2(j_{\nu ,k})}=
\frac{\pi J_{\nu +1}(a\lambda )}{4a^{\mu +1}}\frac{J_\mu (ba)}{J_\nu (a)}
\left[ Y_\nu (a)J_\nu (at)-
J_\nu (a)Y_\nu (at)\right],\label{examp6}\\
& & \qquad \lambda +b\leq t\leq 1, \, \lambda ,b>0,\, {\mathrm{Re}}\mu >-7/2+
\delta _{\lambda +b,t},\nonumber
\end{eqnarray}
where $j_{\nu ,k}$ are zeros of $J_\nu (z)$.

The examples of the series over zeros of Bessel functions we found in
literature (see, e.g., \cite{Erdelyi, Watson, Prudnikov, Magnus}), 
when the corresponding sum was evaluated in finite terms,
are particular cases of the formulae given in this section. 

\section{Summation formulae over zeros of
 $\bar J_\nu (z)\bar Y_\nu (\lambda z)-
 \bar Y_\nu (z)\bar J_\nu (\lambda z)$}

\renewcommand{\theequation}{4.\arabic{equation}} 
\setcounter{equation}{0} 
 
In this section we will consider the series over
zeros of the function
\begin{equation}
C^{AB}_{\nu }(\lambda ,z)\equiv \bar J_\nu (z)\bar Y_\nu (\lambda z)-
\bar Y_\nu (z)\bar J_\nu (\lambda z), \label{bescomb1}
\end{equation} 
where the bared quantities are defined as (\ref{efnot1}).
Series of this type arise in calculations of the vacuum
expectation values in confined regions with boundaries of 
spherical and cylindrical form. To obtain a summation formula 
for these series let us substitute in (\ref{th12})
\begin{equation}
g(z)=\frac{1}{2i}\left[ \frac{\bar H_\nu ^{(1)}(\lambda z)}{
\bar H_\nu ^{(1)}(z)}+\frac{\bar H_\nu ^{(2)}(\lambda z)}{
\bar H_\nu ^{(2)}(z)}\right]\frac{h(z)}{C^{AB}_{\nu }(\lambda ,z)},
\quad f(z)=\frac{h(z)}{\bar H_\nu ^{(1)}(z)\bar H_\nu ^{(2)}(z)},
\label{gefcomb}
\end{equation}
where for definiteness we shall assume that $\lambda >1$. 
The sum and difference of these functions are
\begin{equation}
g(z)-(-1)^kf(z)=-i\frac{\bar H_\nu ^{(k)}(\lambda z)}
{\bar H_\nu ^{(k)}(z)}
\frac{h(z)}{C^{AB}_{\nu }(\lambda ,z)},\quad k=1,2.
\label{gefsumnew}
\end{equation}
The conditions for GAPF written in terms of the function
$h(z)$ are as follows
\begin{equation}
\vert h(z)\vert <\varepsilon _2(x)e^{c_2\vert y\vert }
\quad \textrm{or} \quad 
\vert h(z)\vert <M\vert z\vert ^{-\alpha _2}e^{2(\lambda -1)
\vert y\vert },\quad \vert z\vert \to \infty,
\quad z=x+iy \label{cond31}
\end{equation}
where $c_2<2(\lambda -1)$, $x^{2\delta _{B0}-1}\varepsilon _2(x)\to 0$ 
for $x\to +\infty $, $\alpha _2>2\delta _{B0}$. 
Let $\gamma _{\nu ,k}$ be zeros for the function 
$C^{AB}_{\nu }(\lambda ,z)$ in the right half-plane. In this section 
we will assume values of $\nu $, $A,\, B$ for which all these zeros  
are real and simple, and the function $\bar H_\nu ^{(1)}(z)$ 
($\bar H_\nu ^{(2)}(z)$) has no
zeros in the right half of the upper (lower) half-plane.
As we will see later these conditions are satisfied in
physical problems considered below. For real $\nu $, 
$A$, $B$ the zeros $\gamma _{\nu ,k}$ are simple. To see this note
that the function $J_\nu (tz)\bar Y_\nu (z)-Y_\nu (tz)\bar J_\nu (z)$
is cylinder function with respect to $t$. Using the standard result
for indefinite integrals containing any two cylinder functions 
(see \cite{Watson, abramowiz}) it can be seen that
\begin{equation}
\int_{1}^{\lambda }{t\left[ J_\nu (tz)\bar Y_\nu (z)-
Y_\nu (tz)\bar J_\nu (z)\right] ^2dt}=\frac{2}{\pi ^2zT^{AB}_\nu (
\lambda ,z)}\, ,\quad z=\gamma _{\nu ,k},
\label{tekapositive}
\end{equation}
where we have introduced the notation
\begin{equation}
T_\nu ^{AB}(\lambda ,z)=z\left\{ \frac{\bar J_\nu ^2(z)}
{\bar J_\nu ^2(\lambda z)}\left[ A^2+B^2(\lambda ^2z^2-\nu ^2)
\right] -A^2-B^2(z^2-\nu ^2)\right\} ^{-1}. 
\label{tekaAB}
\end{equation}
On the other hand
\begin{equation}
\frac{\partial }{\partial z}C^{AB}_{\nu }(\lambda ,z)=
\frac{2}{\pi T_\nu ^{AB}(\lambda ,z)}\frac{\bar J_\nu (\lambda z)}
{\bar J_\nu (z)}\, ,\quad z=\gamma _{\nu ,k}.
\label{CABderivative}
\end{equation}
Combining the last two results we deduce that for real $\nu $, 
$A$, $B$ the derivative (\ref{CABderivative}) is nonzero and 
hence the zeros $z=\gamma _{\nu ,k}$ are simple.
By using this it can be seen that
\begin{equation}
{\mathrm{Res}}_{z=\gamma _{\nu ,k}}g(z)=\frac{\pi }{2i}
T_\nu ^{AB}(\lambda ,\gamma _{\nu ,k}).
 \label{rel31}
\end{equation}
Hence if the function $h(z)$ is analytic in the half-plane 
${\mathrm{Re}}z\geq a>0$ except at the 
poles $z_k$ ($\ne \gamma _{\nu ,i}$) and
satisfy to the one of two conditions (\ref{cond31}), the following
formula takes place
\begin{eqnarray}
& & \lim_{b\to +\infty}\left\{ \frac{\pi ^2}{2}\sum_{k=n}^{m}
T_\nu ^{AB}(\lambda ,\gamma _{\nu ,k})h(\gamma _{\nu ,k})
+r_{2\nu }[h(z)]
-{\mathrm{p.v.}}\int_{a}^{b}{\frac{h(x)dx}{\bar J_\nu ^2(x)+
\bar Y_\nu ^2(kx)}}\right\} =
\nonumber\\
& & =\frac{i}{2}\int_{a}^{a+i\infty }{\frac{\bar H_\nu ^{(1)}
(\lambda z)}{\bar H_\nu ^{(1)}(z)}\frac{h(z)}{C^{AB}_{\nu }(\lambda ,z)}
dz}-\frac{i}{2}\int_{a}^{a-i\infty }{\frac{\bar H_\nu ^{(2)}
(\lambda z)}{\bar H_\nu ^{(2)}(z)}\frac{h(z)}{C^{AB}_{\nu }(\lambda ,z)}
dz}.
\label{gapsfcomb}
\end{eqnarray}
Here we assumed that the integral on the left exists, $\gamma _{\nu ,
n-1}<a<\gamma _{\nu ,n}$, $\gamma _{\nu ,m}<b<\gamma _{\nu ,m+1}$,
$a<{\mathrm{Re}}z_k<b$, ${\mathrm{Re}}z_k
\leq {\mathrm{Re}}z_{k+1}$, and the following notation is
introduced
\begin{eqnarray}
r_{2\nu }[h(z)] & = & \pi \sum_{k}{\mathrm{Res}}_
{{\mathrm{Im}}z_k=0}\left[ \frac{\bar J_\nu (z)
\bar J_\nu (\lambda z)+\bar Y_\nu (z)\bar Y_\nu (\lambda z)}
{\bar J_\nu ^2(z)+\bar Y_\nu ^2(z)}
\frac{h(z)}{C^{AB}_{\nu }(\lambda ,z)}\right] +{}\nonumber\\
& & + \pi \sum_{k,l=1,2}
{\mathrm{Res}}_{(-1)^l{\mathrm{Im}}z_k<0}
\left[ \frac{\bar H_\nu ^{(l)}(\lambda z)}{\bar H_\nu ^{(l)}(z)}
\frac{h(z)}{C^{AB}_{\nu }(\lambda ,z)}\right] . \label{r3}
\end{eqnarray}
The general formula (\ref{gapsfcomb}) is a direct consequence of
GAPF and will be as starting point for the further applications 
in this section. In the limit $a\to 0$ one has \cite{Sah1, Sahdis}:

\bigskip

\noindent {\bf Corollary 2.} {\it Let $h(z)$ be analytic function
for ${\mathrm{Re}}z\geq 0$ except the poles $z_k$ ($\ne \gamma _{\nu i}$),
${\mathrm{Re}}z_k>0$ (with possible branch point $z=0$). If it satisfies
one of two conditions (\ref{cond31}) and
\begin{equation}
h(ze^{\pi i})=-h(z)+o(z^{-1}), \quad z\to 0, \label{cor3cond1}
\end{equation}
and the integral
\begin{equation}
{\mathrm{p.v.}}\int_{a}^{b}{\frac{h(x)dx}{\bar J_\nu ^2(x)+
\bar Y_\nu ^2(x)}}
\label{cor2cond2}
\end{equation}
exists, then
\begin{eqnarray}
& & \lim_{b\to +\infty}\left\{ \frac{\pi ^2}{2}\sum_{k=1}^{m}
h(\gamma _{\nu ,k})T_\nu ^{AB}(\lambda ,\gamma _{\nu ,k})
+r_{3\nu }[h(z)]
-{\mathrm{p.v.}}\int_{0}^{b}{\frac{h(x)dx}{\bar J_\nu ^2(x)+
\bar Y_\nu ^2(x)}}\right\}=
\nonumber\\
& & =-\frac{\pi }{2}
{\mathrm{Res}}_{z=0}\left[ \frac{h(z)
\bar H_\nu ^{(1)}(\lambda z)}{C^{AB}_{\nu }(\lambda ,z)
\bar H_\nu ^{(1)}(z)}
\right]-\frac{\pi }{4}
\int_{0}^{\infty }{\frac{\bar K_\nu (\lambda x)}
{\bar K_\nu (x)}\frac{\left[ h(
xe^{\pi i/2})+h(xe^{-\pi i/2})\right] dx}
{\bar K_\nu (x)\bar I_\nu (\lambda x)-
\bar K_\nu (\lambda x)\bar I_\nu (x)}} 
 \label{cor3form}
\end{eqnarray}
}

\bigskip 

\noindent
In the following we shall use this formula to derive the regularized
vacuum energy momentum-tensor for the region between two 
spherical and cylindrical surfaces. Note that (\ref{cor3form}) may 
be generalized for the functions $h(z)$ with purely imaginary poles
$\pm iy_k$, $y_k>0$ satisfying condition
\begin{equation}
h(ze^{\pi i})=-h(z)+o\left( (z\mp iy_k)^{-1}\right) ,
\quad z\to \pm iy_k.
\label{cor3cond1plus2}
\end{equation}
The corresponding formula is obtained from (\ref{cor3form}) by adding
residue terms for $z=iy_k$ in the form of (\ref{cor2form}) (see 
below) and taking the principal value of the integral on the right.
The arguments here are similar to those for Remark after Theorem 3.

By the way similar to (\ref{sumJ1}) one has another result 
\cite{Sah1, Sahdis}:

\bigskip

\noindent {\bf Corollary 3.} {\it Let $h(z)$ be meromorphic function
in the half-plane ${\mathrm{Re}}z\geq 0$ (with exception 
the possible branch 
point $z=0$) with poles $z_k,\, \pm iy_k$ ($\ne \gamma _{\nu ,i}$),
${\mathrm{Re}}z_k,y_k>0$. If this function satisfy condition
\begin{equation}
h(xe^{\pi i/2})=-h(xe^{-\pi i/2}) \label{cor2cond1}
\end{equation}
and the integral (\ref{cor2cond2}) exists then
\begin{eqnarray}
& & \lim_{b\to +\infty}\left\{ \frac{\pi ^2}{2}\sum_{k=1}^{m}
h(\gamma _{\nu ,k})T_\nu ^{AB}(\lambda ,\gamma _{\nu ,k})
+r_{2\nu }[h(z)]
-{\mathrm{p.v.}}\int_{0}^{b}{\frac{h(x)dx}{\bar J_\nu ^2(x)+
\bar Y_\nu ^2(x)}}\right\} ={}
\nonumber\\
& & =-\frac{\pi }{2}\sum_{\eta _k=0,iy_k}\left( 2-
\delta _{0\eta_k}\right)
{\mathrm{Res}}_{z=\eta _k}\left[ \frac{\bar H_\nu ^{(1)}(\lambda z)}
{\bar H_\nu ^{(1)}(z)}
\frac{h(z)}{C^{AB}_{\nu }(\lambda ,z)}\right] ,
 \label{cor2form}
\end{eqnarray}
where in the lhs $\gamma _{\nu ,m}<b<\gamma _{\nu ,m+1}$.
}

\bigskip

Let us consider  a special applications of the formula 
(\ref{cor2form}) for $A=1,\, B=0$. The generalizations
of these results for general $A,B$ under the conditions
given above are straightforward.

\bigskip

\noindent {\bf Theorem 4.} {\it Let the function $F(z)$ be meromorphic 
in the right half-plane ${\mathrm{Re}}z\geq 0$ (with the possible exception
$z=0$) with poles $z_k,\pm iy_k$ ($\ne \gamma _{\nu ,i}$),
$y_k,{\mathrm{Re}}z_k>0$. If it satisfy condition
\begin{equation}
F(xe^{\pi i/2})=(-1)^{m+1}F(xe^{-\pi i/2}), 
\label{th4cond1}
\end{equation}
with an integer $m$, and to the one of two inequalities
\begin{equation}
\vert F(z)\vert <\varepsilon (x)e^{a_1\vert y\vert }
\quad \textrm{or} \quad 
\vert F(z)\vert <M\vert z\vert ^{-\alpha }e^{a_2\vert y\vert },
\quad \vert z\vert \to \infty,
 \label{th4cond2}
\end{equation}
with $a_1<{\mathrm{min}}(2\lambda -\sigma -1,\sigma -1)\equiv a_2$,  
$\sigma >0$, $\varepsilon (x)\to 0$ for $x\to +\infty $, 
$\alpha >1$, then
\begin{eqnarray}
& & \sum_{k=1}^{\infty}
\frac{\gamma _{\nu ,k}F(\gamma _{\nu ,k})}{J_\nu ^2(\gamma _{\nu ,
k})/J_\nu ^2(\lambda \gamma _{\nu ,k})-1}\left[ J_\nu (
\gamma _{\nu ,k})Y_{\nu +m}(\sigma \gamma _{\nu ,k})-
Y_\nu (\gamma _{\nu ,k})J_{\nu +m}(\sigma \gamma _{\nu ,k})\right] ={}
\nonumber\\
& & =\frac{1}{\pi }\sum_{\eta _k=0,iy_k,z_k}\left( 2-
\delta _{0\eta_k}\right)
{\mathrm{Res}}_{z=\eta _k}\frac{Y_\nu (\lambda z)J_{\nu +m}(\sigma z)-
J_\nu (\lambda z)Y_{\nu +m}(\sigma z)}{J_\nu (z)Y_{\nu }(\lambda z)-
J_\nu (\lambda z)Y_{\nu }(z)}F(z).
 \label{th4form}
\end{eqnarray}
}

\bigskip

\noindent {\bf Proof.} As a function $h(z)$ in (\ref{cor2form})
let us choose
\begin{equation}
h(z)=F(z)\left[ J_\nu (z)Y_{\nu +m}(\sigma z)-
Y_\nu (z)J_{\nu +m}(\sigma z)\right] , \label{rel32}
\end{equation} 
which in virtue of (\ref{th4cond2}) satisfy condition (\ref{cond31}).
The condition (\ref{cor2cond1}) is satisfied as well. Hence $h(z)$
satisfy conditions for Corollary 3. The corresponding integral in
(\ref{cor2form}) with $h(z)$ from (\ref{rel32}) can be calculated 
by using the formula (\ref{intJYth65}) (see below). Putting the value of 
this integral into (\ref{cor2form}) after some manipulations we 
receive to (\ref{th4form}). \rule{1.5ex}{1.5ex}

\bigskip

\noindent {\bf Remark.} The formula (\ref{th4form}) may be derived 
also by applying to the contour integral
\begin{equation}
\int_{C_h}{\frac{Y_\nu (\lambda z)J_{\nu +m}(\sigma z)-
J_\nu (\lambda z)Y_{\nu +m}(\sigma z)}{J_\nu (z)Y_{\nu }(\lambda z)-
J_\nu (\lambda z)Y_{\nu }(z)}F(z)dz} \label{altth4}
\end{equation}
the residue theorem, where $C_h$ is a rectangle with 
vertices $\pm ih,\, b\pm ih$.
Here the proof is similar to that for 
Remark to the Corollary 1. \rule{1.5ex}{1.5ex}

\bigskip

Formula similar to (\ref{th4form}) can be obtained also for the series
of type $\sum_{k=1}^{\infty }G(\gamma _{\nu ,k})
J_\mu (\gamma _{\nu ,k}t)$ by using (\ref{cor2form}).

As a function $F(z)$ in (\ref{th4form}) one can choose, for
example,

\begin{itemize}

\item function (\ref{func1}) for $\rho =m-2p$, 
$\sum_{l}^{}b_l<a_2$, $m<2p+\sum_{l}^{}{\mathrm{Re}}\mu _l+n/2+1$,
$p$ - integer;
 
\item function (\ref{func2}) for $\rho =m-2p$, $\sum_{l}^{}b_l<a_2$,
$m<2p+2n+1$;

\item function (\ref{func4}) for $\rho =m-2p$,
$a_l>0$, ${\mathrm{Re}}\mu _l\geq 0$ 
(for ${\mathrm{Re}}\mu _l< 0$, $k_l>|{\mathrm{Re}}\mu _l|$),\\
$\sum_{l=1}^{n}\vert a_l-b_l\vert <a_2$, $m<2p+n-\sum_{l}^{}|k_l|+1$. 

\end{itemize}

\noindent For $F(z)=1/z,\, m=0$ one obtains
\begin{equation}
\sum_{k=1}^{\infty}
\frac{J_\nu (\gamma _{\nu ,k})Y_{\nu }(\sigma \gamma _{\nu ,k})-
Y_\nu (\gamma _{\nu ,k})J_{\nu }(\sigma \gamma _{\nu ,k})}{J_\nu ^2
(\gamma _{\nu ,k})/J_\nu ^2(\lambda \gamma _{\nu ,k})-1}=
\frac{\sigma ^\nu }{\pi }\frac{(\lambda /\sigma )^{2\nu }-
1}{\lambda ^{2\nu}-1}, \quad \lambda \geq \sigma>1.
\label{examp31}
\end{equation}
By similar way it can be seen that
\begin{eqnarray}
& & \! \! \sum_{k=1}^{\infty}
\frac{\gamma _{\nu ,k}^2\left[ J_\nu (\gamma _{\nu ,k})
Y_{\nu }(\sigma \gamma _{\nu ,k})-Y_\nu (\gamma _{\nu ,k})
J_{\nu }(\sigma \gamma _{\nu ,k})\right] }{\left( \gamma _{\nu ,
k}^2-c^2\right)\left[ J_\nu ^2
(\gamma _{\nu ,k})/J_\nu ^2(\lambda \gamma _{\nu ,k})-1\right] }=
\frac{1}{\pi }\frac{Y_\nu (\lambda c)J_{\nu }(\sigma c)-
J_\nu (\lambda c)Y_{\nu }(\sigma c)}{J_\nu (c)Y_{\nu }(\lambda c)-
J_\nu (\lambda c)Y_{\nu }(c)},
\label{examp32}\\
& & \!\! \sum_{k=1}^{\infty}
\frac{J_\nu (\gamma _{\nu ,k})Y_{\nu }(\sigma \gamma _{\nu ,k})-
Y_\nu (\gamma _{\nu ,k})J_{\nu }(\sigma \gamma _{\nu ,k})}{J_\nu ^2
(\gamma _{\nu ,k})/J_\nu ^2(\lambda \gamma _{\nu ,k})-1}
\prod_{l=1}^{p}\gamma _{\nu ,k}^{-\mu _l}J_{\mu _l}(b_l
\gamma _{\nu ,k})={}\nonumber\\
& & =\frac{\sigma ^\nu }{\pi }\frac{(\lambda /\sigma )^{2\nu }-
1}{\lambda ^{2\nu}-1}\prod_{l=1}^{p}\frac{b_l^{\mu _l}}{2^{\mu _l}
\Gamma (\mu _l+1)}, \quad b\equiv \sum_{1}^{p}b_l<\sigma -1,\, 
{\mathrm{Re}}\mu _l+\frac{p}{2}+1>\delta_{b,\sigma -1},
\label{examp33}
\end{eqnarray}
where ${\mathrm{Re}}c\geq 0,\, b_l>0$, 
$\lambda\geq \sigma >1$, $\mu _l\ne -1, -2,\ldots $.

So far in this section we have considered series over zeros of the
function $C^{AB}_\nu (\lambda ,z)$. The similar results can be obtained 
also for the series containing zeros of the function
\begin{equation}
C_{1\nu }(\lambda ,z)=J'_\nu (z)Y_\nu (\lambda z)-
Y'_\nu (z)J_\nu (\lambda z), \label{bescomb2}
\end{equation} 
(on properties of zeroes of these function see
\cite{Watson, Erdelyi, abramowiz}).
The corresponding furmulae for the zeros $\gamma _{1\nu ,k}$
of this function can be obtained from
those for $C^{AB}_{\nu }(\lambda ,z)$ by replacements
\begin{eqnarray}
& & T_\nu ^{AB}(\lambda ,z) \to 
\frac{z}{J_\nu ^2(z)/J_\nu ^2(\lambda z)
-1+\nu ^2/z^2} \label{secomb}\\
& & \bar f(z)\to f'(z), \quad \bar f(\lambda z)\to f(\lambda z), 
\quad f=J_\nu ,Y_\nu ,H_\nu ^{(1,2)}, I_\nu ,K_\nu ,
\quad C^{AB}_\nu \to C_{1\nu }. \nonumber
\end{eqnarray}
The physiacl apllications of the formulae derived in this
section will be considered below in Sections 9 and 12.

\section{Applications to integrals involving Bessel functions}

\renewcommand{\theequation}{5.\arabic{equation}}

\setcounter{equation}{0}

The applications of GAPF to infinite integrals involving some
combinations of Bessel
functions lead to the interesting results \cite{Sah1, Sahdis}. 
First of all one can express integrals over Bessel 
functions through the integrals involving
modified functions. Let us substitute in the formula
(\ref{sumJ1})
\begin{equation}
f(z)=F(z)\bar J_\nu (z). \label{fint41}
\end{equation}
For the function $F(z)$ having no poles at $z=\lambda _{\nu ,k}$ 
the sum over zeros of $\bar J_\nu (z)$ is zero. The
conditions (\ref{condf}) and (\ref{case21}) may be written in 
terms of $F(z)$ as
\begin{equation}
\vert F(z)\vert <\varepsilon _1(x)e^{c_1\vert y\vert }
\quad \textrm{or} \quad 
\vert F(z)\vert <M\vert z\vert ^{-\alpha _1}e^{\vert y\vert },
\quad \vert z\vert \to \infty,
 \label{cond42}
\end{equation}
with $c_1<1$, $x^{1/2-\delta _{B0}}\varepsilon _1(x)\to 0 $
for $x\to \infty $, $\alpha _1>\alpha _0=3/2-\delta _{B0}$, and
\begin{equation}
F(ze^{\pi i})=-e^{\nu \pi i}F(z)+o\left( z^{|{\mathrm{Re}}
\nu |-1}\right) .
\label{cond43}
\end{equation}
Hence for the function $F(z)$ satisfying conditions (\ref{cond42})
and (\ref{cond43}) it follows from (\ref{sumJ1}) that
\begin{eqnarray}
& & {\mathrm{p.v.}}\int_{0}^{\infty }{F(x)\bar J_\nu (x)dx}=r_{1\nu}\left[ F(z)
\bar J_\nu (z)\right] +\frac{\pi }{2}
{\mathrm{Res}}_{z=0}F(z)\bar Y_\nu (z)+{}
\nonumber\\
& & +\frac{1}{\pi}\int_{0}^{\infty }{\bar K_\nu (x)\left[ 
e^{-\nu \pi i/2}F(xe^{\pi i/2})+e^{\nu \pi i/2}F(xe^{-\pi i/2})
\right] dx}. \label{intJform41} 
\end{eqnarray}
In expression (\ref{r1}) for $r_{1\nu }$ 
the points $z_k$ are poles of the 
meromorpic function $F(z),\, {\mathrm{Re}}z_k>0$.
On the base of Remark after Theorem 3 the 
formula (\ref{intJform41}) may 
be generalized for the functions $F(z)$ with purely imaginary poles
$\pm iy_k$, $y_k>0$ and satisfying condition
\begin{equation}
F(ze^{\pi i})=-e^{\nu \pi i}F(z)+o\left( (z\mp iy_k)^{-1}\right) ,
\quad z\to \pm iy_k.
\label{cor3cond1plus2n}
\end{equation}
The corresponding formula is obtained from (\ref{intJform41}) by adding
residue terms for $z=iy_k$ in the form of (\ref{intJform42}) (see 
below) and taking the principal value of the integral on the right.

The same substitution (\ref{fint41}) with the function $F(z)$
satisfying the conditions (\ref{cond42}) and
\begin{equation}
F(xe^{\pi i/2})=-e^{\nu \pi i}F(xe^{-\pi i/2})
\label{cond44}
\end{equation}
for real $x$, into the formula (\ref{sumJ2}) yields the following
result
\begin{equation}
{\mathrm{p.v.}}\int_{0}^{\infty }{F(x)\bar J_\nu (x)dx}=r_{1\nu}\left[ F(z)
\bar J_\nu (z)\right] +\frac{\pi i}{2}\sum_{\eta _k=
0,iy_k}\left( 2-\delta _{0\eta _k}\right){\mathrm{Res}}_{z=\eta _k}F(z)
\bar H^{(1)}_\nu (z).
\label{intJform42} 
\end{equation}
In (\ref{r1}) now summation is over the poles $z_k$,
${\mathrm{Re}}z_k>0$ of the meromorphic 
function $F(z)$, and $\pm iy_k, \, y_k>0$
are purely imaginary poles of this function. Recall that the 
possible real poles of $F(z)$ are such, that integral on the 
left of (\ref{intJform42}) exists.

For the functions $F(z)=z^{\nu +1}\tilde F(z)$, with $\tilde F(z)$
being analytic in the right half-plane and even along the imaginary
axis, $\tilde F(ix)=\tilde F(-ix)$, one obtains
\begin{equation}
\int_{0}^{\infty }{x^{\nu +1}\tilde F(x)\bar J_\nu (x)dx}=0.
\label{examp41}
\end{equation}
This result for $B=0$ (see (\ref{efnot1})) have been given
previously in \cite{Schwartz}. The another result of \cite{Schwartz}
is obtained from (\ref{intJform42}) choosing $F(z)=z^{\nu +1}
\tilde F(z)/(z^2-a^2)$.

Formulae similar to (\ref{intJform41}) and (\ref{intJform42})
can be derived for Neumann function $Y_\nu (z)$.
Let for the function $F(z)$ the integtral 
${\mathrm{p.v.}}\int_{0}^{\infty }{F(x)\bar Y_\nu (x)dx}$ exists. 
Let us substitute in the formula (\ref{th12})
\begin{equation}
f(z)=Y_\nu (z)F(z), \quad g(z)=-iJ_\nu (z)F(z)
\label{fgneum41}
\end{equation}
and consider the limit $a\to +0$. The summands containing residues 
may be presented in the form
\begin{eqnarray}
R[f(z),g(z)] & = & \pi \sum_{k}
{\mathrm{Res}}_{{\mathrm{Im}}z_k>0}H^{(1)}_\nu (z)
F(z)+ \pi \sum _{k}{\mathrm{Res}}_
{{\mathrm{Im}}z_k<0}H^{(2)}_\nu (z)F(z) +{}\nonumber\\
& + & \pi \sum_{k}^{}
{\mathrm{Res}}_{{\mathrm{Im}}z_k=0}J_\nu (z)F(z) \equiv r_{3
\nu }[F(z)], \label{r2}
\end{eqnarray}
where $z_k$ (${\mathrm{Re}}z_k>0$) are the poles of $F(z)$ in 
the right half-plane.
Now the following results can be prooved by using (\ref{th12}):

\bigskip

{\it 1) If the meromorphic function $F(z)$ has no poles on the 
imaginary axis and satisfy the condition (\ref{cond42}) then
\begin{equation}
{\mathrm{p.v.}}\int_{0}^{\infty }{F(x)Y_\nu (x)dx}=r_{3\nu}[F(z)] -
\frac{i}{\pi}\int_{0}^{\infty }{K_\nu (x)\left[ 
e^{-\nu \pi i/2}F(xe^{\pi i/2})-e^{\nu \pi i/2}F(xe^{-\pi i/2})
\right] dx} \label{intYform41} 
\end{equation}
}
 
\noindent and 

\bigskip

{\it 2) If the meromorphic function $F(z)$ satisfy the conditions
\begin{equation}
 F(xe^{\pi i/2})=e^{\nu \pi i}F(xe^{-\pi i/2})
 \label{cond45}
\end{equation}
and (\ref{cond42}) then one has
\begin{equation}
{\mathrm{p.v.}}\int_{0}^{\infty }{F(x)Y_\nu (x)dx}=r_{3\nu}[F(z)] +
\pi \sum_{k}{\mathrm{Res}}_{z=iy_k}H^{(1)}_\nu (z)F(z), 
\label{intYform42} 
\end{equation}  
where $\pm iy_k,\, y_k>0$ are purely imaginary poles of $F(z)$.}

\bigskip

From (\ref{intYform42}) it directly follows that for
$F(z)=z^\nu \tilde F(z)$, with $\tilde F(z)$ being even along
the imaginary axis, $\tilde F(ix)=\tilde F(-ix)$, and analytic
in the right half-plane \cite{Schwartz}
\begin{equation}
{\mathrm{p.v.}}\int_{0}^{\infty }{x^\nu \tilde F(x)Y_\nu (x)dx}=0,
 \label{examp42} 
\end{equation}
if the condition (\ref{cond42}) takes place.

Let us consider more general case. Let the function $F(z)$
satisfy the condition 
\begin{equation}
 F(ze^{-\pi i})=-e^{-\lambda \pi i}F(z)
 \label{case42}
\end{equation}
for ${\mathrm{arg}}z=\pi /2$. 
In GAPF as functions $f(z)$ and $g(z)$ we choose
\begin{eqnarray}
f(z) & = & F(z)\left[ J_\nu (z)\cos \delta +Y_\nu (z)\sin 
\delta \right]\nonumber\\
g(z) & = & -iF(z)\left[ J_\nu (z)\sin \delta -Y_\nu (z)
\cos \delta ,\quad \delta=(\lambda -\nu )\pi /2\right] ,
\label{fgcomb42}
\end{eqnarray} 
with $g(z)-(-1)^kf(z)=H^{(k)}_\nu (z)F(z)\exp [(-1)^ki\delta ]$,
 $k=1,2$. It can be seen that for such a choice the integral on
 rhs of (\ref{th12}) for $a\to 0$ is equal to
\begin{equation}
\pi i\sum_{\eta _{k}=iy_k}
{\mathrm{Res}}_{z=\eta _k}H^{(1)}_\nu (z)F(z)e^{-i\delta } ,
\label{rel42}
\end{equation}
where $\pm iy_k,\, y_k>0$, as above, are purely imaginary poles
of $F(z)$. Substituting (\ref{fgcomb42}) into (\ref{th12}) and 
using (\ref{cor13}) we obtain \cite{Sah1, Sahdis}

\bigskip

\noindent {\bf Corollary 4.} {\it Let $F(z)$ be meromorphic function
for ${\mathrm{Re}}z\geq 0$ (except possibly at $z=0$) with
poles $z_k,\, \pm iy_k$; $y_k,{\mathrm{Re}}z_k>0$. If this function 
satisfies conditions (\ref{cond42}) (for $B=0$) and (\ref{case42})
then
\begin{eqnarray}
& & {\mathrm{p.v.}}\int_{0}^{\infty }{F(x)\left[ J_\nu (x)\cos \delta +
Y_\nu (x)\sin \delta \right] dx}=\pi i
\left\{ \sum_{z_k}{\mathrm{Res}}_{{\mathrm{Im}}z_k>0}
H^{(1)}_\nu (z)F(z)e^{-i\delta }\right.-\nonumber\\
& & -\sum_{z_k}{\mathrm{Res}}_
{{\mathrm{Im}}z_k<0}H^{(2)}_\nu (z)F(z)e^{i\delta }-
i\sum _{z_k}{\mathrm{Res}}_
{{\mathrm{Im}}z_k=0}\left[ J_\nu (z)\sin \delta -
Y_\nu (z)\cos \delta \right] F(z)+\nonumber\\
& & +\left.\sum_{\eta _k=iy_k}^{}
 {\mathrm{Res}}_{z=\eta _k}H^{(1)}_\nu (z)F(z)e^{-i\delta }\right\} ,
\label{intJYform43}
\end{eqnarray}
where it is assumed that integral on the left exists.}

\bigskip

\noindent In particular for $\delta =\pi n,\, n=0,1,2...$ the 
formula (\ref{intJform42}) follows from here in the case $B=0$.

One will find a great many particular cases of the formulae
(\ref{intJform42}) and (\ref{intJYform43}) looking at the 
standard books and tables of known integrals with Bessel 
functions (see, e.g., \cite{Erdelyi, Watson, Magnus, 
Prudnikov, Erdelyi2, Luke, Gradshteyn, Wheelon, Oberhettinger}).
Some special examples are given in the next section.

\section{Integrals involving Bessel functions: Illustrations of
general formulae}

\renewcommand{\theequation}{6.\arabic{equation}}

\setcounter{equation}{0}

To illustrate the applications of the general formulae from previous
section first of all consider integrals involving the function
$\bar J_\nu (z)$. Let us introduce the functional
\begin{equation}
A_{\nu m}[G(z)]\equiv {\mathrm{p.v.}}\int_{0}^{\infty}{z^{\nu -2m-1}G(z)
\bar J_\nu (z)dz} \label{Anjuem}
\end{equation}
with $m$ being an integer. Let $F_1(z)$ be an analytic function in
the right half-plane satisfying condition
\begin{equation}
 F_1(xe^{\pi i/2})=F_1(xe^{-\pi i/2}), \quad F_1(0)\ne 0
 \label{cond5F1}
\end{equation}
(the case when $F_1(z)\sim z^q,\, z\to 0$ with an integer $q$ can 
be reduced to this one by redefinitions of $F_1(z)$ and $m$). 
From (\ref{intJform42}) the following results can be obtained 
\cite{Sah1, Sahdis}
\begin{eqnarray}
A_{\nu m}[F_1(z)] & = & A^{(0)}_{\nu m}[F_1(z)]\equiv 
-\frac{\pi (1+{\mathrm{sgn}}m)}{4(2m)!}\left( \frac{d}{dz}\right) ^{2m}
\left[ z^\nu \bar Y_\nu (z)F_1(z)\right]\left\vert _{z=0} \right.
 \label{examp51}\\
A_{\nu m}\left[ \frac{F_1(z)}{z^2-a^2}\right] & = & -\frac{\pi }{2}
a^{\nu -2m-2}\bar Y_\nu (a)F_1(a)+A^{(0)}_{\nu m}
\left[ \frac{F_1(z)}{z^2-a^2}\right], \quad 
\label{examp52}\\
A_{\nu m}\left[ \frac{F_1(z)}{z^4-a^4}\right] & = &
-\frac{a^{\nu -2m-4}}{2}\left[ \frac{\pi}{2}\bar Y_\nu (a)F_1(a)-
(-1)^m\bar K_\nu (a)F_1(ia)\right]+\nonumber\\
& & +A^{(0)}_{\nu m}\left[ \frac{F_1(z)}{z^4-a^4}\right] ,
\label{examp53}\\
A_{\nu m}\left[ \frac{F_1(z)}{\left( z^2-c^2\right) ^{p+1}}
\right] & = & \frac{\pi i}{2^{p+1}p!}\left( \frac{d}{cdc}
\right) ^{p}\left[ c^{\nu-2m-2}
F_1(c)H_\nu ^{(1)}(c)\right] 
+A^{(0)}_{\nu m}\left[ \frac{F_1(z)}{\left( z^2-
c^2\right) ^{p+1}}\right] 
\label{examp54}\\
A_{\nu m}\left[ \frac{F_1(z)}{\left( z^2+a^2\right) ^{p+1}}
\right] & = & \frac{(-1)^{m+p+1}}{2^{p}\cdot p!}\left( \frac{d}{ada}
\right) ^{p}\left[ a^{\nu-2m-2}
K_\nu (a)F_1(ae^{\pi i/2})\right] +\nonumber\\ 
&  & +A^{(0)}_{\nu m}\left[ \frac{F_1(z)}{\left( z^2+
a^2\right) ^{p+1}}\right] ,
\label{examp54ad}
\end{eqnarray}
and etc.(note that $A^{(0)}_{\nu m}=0$ for $m<0$). Here $a>0$, 
$0<{\mathrm{arg}}c<\pi /2$, and we have assumed that 
${\mathrm{Re}}\nu >0$. To secure convergence at the origin
the condition ${\mathrm{Re}}\nu >m$ is necessary.
In the last two formulae
we have used the identity 
\begin{equation}
\left( \frac{d}{dz}\right) ^p\left[ \frac{zF(z)}
{(z+b)^{p+1}}\right] _{z=b}=\frac{1}{2^{p+1}}\left( \frac{d}{bdb}
\right) ^pF(b).
\label{ident1}
\end{equation}
Note that (\ref{examp54ad}) can be obtained also from (\ref{examp54}) 
in the limit ${\mathrm{Re}}c\to 0$. For the case $F_1=1,\, m=-1$
of (\ref{examp54ad}) see, for example, \cite{Watson}. In
(\ref{examp51})-(\ref{examp54ad})
as a function $F_1(z)$ we can choose:

\begin{itemize}

\item function (\ref{func1})
for $\rho =1$, ${\mathrm{Re}}\nu <
\sum {\mathrm{Re}}\mu _l+2m+(n+1)/2-\delta _{b1}+
\delta _{B0}$, $b=\sum b_l\leq 1,\, b_l>0$; 

\item function (\ref{func2})
with $\rho =1$, ${\mathrm{Re}}\nu<2(m+n)+1/2-\delta _{b1}+
\delta _{B0}$, $b=\sum b_l\leq 1$; 

\item function (\ref{func3}) for $\rho =1$, 
${\mathrm{Re}}\nu <2m-\sum {\mathrm{Re}}\mu _l+(n+1)/2-\delta _{b1}+
\delta _{B0}$, $\mu _l>0$ is half of an odd integer, 
$b=\sum b_l\leq 1$; 

\item function (\ref{func4}) for 
${\mathrm{Re}}\nu <2m+n-\sum |k_l|+1/2+\delta _{B0}-
\delta _{\tilde a1}$, $\tilde a=\sum |a_l-b_l|\leq 1$, $a_l\geq 0$,
$k_l$ - integer. 

\end{itemize}

\noindent Here we have written the conditions for
(\ref{examp51}). The corresponding ones for (\ref{examp52}),
(\ref{examp53}), (\ref{examp54}),(\ref{examp54ad}) 
are obtained from these by adding 
on the rhs of inequalities for ${\mathrm{Re}}\nu $, respectively 2, 4,
$2(p+1)$, $2(p+1)$. In (\ref{examp51})-(\ref{examp54ad}) we can choose also
any combinations of the functions (\ref{func1})-(\ref{func4}) 
with appropriate conditions.

For concrete evaluations of $A^{(0)}_{\nu m}$ in special cases
it is useful the following formula 
\begin{equation}
\lim_{z\to 0}\left( \frac{d}{dz}\right) ^{2m}f_1(z)=(2m-1)!!
\lim_{z\to 0}\left( \frac{d}{zdz}\right) ^{m}f_1(z),
\label{rel51}
\end{equation}
valid for the function $f_1(z)$ satisfying condition
$f_1(-z)=f_1(z)+o(z^{2m})$, $z\to 0$. From here, for instance, it
follows that for $z\to 0$
\begin{equation}
\left( \frac{d}{dz}\right) ^{2m}
\left[ z^\nu Y_\nu (bz)F_1(z)\right] =-(2m-1)!!
\frac{2^{\nu -m}}{\pi b^{\nu -m}}\sum_{k=0}^{m}{m\choose k}2^k
\frac{\Gamma (\nu -m+k)}{b^{2k}}\left( \frac{d}{zdz}\right) ^{k}
F_1(z), 
\label{rel52}
\end{equation}
where we have used the standard formula for the derivative
$(d/zdz)^n$ of cylinder functions (see \cite{abramowiz}). From
(\ref{examp51}) one obtains ($B=0$)
\begin{eqnarray}
& & \int_{0}^{\infty }{z^{\nu -2m-1}J_\nu (z)
\prod_{l=1}^{n}\left( z^2+z_l^2\right) ^{-\mu _{l}/2} 
J_{\mu _l}(b_l\sqrt{z^2+z_l^2})dz}= \nonumber\\
& & =-\frac{\pi }{2^{m+1}m!}\left( \frac{d}{zdz}\right) ^{m}
\left[ z^\nu Y_\nu (z) \prod_{l=1}^{n}\left( z^2+z_l^2
\right) ^{-\mu _{l}/2}
J_{\mu _l}(b_l\sqrt{z^2+z_l^2}) \right] _{z=0},
\label{examp55}
\end{eqnarray}
for $m\geq 0$ and the integral is zero for $m<0$.
Here ${\mathrm{Re}}\nu >0$, $b\equiv \sum_{1}^{n}b_l\leq 1$, $b_l>0$,
$m<{\mathrm{Re}}\nu <\sum_{l=1}^{n}
{\mathrm{Re}}\mu _l+2m+(n+3)/2-\delta _{b1}$.
In particular case $m=0$ the Gegenbauer integral
follows from here \cite{Watson, Erdelyi}. In
the limit $z_l\to 0$ from (\ref{examp55})  the 
value of integral $\int_{0}^{\infty }{z^{\nu -2m-1}J_\nu (z)
\prod_{l=1}^{n}z^{-\mu _{l}}J_{\mu _l}(z)dz}$ can be obtained. 

 By using (\ref{intJYform43}) the formulae similar to 
(\ref{examp51})-(\ref{examp54ad}) may be 
derived for the integrals of type
\begin{equation}
B_{\nu }[G(z)]\equiv {\mathrm{p.v.}}\int_{0}^{\infty}{G(z)
\left[ J_\nu (z)\cos \delta +Y_\nu (z)
\sin \delta \right] dz}, \quad \delta =(\lambda -\nu )
\pi /2 .\label{Bnju}
\end{equation}
 It directly follows 
from Corollary 4 that for function $F(z)$ analytic
for ${\mathrm{Re}}z\geq 0$ and satisfying conditions (\ref{cond42}) and
(\ref{case42}) the following formulae take place
\begin{eqnarray}
B_{\nu }[F(z)] & = & 0 \label{examp56}\\
B_{\nu }\left[ \frac{F(z)}{z^2-a^2}\right] & = & \pi F(a)
\left[ J_\nu (a)\sin \delta -Y_\nu (a)
\cos \delta \right]  /2, \quad  \label{examp57}\\
B_{\nu }\left[ \frac{F(z)}{z^4-a^4}\right] & = & \frac{\pi }{4a^3}F(a)
\left[ J_\nu (a)\sin \delta -Y_\nu (a)
\cos \delta \right]  +\frac{i}{2a^3}K_\nu (a)F(ia)e^{-i\lambda \pi /2}
, \label{examp58}\\
B_{\nu }\left[ \frac{F_1(z)}{\left( z^2-c^2\right) ^{p+1}}
\right] & = & \frac{\pi i}{2^{p+1}\cdot p!}\left( \frac{d}{cdc}
\right) ^{p}\left[ c^{-1}F(c)H_\nu ^{(1)}(c)\right] e^{-i\delta }
\label{examp59}\\
B_{\nu }\left[ \frac{F_1(z)}{\left( z^2+a^2\right) ^{p+1}}
\right] & = & \frac{(-1)^{p+1}}{2^{p}\cdot p!}\left( \frac{d}{ada}
\right) ^{p}\left[ a^{-1}F(ae^{\pi i/2})K_\nu (a)\right] e^{-i
\pi \lambda /2 },
\label{examp59ad}
\end{eqnarray}
where $a>0$, $0<{\mathrm{arg}}c\leq \pi/2$. To obtain the last two 
formulae we have used the identity (\ref{ident1}). The formula 
(\ref{examp56}) generalizes the result of \cite{Schwartz} (the cases
$\lambda =\nu$ and $\lambda =\nu +1$). From the last formula
taking $F(z)=z^{\lambda -1}$ we obtain result given in 
\cite{Watson}. In (\ref{examp56}) -
(\ref{examp59ad}) as a function $F(z)$ one can choose (the
constraints on parameters are written for the formula 
(\ref{examp56}); the corresponding constraints for (\ref{examp57}),
(\ref{examp58}), (\ref{examp59}), (\ref{examp59ad}) are 
obtained from given ones by adding the summands 2, 4, 
$2(p+1)$, $2(p+1)$ to the rhs of inequalities,
 correspondingly):

\begin{itemize}

\item function (\ref{func1}) for $\rho =\lambda $, $|{\mathrm{Re}}\nu |<
{\mathrm{Re}}\rho <\sum {\mathrm{Re}}\mu _l+(n+3)/2-\delta _{b1}$,
 $b=\sum_{l}^{}b_l\leq 1$;

\item function (\ref{func2}) for $\rho =\lambda $, $|{\mathrm{Re}}\nu |<
{\mathrm{Re}}\rho <3/2-\delta _{b1}$, $b=\sum_{l}^{}b_l\leq 1$;

\item function
\begin{eqnarray}
& & \! \! z^{\rho -1}\prod_{l=1}^{n}\left[ J_{\mu _l+k_l}(a_lz)
Y_{\mu _l}(b_lz)-Y_{\mu _l+k_l}(a_lz)J_{\mu _l}(b_lz)\right] ,
\quad \lambda =\rho +\sum_{l=1}^{n}k_l,\, a_l>0,\label{func5new}\\
& & \quad |{\mathrm{Re}}\nu |+\sum |k_l|<
{\mathrm{Re}}\rho <n+3/2-\delta_{c1},\, c=\sum |a_l
-b_l|\leq 1,\, {\mathrm{Re}}\mu _l\geq 0 \nonumber
\end{eqnarray}
(for ${\mathrm{Re}}\mu _l< 0$ one has $k_l>|{\mathrm{Re}}\mu _l|$). 

\end{itemize}

\noindent Any combination of these functions with appropriate 
conditions on parameters can be choosed as well.

 Now consider integrals which can be expressed via series by
 using (\ref{intJform42}) and (\ref{intJYform43}). In (\ref{intJform42})
 let us choose the function 
\begin{equation}
F(z)=\frac{z^{\nu -2m}F_1(z)}{\sinh \pi z}, \label{examp510}
\end{equation}
where $F_1(z)$ is the same as in the formulae (\ref{examp51}) -
(\ref{examp54}). As the points $\pm i,\pm 2i, \ldots $ are simple 
poles for $F(z)$ from (\ref{intJform42}) one obtains
\begin{equation}
\int_{0}^{\infty}{\frac{z^{\nu -2m}}{\sinh{(\pi z)}}F_1(z)
\bar J_\nu (z)dz}=A_{\nu m}^{(0)}\left[ \frac{zF_1(z)}{\sinh{(\pi z)}}
\right] +\frac{2}{\pi}\sum_{k=1}^{\infty}(-1)^{m+k}k^{\nu -2m}
\bar K_{\nu }(k)F_1(ik), 
\label{intJ5sum}
\end{equation}
where $A_{\nu m}^{(0)}[f(z)]$ is defined by (\ref{examp51})
and ${\mathrm{Re}}\nu >m$. The corresponding costraints on 
$F_1(z)$ follow directly from (\ref{cond42}). The 
particular case of this formula when $F_1(z)=\sinh (az)/z$ and 
$m=-1$ is given in \cite{Watson}. As a function $F_1(z)$ here 
one can choose any of functions (\ref{func1})-(\ref{func4}) with
$\rho =1$ and $\tilde a,\, \sum_{l}^{}b_l<1$. From 
(\ref{intJ5sum}) it follows that
\begin{eqnarray}
& & \int_{0}^{\infty}{\frac{z^{\nu -2m}}{\sinh{(\pi z)}}
J_\nu (z)\prod_{l=1}^{n}z^{-\mu _l}I_{\mu _l}(b_lz)dz}=
A_{\nu m}^{(0)}\left[ \frac{z}{\sinh{(\pi z)}}\prod_{l=1}^{n}z^{-
\mu _l}I_{\mu _l}(b_lz)\right] +\nonumber\\
& & +\frac{2}{\pi}\sum_{k=1}^{\infty}(-1)^{m+k}K_\nu (k)
\prod_{l=1}^{n}k^{-\mu _l}J_{\mu _l}(b_lk), \quad b_l>0,\, 
\pi -\sum_{l=1}^{n}b_l>0,\, {\mathrm{Re}}\nu >m.
 \label{examp511}
\end{eqnarray}
In similar way from (\ref{intJYform43}) it can be derived the following
formula
\begin{equation}
\int_{0}^{\infty}{\frac{zF(z)}{\sinh{(\pi z)}}\left[ J_\nu (z)
\cos \delta +Y_\nu (z)\sin \delta \right] dz}=
\frac{2i}{\pi}e^{-i\lambda \pi /2}\sum_{k=1}^{\infty}(-1)^{k}k
K_{\nu }(k)F(ik). \label{intJY5sum}
\end{equation}
Constraints on the function $F(z)$ immediately follow from Corollary 4.
Instead of this function we can choose the functions (\ref{func1}),
(\ref{func2}), (\ref{func4}). 

As it have been mentioned above adding residue terms 
$\pi i {\mathrm{Res}}_{z=iy_k}F(z)\bar H_\nu ^{(1)}(z)$ to the rhs
of (\ref{intJform41}) this formula may be generalized for the 
functions having purely imaginary poles $\pm iy_k$, $y_k>0$, 
provided the condition (\ref{cor3cond1plus2n}) is satisfied. As an
application let us choose
\begin{equation}
F(z)=\frac{z^\nu F_1(z)}{e^{2\pi z/b}-1}, \quad F_1(-z)=F_1(z),
\quad b>0 \label{exampforpole}
\end{equation}
with an analytic function $F_1(z)$. The function (\ref{exampforpole})
satisfy condition (\ref{cor3cond1plus2n}) and have poles 
$\pm ikb$, $k=0,1,2\ldots $. The additional constraint directly 
follows from (\ref{cond42}). Then one obtains
\begin{equation}
\int_{0}^{\infty }{\frac{x^\nu J_\nu (x)}{e^{2\pi x/b}-1}F_1(x)dx}=
\frac{2}{\pi }\sum_{k=0}^{\infty }{'}(bk)^\nu K_\nu (bk)F_1(ibk)-
\frac{1}{\pi }\int_{0}^{\infty }{x^\nu K_\nu (x)F_1(ix)dx},
\label{exampforpoleform}
\end{equation}
where the prime indicates that the $m=0$ term is to be halved.
For the particular case $F_1(z)=1$, using the relation
\begin{equation}
\sum_{k=0}^{\infty }{'}(bk)^\nu K_\nu (bk)=\frac{\sqrt{\pi }}{b}
2^\nu \Gamma (\nu +1/2)\sum_{n=0}^{\infty }{'}
\left[ \left( \frac{2\pi n}{b}\right) ^2+1\right] ^{-\nu -1/2}
\label{relipole}
\end{equation}
and the known value for the integral on the right,
we immediately obtain the result given in \cite{Watson}. The 
relation (\ref{relipole}) can be proved by using the formulae
\begin{equation}
K_\nu (z)=\frac{2^\nu \Gamma (\nu +1/2)}{\sqrt{\pi }z^\nu }
\int_{0}^{\infty }{\frac{\cos zt\, dt}{(t^2+1)^{\nu +1/2}}},
\quad \sum_{k=-\infty }^{+\infty }e^{ikz}=2\pi 
\sum_{n=-\infty }^{+\infty }\delta (z-2\pi n),
\label{relipole1}
\end{equation}
(for the integral representation of Macdonald's function see 
\cite{Watson}), $\delta (z)$ is the Dirac delta function.

\section{Formulae for integrals involving
$J_\nu (z)Y_\mu (\lambda z)-Y_\nu (z)J_\mu (\lambda z)$}

\renewcommand{\theequation}{7.\arabic{equation}}

\setcounter{equation}{0}

In this section we shall consider the applications of GAPF to the 
integrals involving the function $J_\nu (z)Y_\mu (\lambda z)-
Y_\nu (z)J_\mu (\lambda z)$. In the formula (\ref{th12}) we substitute
\begin{equation}
f(z)=-\frac{1}{2i}F(z)\sum_{l=1}^{2}(-1)^{l}
\frac{H^{(l)}_{\mu }(\lambda z)}{H^{(l)}_{\nu }(z)},\quad
g(z)=\frac{1}{2i}F(z)\sum_{l=1}^{2}
\frac{H^{(l)}_{\mu }(\lambda z)}{H^{(l)}_{\nu }(z)}\, .
\label{fg61}
\end{equation}
For definiteness we consider the case $\lambda >1$ (for $\lambda <1$
the expression for $g(z)$ have to be choosen with opposite sign).
The conditions (\ref{cor11}) and (\ref{th11}) are satisfied if the function
$F(z)$ is constrained by the one of the following two inequalities
\begin{equation}
\vert F(z)\vert<\varepsilon (x)e^{c\vert y\vert},\quad c<\lambda -1,
\, \varepsilon (x)\to 0,\, x\to +\infty
\label{cond61}
\end{equation} 
or
\begin{equation}
\vert F(z)\vert<M\vert z\vert ^{-\alpha }e^{(\lambda -
1)\vert y\vert},\quad \alpha >1,\, \vert z
\vert \to \infty ,\, z=x+iy.
\label{cond62}
\end{equation}
Then from (\ref{th12}) it follows that for the 
function $F(z)$ meromorphic in ${\mathrm{Re}}z\geq a>0$ one has
\begin{eqnarray}
& & {\mathrm{p.v.}}\int_{0}^{\infty}{\frac{J_\nu (x)Y_\mu (\lambda x)-
J_\mu (\lambda x)Y_\nu (x)}{J_\nu ^2(x)+Y_\nu ^2(x)}F(x)dx}=
r_{1\mu \nu }[F(z)]+\nonumber\\
& & +\frac{1}{2i}\left[ \int_{a}^{a+i\infty}{F(z)
\frac{H^{(1)}_{\mu }(\lambda z)}{H^{(1)}_{\nu }(z)}dz}-
\int_{a}^{a-i\infty}{F(z)
\frac{H^{(2)}_{\mu }(\lambda z)}{H^{(2)}_{\nu }(z)}dz}\right] ,
\label{intJY61}
\end{eqnarray}
where we have introduced the notation
\begin{equation}
r_{1\mu \nu }[F(z)]=\frac{\pi}{2}\sum_{k}
{\mathrm{Res}}_{{\mathrm{Im}}z_k=0}\left[ F(z)
\sum_{l=1}^{2}\frac{H^{(l)}_{\mu }(\lambda z)}{H^{(l)}_{\nu }(z)}
\right] +\pi \sum_{k}\sum_{l=1}^{2}
{\mathrm{Res}}_{(-1)^l{\mathrm{Im}}z_k<0}
\left[ F(z)\frac{H^{(l)}_{\mu }(\lambda z)}{H^{(l)}_{\nu }(z)}
\right] .\label{r1mn}
\end{equation} 
The most important case for the applications is the limit $a\to 0$.
The following statements take place \cite{Sah1, Sahdis}:

\bigskip

\noindent {\bf Theorem 5.} {\it Let the function $F(z)$ be meromorphic for
${\mathrm{Re}}z\geq 0$ (except the possible branch point $z=0$) with poles 
$z_k,\, \pm iy_k$ ($y_k,{\mathrm{Re}}z_k>0$). If this function satisfy
conditions (\ref{cond61}) or (\ref{cond62}) and
\begin{equation}
F(xe^{\pi i/2})=-e^{(\mu -\nu )\pi i}F(xe^{-\pi i/2}),
\label{condth65}
\end{equation}
then for values of $\nu $ for which the function $H^{(1)}_{\nu }(z)$
($H^{(2)}_{\nu }(z)$) have no zeros for $0\leq argz\leq \pi /2$
($-\pi /2\leq argz\leq 0$) the following formula is valid
\begin{eqnarray}
& & {\mathrm{p.v.}}\int_{0}^{\infty}{\frac{J_\nu (x)Y_\mu (\lambda x)-
Y_\nu (x)J_\mu (\lambda x)}{J_\nu ^2(x)+Y_\nu ^2(x)}F(x)dx}=
r_{1\mu \nu }[F(z)]+\nonumber\\
& & +\frac{\pi}{2}\sum_{\eta _k=0,iy_k}\left( 2-\delta _{0\eta _k}\right)
{\mathrm{Res}}_{z=\eta _k} F(z)
\frac{H^{(1)}_{\mu }(\lambda z)}{H^{(1)}_{\nu }(z)},
\label{intJYth65}
\end{eqnarray}
where it is assumed that the integral on the left exists.}

\bigskip

\noindent {\bf Proof.} From the condition (\ref{condth65}) it follows
that for ${\mathrm{arg}} z=\pi /2$
\begin{equation}
\frac{H^{(1)}_{\mu }(\lambda z)}{H^{(1)}_{\nu }(z)}F(z)=
\frac{H^{(2)}_{\mu }(\lambda z_1)}{H^{(2)}_{\nu }(z_1)}F(z_1),
\, z_1=e^{-\pi i}, \label{rel61}
\end{equation}
and that the possible purely imaginary poles of $F(z)$ are conjugate:
$\pm iy_k,\, y_k>0$. Hence in rhs of (\ref{intJY61}) in the limit
$a\to 0$ the term in the square brackets may be presented in the form
(it can be seen similarly to (\ref{rel11}))
\begin{equation}
\left( \int_{\gamma _\rho^+}+\sum_{k}\int_{C_\rho (iy_k)}\right) 
\frac{H^{(1)}_{\mu }(\lambda z)}{H^{(1)}_{\nu }(z)}F(z)dz+
\left( \int_{\gamma _\rho^-}+\sum_{k}\int_{C_\rho (-iy_k)}\right) 
\frac{H^{(2)}_{\mu }(\lambda z)}{H^{(2)}_{\nu }(z)}F(z)dz
\label{rel62}
\end{equation}
with the same notations as in (\ref{rel11}). By using (\ref{rel61})
and the condition that the integral converges at the origin we obtain
\begin{equation}
\int_{\Omega _\rho ^+(\eta _k)} 
\frac{H^{(1)}_{\mu }(\lambda z)}{H^{(1)}_{\nu }(z)}F(z)dz+
\int_{\Omega _\rho ^-(\eta _k)} 
\frac{H^{(2)}_{\mu }(\lambda z)}{H^{(2)}_{\nu }(z)}F(z)dz=
\left( 2-\delta _{0\eta _k}\right) \pi i\, {\mathrm{Res}}_{z=\eta _k}
\frac{H^{(1)}_{\mu }(\lambda z)}{H^{(1)}_{\nu }(z)}F(z),
\label{rel63}
\end{equation}
where $\Omega _\rho ^{\pm }(0)=\gamma _\rho ^{\pm }$,
$\Omega _\rho ^{\pm }(iy_k)=C_\rho (\pm iy_k)$. By using this relation
from (\ref{intJY61}) we receive the formula (\ref{intJYth65}).
\rule{1.5ex}{1.5ex}

\bigskip

Note that one can write the residue at $z=0$ in the form
\begin{equation}
{\mathrm{Res}}_{z=0}\frac{H^{(1)}_{\mu }(\lambda z)}{H^{(1)}_{\nu }(z)}F(z)
={\mathrm{Res}}_{z=0}\frac{J_\nu (z)J_\mu (\lambda z)+
Y_\nu (z)Y_\mu (\lambda z)}{J_\nu ^2(z)+Y_\nu ^2(z)}F(z)
\label{rel64}
\end{equation}
as well. Integrals of type (\ref{intJYth65}) we have been able to 
find in literature (see, e.g., \cite{Erdelyi, Prudnikov, Erdelyi2, 
Gradshteyn}) 
are special cases of this
formula. For example, taking $F(z)=J_\nu (z)Y_{\nu +1}(\lambda ' z)-
Y_\nu (z)J_{\nu +1}(\lambda 'z)$ for the integral on the left in 
(\ref{intJYth65}) we obtain $-\lambda ^{-\nu}\lambda '^{-\nu -1}$ for 
$\lambda '<\lambda $ and $\lambda ^{\nu }\lambda '^{-\nu -1}-
\lambda ^{-\nu }\lambda '^{-\nu -1}$ for $\lambda '>\lambda $
\cite{Erdelyi2}. By taking $z^{2m+1}/(z^2+a^2)$, $z^{2m+1}/(z^2-c^2)$
 as $F_1(z)$ for $\mu =\nu $ and integer $m\geq 0$ one receive
\begin{eqnarray}
\int_{0}^{\infty}{\frac{J_\nu (x)Y_\nu (\lambda x)-
Y_\nu (x)J_\nu (\lambda x)}{J_\nu ^2(x)+Y_\nu ^2(x)}
\frac{x^{2m+1}}{x^2+a^2}dx} & = & (-1)^ma^{2m}\frac{\pi}{2}
\frac{K_\nu (\lambda a)}{K_\nu (a)},\, {\mathrm{Re}}a>0\label{examp61}\\
{\mathrm{p.v.}}\int_{0}^{\infty}{\frac{J_\nu (x)Y_\nu (\lambda x)-
Y_\nu (x)J_\nu (\lambda x)}{J_\nu ^2(x)+Y_\nu ^2(x)}
\frac{x^{2m+1}}{x^2-c^2}dx} & = & \frac{\pi}{2}c^{2m}
\frac{J_\nu (c)J_\nu (\lambda c)+
Y_\nu (c)Y_\nu (\lambda c)}{J_\nu ^2(c)+Y_\nu ^2(c)}
\label{examp62}
\end{eqnarray}
where $c>0,\, \lambda >1$. The particular cases of this formula for 
$\nu =m=0$  are given in \cite{Erdelyi2}. In (\ref{examp61}) taking
the limit $a\to 0$ and choosing $m=0$ we obtain the 
integral of this type given in \cite{Prudnikov}. In (\ref{intJYth65})
as a function $F(z)$ we can choose (\ref{func1}), (\ref{func2}),
(\ref{func4}) (the corresponding conditions for paremeters 
directly follow from (\ref{cond61}) or (\ref{cond62})) with
$\rho =\mu -\nu -2m$ ($m$ - integer), as well as any products
between them and with $\prod_{l=1}^{n}(z^2-c_l^2)^{-k_l}$. For
instance,
\begin{eqnarray}
& & \int_{0}^{\infty}{\frac{J_\nu (x)Y_\nu (\lambda x)-
Y_\nu (x)J_\nu (\lambda x)}{J_\nu ^2(x)+Y_\nu ^2(x)}
\prod_{l=1}^{n}\frac{J_{\mu _l}(b_l\sqrt{x^2+z_l^2})}
{\left( x^2+z_l^2\right)^{\mu _l/2}}
\frac{dx}{x}}=\frac{\pi}{2\lambda ^\nu }\prod_{l=1}^{n}z^{-\mu _l}
J_{\mu _l}(b_lz_l), \label{examp63}\\
& & \quad b_l, \, {\mathrm{Re}}\nu >0,\, {\mathrm{Re}}z_l\geq 0,\, 
 \lambda >1,\, \sum_{l=1}^{n}{\mathrm{Re}}\mu _l+n/2+1>
 \delta _{b,\lambda -1},\, b\equiv \sum_{l=1}^{n}b_l\leq \lambda -1
\nonumber
\end{eqnarray}

As another consequence of (\ref{intJY61}) one has:

\bigskip

\noindent {\bf Theorem 6.} {\it  Let $F(z)$ be meromorphic in the 
right half-plane (with possible exception $z=0$) with poles 
$z_k,\, {\mathrm{Re}}z_k>0$, and satisfy conditions (\ref{cond61}) or 
(\ref{cond62}) and
\begin{equation}
F(ze^{\pi i})=-e^{(\mu -\nu )\pi i}F(z)+o\left( z^{|{\mathrm{Re}}\mu |-
|{\mathrm{Re}}\nu |-1}\right) ,\quad z\to 0,
\label{condth61}
\end{equation}
then for values of $\nu $ for which the function $H^{(1)}_{\nu }(z)$
($H^{(2)}_{\nu }(z)$) have no zeros for $0\leq argz\leq \pi /2$
($-\pi /2\leq argz\leq 0$) the following formula takes place
\begin{eqnarray}
& & {\mathrm{p.v.}}\int_{0}^{\infty}{\frac{J_\nu (x)Y_\mu (\lambda x)-
Y_\nu (x)J_\mu (\lambda x)}{J_\nu ^2(x)+Y_\nu ^2(x)}F(x)dx}=
r_{1\mu \nu }[F(z)]+\frac{\pi }{2}{\mathrm{Res}}_{z=0}\frac{H_\mu ^{(1)}
(\lambda z)}{H_\nu ^{(1)}(z)}F(z)+\nonumber\\
& & +\frac{1}{2}\int_{0}^{\infty}{\frac{K_\mu (\lambda x)}{K_\nu (x)}
\left[ e^{(\nu -\mu )\pi i/2}F(xe^{\pi i/2})+e^{(\mu -\nu )
\pi i/2}F(xe^{-\pi i/2})\right] dx},\quad \lambda >1,
\label{th6form}
\end{eqnarray}
provided the integral on the left exists.}

\bigskip

\noindent {\bf Proof.} This result immediately follows 
from (\ref{intJY61}) in the 
limit $a\to 0$ and from (\ref{rel63}) with $\eta _k=0$.
\rule{1.5ex}{1.5ex}

\bigskip

For example, by using (\ref{th6form}) one obtains \begin{eqnarray}
& & \int_{0}^{\infty}{\frac{J_\nu (x)Y_\mu (\lambda x)-
Y_\nu (x)J_\mu (\lambda x)}{J_\nu ^2(x)+Y_\nu ^2(x)}
\prod_{l=1}^{n}J_{\mu _l}(b_lx)dx}=\cos 
\mu _s \int_{0}^{\infty}{\frac{K_\mu (\lambda x
)}{K_\nu (x)}\prod_{l=1}^{n}I_{\mu _l}(b_lx)dx},\label{examp64}\\
& & \sum_{l=1}^{n}{\mathrm{Re}}\mu _l+|
{\mathrm{Re}}\nu |>|{\mathrm{Re}}\mu |-1,\, b=\sum_{l=1}^{n}
b_l\leq \lambda -1,\, b_l>0,\, n>\delta _{b,\lambda -1},\, 
\mu _s\equiv \nu -\mu +\sum_{l=1}^{n}\mu _l.
\nonumber
\end{eqnarray} 
Such relations are convenient in numerical calculations of 
integrals on the left as the subintegrand on the right 
 at infinity goes to zero exponentially fast.

We have considered the formulas containing 
$J_\nu (z)Y_\mu (\lambda z)-Y_\nu (z)J_\mu (\lambda z)$.
 The similar results can
be obtained for integrals containing the functions 
$J'_\nu (z)Y_\mu (\lambda z)-Y'_\nu (z)J_\mu (\lambda z)$ and
$J'_\nu (z)Y'_\mu (\lambda z)-Y'_\nu (z)J'_\mu (\lambda z)$.

\section{Applications to the Casimir effect. Vacuum energy
density and stress inside a
perfectly conducting spherical shell}

\renewcommand{\theequation}{8.\arabic{equation}}

\setcounter{equation}{0}

In this and next sections we shall consider applications of
the summations formulae obtained in previous sections to the 
physical problem, namely the Casimir effect. In what follows
on the example of spherical and cylindrical geometries we
will show that the using of GAPF allows to obtain the regularized 
values of physical quantities in cases then the explicit dependence
of eigenmodes on quantum numbers is complicated and irregular.

Historically the investigation of the Casimir effect for a perfectly
conducting spherical shell was motivated by Casimir semiclassical
model of an electron. In this model Casimir suggested that 
Poincare stress to stabilize the charged particle could arise
from vacuum quantum fluctuations and the fine structure constant
can be determined by a balance between the Casimir force 
(assumed attractive) and the Coulomb repulsion. However,
 as it have been shown by
Boyer \cite{Boyer}, the Casimir energy for the sphere is positive,
implying a repulsive force. This result has later been 
reconsidered by a number of authors \cite{DaviesSph, Balian, MiltonSph}.
More recently new methods have been developed for this problem
including a direct mode summation \cite{Nesterenko, Bowers} 
and zeta function \cite{Leseduarte, Cognola, Lambiase} approaches. 
However the main part of studies have 
focused on global quantities such as total energy. The 
investigation of the energy distribution inside a perfectly
reflecting spherical shell was made in \cite{Olaussen1} in the 
case of QED and in \cite{Olaussen2} for QCD. The distribution
of the other components for the electromagnetic EMT inside as 
well as outside the shell can be obtained from the results 
of \cite{Brevik1, Brevik2}. In these papers the consideration
was carried out in terms of Schwinger's source theory. In
\cite{Grig1, Grig2, Grig3} the calculations of the regularized 
values for vev of the EMT 
components inside and outside the perfectly conducting
spherical shell are based on the generalized Abel-Plana 
summation formula.
Our consideration below is based on this approach.

The main quantities we will consider here are vacuum expectation
values (vev) of the energy-momentum tensor (EMT) 
for the electromagnetic field inside a 
perfectly conducting spherical shell of radius $a$. It may be 
obtained by using the standard formula of mode summation
\cite{Mostepanenko, Birrel}
\begin{equation}
\langle 0\vert T_{ik}\vert 0\rangle =\sum_{\alpha }T_{ik}(x)
\left\{ \Psi _{\alpha }^{(-)}(x), \Psi _{\alpha }^{(+)}(x)
\right\}, \label{emtgform}
\end{equation}
where bilinear form $T_{ik}\{ f,g\} $ for a field $\Psi $
is given by the classical EMT. Here $\vert 0\rangle$ is the amplitude 
of the vacuum state, $\left\{ \Psi _{\alpha }^{(\pm )}(x)\right\} $ 
is a complete set of the 
positive and negative frequency solutions to the field equations,
satisfying the boundary conditions, and subscript $\alpha $ may 
contain discrete and continous components.

In the case of the electromagnetic field inside the perfectly conducting 
sphere, by using Coulomb gauge for vector potential ${\mathbf{A}}$, the 
corresponding system of solutions, regular at $r=0$, can be 
presented in the form
\begin{equation}
{\mathbf{A}}_\alpha =\omega ^{-1}\beta _l(a,\omega )
\left\{ \begin{array}{ll}
j_l(\omega r){\mathbf{X}}_{lm}e^{-i\omega t} & \textrm{if $\quad \lambda =0$}\\
\omega ^{-1}\nabla \times \left[ j_l(\omega r)
{\mathbf{X}}_{lm}\right] e^{-i\omega t}
& {\textrm{if $\quad \lambda =1$}} \end{array} \right.,
\quad \alpha =({\omega lm\lambda }),
\label{eigfuncins}
\end{equation}
where $\lambda =0$ and 1 correspond to the spherical waves of
electric and magnetic type (TM and TE - modes). They describe photon with
definite values of total momentum $l$, its projection $m$,
energy $\omega $ and parity $(-1)^{l+\lambda +1}$ (units $\hbar =c=1$
are used). Here the vector spherical harmonics have the form
\begin{equation}
{\mathbf{X}}_{lm}(\theta , \varphi )=-i\frac{{\mathbf{r}}\times \nabla }
{\sqrt{l(l+1)}}Y_{lm}(\theta , \varphi ),\quad l\neq 0,
\label{vecspharm}
\end{equation}
with $Y_{lm}$ being spherical function, and 
 $j_l(x)=\sqrt {\pi /2x}J_{l+1/2}(x)$ is 
spherical Bessel function. The coefficients 
$\beta _l(a,\omega )$ are determined by the normalization condition
\begin{equation}
\int {dV{\mathbf{A}}_\alpha \cdot {\mathbf{A}}^{\ast }_{\alpha '}}=
\frac{2\pi }{\omega }\delta _{\alpha \alpha '},
\label{normcond}
\end{equation}
where the integration goes over the region inside the sphere. Using
the standard formulae for vector spherical harmonics and 
spherical Bessel functions (see, for example, \cite{Jackson})
one finds
\begin{equation}
\beta _l^2(a,\omega )=8\omega ^3T_{\nu }(\omega a)/a ,\quad
\nu =l+1/2,
\label{normcoef}
\end{equation}
where $T_{\nu }(z)$ is defined in (\ref{teka}).

Inside the perfectly conducting sphere the photon energy levels
are quantized by standard boundary conditions:
\begin{equation}
{\mathbf{r}}\times {\mathbf{E}} =0,\quad 
{\mathbf{r}}\cdot {\mathbf{B}}=0,\quad r=a,
\label{boundcond}
\end{equation}
where ${\mathbf{E}}$ and ${\mathbf{B}}$ are the electric and 
magnetic fields. They lead to the 
following eigenvalue equations with respect to $\omega $
\begin{eqnarray}
j_l(\omega r)\vert _{r=a} & = & 0 \quad 
\textrm{if $\lambda =0$} \label{TMmode}\\
\frac{d}{dr}\left[ rj_l(\omega r)\right]_{r=a} & = & 0 \quad
 \textrm{if $\lambda =1$} \label{TEmode}
\end{eqnarray}
It is well known that these equations have infinite number of
real simple roots \cite{Watson, abramowiz}.

By substituting the eigenfunctions into (\ref{emtgform}) with
the standard expression of the electromagnetic EMT and after the 
summation over $m$ by using the standard formulae 
for vector spherical harmonics (see, for example, \cite{Jackson})
one obtains
\begin{equation}
\langle 0\vert T_{ik}\vert 0\rangle ={\mathrm{diag}}
\left( \varepsilon ,\, -p,
\, -p_{\bot},\, -p_{\bot} \right)
\label{emt1}
\end{equation}
(here index values 1,2,3 correspond to the spherical 
coordinates $r,\theta ,\varphi $ with origin at the sphere centre). 
Energy density, $\varepsilon $, pressures in transverse, 
$p_{\perp }$, and radial, $p$, directions are determined by relations
\begin{eqnarray}
q(a,r) & = & \frac{1}{4\pi ^2a}\sum_{\omega l\lambda}
\omega ^3T_\nu (\omega a)D_{l}^{(q)}(\omega r),
\quad q=\varepsilon ,\, p,\, p_\perp ,
\label{q1}\\
p(a,r) & = & \varepsilon -2p_\perp , \label{tr0}
\end{eqnarray}
where the following notations are introduced
\begin{equation}
D_{l}^{(q)}(y)=\left\{ \begin{array}{l}
lj_{l+1}^2(y)+(l+1)j_{l-1}^2(y)+(2l+1)j_{l}^2(y),\quad q=\varepsilon \\
l(l+1)(2l+1)j_{l}^2(y)/y^2, \quad q=p_\perp \end{array}\right. 
\label{Dl}
\end{equation}
In the sum (\ref{q1}) $\omega $ takes discrete set of values determined by
the equations (\ref{TMmode}) and (\ref{TEmode}). The relation (\ref{tr0})
corresponds to the zero trace of the EMT.

The vev (\ref{q1}) are infinite. The renormalization of 
$\langle 0\vert T_{ik}\vert 0\rangle $ in flat spacetime is affected 
by subtracting from this quantity its singular part 
$\langle \bar 0\vert T_{ik}\vert \bar 0\rangle $, which is precisely the
value it would have if the boundary were absent. Here 
$\vert \bar 0\rangle $ is the amplitude for Minkowski vacuum state. To 
evaluate the finite difference between these two infinities we will 
introduce a cutoff function $\psi _{\mu }(\omega )$, which decreases
with increasing $\omega $ and satisfies the condition
$\psi _{\mu }(\omega )\to 1,\, \mu \to 0$, and makes the sums finite.
After subtracting we will allow $\mu \to 0$ and will show that the 
result does not depend on the form of cutoff:
\begin{equation}
{\mathrm{reg}}\langle 0\vert T_{ik}\vert 0\rangle =\lim_{\mu \to 0}
\left[ \langle 0\vert T_{ik}\vert 0\rangle - 
\langle \bar 0\vert T_{ik}\vert \bar 0\rangle \right] .
\label{reg}
\end{equation} 
Hence we consider the following finite quantities
\begin{equation}
q(\mu ,a,r)=\frac{1}{4\pi ^2a^4}\sum_{l=1}^{\infty}
\sum_{k=1}^{\infty} \sum_{\lambda =0}^{1}j_{\nu ,k}^{(\lambda )3}
T_\nu (j_{\nu ,k}^{(\lambda )})\psi _\mu (j_{\nu ,k}^{(\lambda )}/a)
D_l^{(q)}(j_{\nu ,k}^{(\lambda )}x), \quad x=r/a, \label{q2}
\end{equation}
where $\omega =j_{\nu ,k}^{(\lambda )}/a$  
are solutions to the eigenvalue equations
(\ref{TMmode}) and (\ref{TEmode}) for $\lambda =0,1$, 
respectively. The summations over $k$ in (\ref{q2})
can be done by using the formula (\ref{sumJ1anal}) and taking
$A=1,B=0$  for TM-modes ($\lambda =0$) and $A=1,B=2$ 
for TE-modes ($\lambda =1$) (recall that in (\ref{sumJ1}) 
$\lambda _{\nu ,k}$ are zeros of $\bar J_\nu (z)$ with
bared quantities defined as (\ref{efnot1})). Note that the
resulting sums are of type (\ref{sumJ1analcas}). 
Let us substitute in formula (\ref{sumJ1anal})
\begin{equation}
f(z)=z^3\psi _\mu (z/a)D_l^{(q)}(zx), \label{ftoAPF}
\end{equation}
with $D_l^{(q)}(y)$ defined from (\ref{Dl}). 
We will assume the class of cutoff functions
for which the function (\ref{ftoAPF}) satisfies conditions for 
Theorem 2, uniformly with respect to $\mu $ (the corresponding
restrictions for $\psi _\mu $ can be easily found from these
conditions using the asymptotic formulae for Bessel 
functions). Below for simplicity 
we will consider the functions with no poles. In this case
(\ref{ftoAPF}) is analytic on the right-half
plane of the complex variable $z$. The discussion on
the conditions to cutoff functions under which the difference
between divergent sum and integral exists and has a finite
value independent any further details of cutoff function see
\cite{Barton1}. For TE- and TM-modes by choosing the constants
$A$ and $B$ as mentioned above one obtaines
\begin{equation}
q=\frac{1}{8\pi ^2}\sum_{l=1}^{\infty}\left\{ 2\int_{0}^{\infty}
{\omega ^3\psi _\mu (\omega )D_l^{(q)}(\omega r)
d\omega }-
\frac{1}{a^4}\int_{0}^{\infty}{\chi _\mu (z/a)
 F_l^{(q)}(z,x)dz}\right\} ,\quad q=\varepsilon ,p_\bot ,p,
 \label{q3}
\end{equation}
where for $x<1$ the functions $F_l^{(q)}(z,x)$ are defined as
\begin{eqnarray}
F_l^{(\varepsilon )}(z,x) & = & \frac{z}{x^2}
\left[ \frac{e_l(z)}{s_l(z)}+\frac{e'_l(z)}{s'_l(z)}\right] \left[ l
s_{l+1}^2(zx)+(l+1)s_{l-1}^2(zx)-(2l+1)s_{l}^2(zx)\right] ,
\label{Feps}\\
F_l^{(p_\perp )}(z,x) & = & (2l+1)\frac{l(l+1)}{zx^4}
\left[ \frac{e_l(z)}{s_l(z)}+\frac{e'_l(z)}{s'_l(z)}\right] s_{l}^2(zx),
\label{Fp}\\
F_l^{(p)}(z,x) & = & F_l^{(\varepsilon )}-2F_l^{(p_\perp )},\quad
\chi _\mu (y)=\left[ \psi _\mu (iy)+\psi _\mu (-iy)\right] /2.
\label{chisph}
\end{eqnarray}
In these expressions we have introduced Ricatti-Bessel functions
of imaginary argument,
\begin{equation}
s_l(z)=\sqrt{\frac{\pi z}{2}}I_\nu (z),\quad
e_l(z)=\sqrt{\frac{2 z}{\pi }}K_\nu (z),\, \nu =l+1/2 .
\label{RicBes}
\end{equation}
As $\langle \bar 0\vert T_{ik}\vert \bar 0\rangle =\lim_{a\to \infty }
\langle 0\vert T_{ik}(\mu ,a,r)\vert 0\rangle$ the first integral
in (\ref{q3}) represents the vacuum EMT for empty Minkowski
spacetime:
\begin{equation}
q = \frac{1}{4\pi ^2}\sum_{l=1}^{\infty}
\int_{0}^{\infty}{\omega ^3\psi _\mu (\omega )D_l^{(q)}(\omega r)
d\omega }. \label{EMTMin}
\end{equation}
This expression can be further simplified. For example 
in the case of the energy density one has
\begin{eqnarray}
\varepsilon ^{(0)} & = & \frac{1}{4\pi ^2}\sum_{l=1}^{\infty }
\int_{0}^{\infty }{\omega ^3\psi _\mu (\omega )
\left[ lj_{l+1}^{2}(\omega r)+(l+1)j_{l-1}^{2}(\omega r)+
(2l+1)j_l^{2}(\omega r)\right] d\omega }= {} \nonumber\\
&  & = \frac{1}{2\pi ^2}
\int_{0}^{\infty }{\omega ^3\psi _\mu (\omega )
\sum_{l=0}^{\infty }(2l+1)j_{l}^{2}(\omega r)d\omega }
=\int_{0}^{\infty}{\omega ^3\psi _\mu (\omega )d\omega }.
\label{epsMin}
\end{eqnarray} 

As we see the using of GAPF allows us to extract from 
infinite quantities the divergent part without specifying the 
form of cutoff function. Now the regularization of the EMT is equivalent
to the omitting the first summand in (\ref{q3}), which as we saw 
corresponds to the contribution of the spacetime without boundaries.
For the regularized components one obtains
\begin{equation}
{\mathrm{reg}}\, q(a,r)=-\frac{1}{8\pi ^2a^4}
\sum_{l=1}^{\infty}\int_{0}^{\infty}
{\chi _\mu (z/a)F_l^{(q)}(z,x)dz},\quad r<a,\quad 
q=\varepsilon , p,p_\perp .
\label{regq}
\end{equation}
By using the recurrence relations for Riccati-Bessel modified
functions the expressions for the regularized vacuum energy 
density and radial pressure may be presented in the form
\begin{equation}
q=\frac{-1}{8\pi ^2a^2r^2}
\sum_{l=1}^{\infty}(2l+1)\int_{0}^{\infty}
{dzz\chi _\mu (z/a) \left[ \frac{e_l(z)}{s_l(z)}+
\frac{e'_l(z)}{s'_l(z)}\right] \left\{ s_l^{'2}(zx)-
s_l^{2}(zx)\left[ 1-\frac{l(l+1)}{(-1)^iz^2x^2}\right] \right\} },
\label{regqnew}
\end{equation}
where $i=0$ for $q=\varepsilon $ and $i=1$ for $q=p$.
Here we keep the cutoff factor because it plays an 
important role in the calculations of the total Casimir energy
for the spherical shell (see below). The derivation of the 
vacuum densities (\ref{regq}), (\ref{regqnew}) given above
uses GAPF to summarize mode sums and is based on \cite{Grig1, Grig2}.
One can see that this formulae for the case of exponential 
cutoff function may be obtained also from the results of 
\cite{Brevik1, Brevik2}, where Green function method is used.

We obtained the regularized values (\ref{regq}) by introducing 
a cutoff function and susequent subtracting the contribution 
due to the unbounded space. The GAPF in the form (\ref{sumJ1anal})
allows to obtain immediately this finite difference. 
However it should be noted that by using GAPF in the form 
(\ref{sumJ1}) we can derive the expressions for the regularized 
azimuthal pressure without introducing any special cutoff function.
To see this note that for $x<1$ the function (\ref{ftoAPF}) with 
$q=p_\perp $ and $\psi _\mu =1$ satisfies conditions to 
Theorem 2. It follows from here that we can apply the formula
(\ref{sumJ1}) directly to the corresponding 
sum over $\omega $ in (\ref{q1})
or over $k$ in (\ref{q2}) (with $\psi _\mu =1$) without
introducing the cutoff function. This immediately yields to 
the formula (\ref{regq}) for $q=p_\perp $ with $\psi _\mu =1$.

Let us consider the behaviour of the functions $F_l^{(q)}(z,x)$
in various limiting cases. By using the corresponding formulae
for Bessel functions one obtains:

(a) When $l$ is fixed and $z\to 0$
\begin{equation}
F_l^{(\varepsilon )}(z,x)\sim \pi (l+1/2)x^{2(l-1)},
\quad F_l^{(p_\perp )}(z,x)\sim \pi lx^{2(l-1)}/2.
\label{asympcentre}
\end{equation}

(b) When $l$ is fixed and $zx$ is large
\begin{equation}
F_l^{(\varepsilon )}\sim 2F_l^{(p_\perp )}\sim l^2(l+1)^2
(2l+1)\frac{\pi }{2(zx)^4}\exp [-2z(1-x)],
\label{asympzinf}
\end{equation}
and the integral over $z$ in (\ref{regq}) converges for all
$x\leq 1$;

(c) For large $l$ by using the uniform asymptotic expansions for
Bessel functions \cite{abramowiz} one finds
\begin{equation}
F_l^{(q)}(\nu z,x)\sim \Phi _l^{(q)}(z,x)
 \exp \left\{ -2\nu [\eta (z)-
\eta (zx)]\right\} ,\quad \nu =l+1/2
\label{phiasymp}
\end{equation}
with
\begin{eqnarray}
\Phi _l^{(\varepsilon )}(z,x) & = & \frac{\pi \nu }{x^3}t(zx)t^3(z)
\left\{ 1-\frac{1}{12\nu}\left[ t(z)(t^2(z)+3)+t(zx)
(5t^2(zx)-9)\right] \right\} , \label{phiasympeps}\\
\Phi _l^{(p)}(z,x) & = & \frac{\pi }{2x^3}t^2(zx)t^3(z),
\label{phiasymppe}
\end{eqnarray}
where the standard notations are used:
\begin{equation}
t(z)=\frac{1}{\sqrt{1+z^2}},\quad \eta (z)=\sqrt{1+z^2}+
\ln \frac{z}{1+\sqrt{1+z^2}}. \label{tezeteta}
\end{equation}

From these asymptotic formulae it follows that in (\ref{regq}) 
and  (\ref{regqnew}) the rhs is finite for $x<1$ and the cutoff 
may be removed by putting $\chi _\mu =1, \mu \to 0$.
From here it is obvious the independence of the 
regularized quantities on the specific form of the 
cutoff, on class of functions for which (\ref{ftoAPF})
satisfy conditions for (\ref{sumJ1anal}).

From (\ref{phiasymp}), (\ref{phiasympeps}) and (\ref{phiasymppe}) 
it follows that the vev of the EMT diverge at sphere 
surface, $x\to 1$, due to the contribution of large $l$
(note that, as it follows from (\ref{phiasympeps}) and 
(\ref{phiasymppe}), in (\ref{regq}) with $\chi _\mu =1$ 
the integral over $z$ converges at $x=1$). 
The corresponding asymptotic behaviour can be 
found by using the uniform asymptotic expansions given
above and the leading terms have the form
\begin{equation}
\varepsilon \sim 2p_\perp \sim \frac{-1}{30\pi ^2a(a-r)^3},
\quad p \sim \frac{-1}{60\pi ^2a^2(a-r)^2} \, .
\label{nearsurface}
\end{equation}
These surface divergences originate in the unphysical nature
of perfect conductor boundary conditions and are well known 
in quantum field theory with boundaries.
They are investigated in detail for various types
of fields and general shape of the boundary 
\cite{Deutsch, Kennedy}. Eqs. (\ref{nearsurface}) are 
particular cases of the asymptotic expansions for EMT vev
near the smooth boundary given in these papers. 
In reality the expectation values 
for the EMT components will attain a limiting value on 
the conductor surface, which will depend on the molecular 
details of the conductor. From the asymptotic expansions
given above it follows that the main contributions to 
$q(r)$ are due to the frequencies $\omega <(a-r)^{-1}$. 
Hence we expect that the formulae (\ref{regq}) are valid for
real conductors up to distances $r$ for which 
$(a-r)^{-1}\ll \omega _0$, with $\omega _0$ being the 
characteristic frequency, such that for 
$\omega >\omega _0$ the conditions for perfect conductivity  
are failed.

At the sphere centre in (\ref{regq}) $l=1$ multipole contributes
only and we obtain \cite{Olaussen1, Grig2}
\begin{eqnarray}
\varepsilon (0) & = & -\frac{1}{2\pi ^2a^4}\int_{0}^{\infty}
dzz^3
{\left[ \left( \frac{z-1}{z+1}e^{2z}+1\right) ^{-1}-
\left( \frac{z^2-z+1}{z^2+z+1}e^{2z}-1\right) ^{-1}\right] }=-
0.0381a^{-4},\nonumber\\
p(0) & = & p_\perp (0)=\varepsilon (0)/3. \label{atcenter}
\end{eqnarray}
At centre the equation of state for the electromagnetic vacuum
is the same as that for blackbody radiation. Note that the 
corresponding results obtained using the uniform asymptotic
expansions for Bessel functions \cite{Brevik1, Brevik2} are 
in good agreement with (\ref{atcenter}).

The components of the regularized EMT satisfy continuity equation
$T^k_{i;k}=0$, which for the spherical geometry takes the form
\begin{equation}
p'(r)+\frac{2}{r}(p-p_\perp )=0. \label{conteq}
\end{equation}
From here by using the zero trace condition the following
integral relations may be obtained
\begin{equation}
p(r)=\frac{1}{r^3}\int_{0}^{r}{\varepsilon (t)t^2dt}=
\frac{2}{r^2}\int_{0}^{r}{p_\perp (t)tdt}, \label{intrel7}
\end{equation}
where the integration constant is determined from the 
relations (\ref{atcenter}) at the sphere centre.
It follows from the first relation that the total energy 
within a sphere with radius $r$ is equal to
\begin{equation}
E(r)=4\pi \int_{0}^{r}{\varepsilon (t)t^2dt}=3V(r)p(r),
\label{thermrel}
\end{equation}
where $V(r)$ is the corresponding volume.

The distribution for the
vacuum energy density and pressures inside the perfectly
conducting sphere can be obtained from the 
results of the numerical calculations given in
\cite{Brevik1, Brevik2}. In their calculations Brevik and
Kolbenstvedt use the uniform asymptotic expansions of Ricatti-Bessel
functions for large values of order. In \cite{Grig1, Grig2} 
(see also \cite{Sahdis}) the corresponding
quantities are calculated on the base of the exact relations
for these functions and the accuracy of the numerical results
in \cite{Brevik1, Brevik2} is estimated 
($\approx 5\% $). The simple approximataion formulae are 
presented with the same accuracy as asymptotic expressions.
Note that inside the sphere all quantities  
$\varepsilon ,\, p,\, p_\perp $ are negative and corresponding
vacuum forces tend to contract sphere.

\section{Electromagnetic vacuum EMT outside a 
spherical shell}

\renewcommand{\theequation}{9.\arabic{equation}}

\setcounter{equation}{0}

Now let us consider the electromagnetic vacuum in the region 
outside of a perfectly conducting sphere. To deal with
discrete modes we firstly consider vacuum fields in the region
between two cocentric conducting spherical shells with radii
$a$ and $b$, $a<b$. Letting $b\to \infty$ we will obtain from 
here the result for the region under question.

By using Coulomb gauge the complete set of solutions to the 
Maxwell equations
can be written in the form similar to (\ref{eigfuncins})
\begin{equation}
{\mathbf{A}}_{\omega lm\lambda}({\mathbf{r}}, t)=
\frac{e^{-i\omega t}}{\sqrt{4\pi }}\beta _{\lambda l}(a,b,\omega )
\left\{ \begin{array}{ll}
\omega g_{0l}(\omega a,\omega r){\mathbf{X}}_{l
m} & \textrm{if $\quad \lambda =0$}\\
\nabla \times 
\left[ g_{1l}(\omega a,\omega r){\mathbf{X}}_{lm}\right] 
& \textrm{if $\quad \lambda =1$} \end{array} \right.,
\label{eigfuncab}
\end{equation}
where as above the values $\lambda =0$ and $\lambda =1$ correspond 
to the waves of magnetic (TE-modes) and electric (TM-modes) type,
\begin{equation}
g_{\lambda l}(x,y)=
\left\{ \begin{array}{ll}
j_l(y)n_l(x)-j_l(x)n_l(y) & \textrm{if $\quad \lambda =0$}\\
j_l(y)[xn_l(x)]^{'}-[xj_l(x)]^{'}n_l(y),
& \textrm{if $\quad \lambda =1$} \end{array} \right.,
\label{gl8}
\end{equation}
with $n_l(x)$ being Neumann spherical function. From the standard
boundary conditions at surfaces $r=a$ and $r=b$ one finds that
possible energy levels of photon are solutions to the 
following equations
\begin{equation}
\left( \frac{d}{dr}\right) ^\lambda \left[ rg_{\lambda l}
(\omega a,\omega r)\right] _{r=b}=0,\quad \lambda =0,1.
\label{modesab}
\end{equation}
All roots of these equations are real and simple \cite{abramowiz}.

The coefficients $\beta _{\lambda l}$ in (\ref{eigfuncab}) are determined
from the normalization condition (\ref{normcond}), where now 
the integration goes over the region between spherical shells,
$a\leq r\leq b$. By using the standard relations for spherical
Bessel functions they can be presented in the form
\begin{eqnarray}
\beta _{0l}^2 & = & 
\omega a\left[ \frac{aj_l^2(\omega a)}{bj_l^2(\omega b)}-1
\right] ^{-1}, \quad \lambda =0 \label{normab}\\
\beta _{1l}^2 & = & 
\frac{1}{\omega a}\left\{ \frac{b\left[ \omega aj_l(\omega a)\right] ^{'2}}
{a\left[ \omega bj_l(\omega b)\right] ^{'2}}\left[ 1-
\frac{l(l+1)}{\omega ^2b^2}\right] -1+\frac{l(l+1)}{\omega ^2a^2}
\right\} ^{-1}, \quad \lambda =1 .
\label{normab1}
\end{eqnarray}
From (\ref{emtgform}) with the electromagnetic field EMT and functions
(\ref{eigfuncab}) as a complete set of solutions one obtains the 
vev in the form (\ref{emt1}) with
\begin{equation}
q(a,b,r)=\frac{1}{8\pi }\sum_{\omega l\lambda }(2l+1)
\omega ^4\beta ^2_{\lambda l}f_{\lambda l}^{(q)}(\omega a,\omega r),
\quad q=\varepsilon,\, p,\, p_\perp ,
\label{epspeab}
\end{equation}
where the frequencies $\omega $ are solutions to the equations 
(\ref{modesab}), and
\begin{eqnarray}
f_{\lambda l}^{(\varepsilon )}(\omega a,\omega r) & = & 
\left[ 1+\frac{l(l+1)}
{\omega ^2r^2}\right] g^2_{\lambda l}(\omega a,\omega r) +
\frac{1}{\omega ^2r^2}\left[ \frac{d}{d(\omega r)}
\left( \omega r g_{\lambda l}(\omega a,\omega r)
\right) \right] ^{2}, \label{fab}\\
f_{\lambda l}^{(p_\perp )}(\omega a,\omega r) & = &
\frac{l(l+1)}{\omega ^2r^2}g^2_{\lambda l}(\omega a,\omega r).
\label{fabperp}
\end{eqnarray}
 It is easy to see that the eigenvalue equations (\ref{modesab})
can be written in terms of the function $C_\nu ^{AB}$, defined by
(\ref{bescomb1}), as
\begin{equation}
C_\nu ^{AB}(\eta ,\omega a)=0,\quad \nu =l+1/2,\, \eta =b/a,\, 
A=1/(1+\lambda),\, B=\lambda,\, \lambda =0,1.
\label{modesabC}
\end{equation}
By this choice of constants $A$ and $B$ the normalization coefficients
(\ref{normab}) and (\ref{normab1}) are related with the function
$T_\nu ^{AB}$ from (\ref{tekaAB}) as:
\begin{equation}
\beta _{\lambda l}^2=T_\nu ^{AB}(\eta ,\omega a).
\label{normabT}
\end{equation}
This allows to use the formulae from section 4 for the summation over
eigenmodes.

As above to regularize the infinite quantities (\ref{epspeab})
we introduce a cutoff function $\psi _\mu (\omega )$ and consider the 
difference
\begin{equation}
{\mathrm{reg}}\langle 0\vert T_{ik}\vert 0\rangle =\lim_{\mu \to 0}
\left[ \langle 0\vert T_{ik}(\mu , a,b)\vert 0\rangle - 
\lim_{a \to 0}\lim_{b\to \infty}\langle 0\vert T_{ik}(\mu , a,b)
\vert 0\rangle \right] .
\label{regab}
\end{equation}
This procedure is equivalent to the subtraction of Minkowskian
part without boundaries. 

Hence instead of (\ref{epspeab}) we consider 
the finite quantities
\begin{equation}
q=\frac{1}{8\pi a^4}\sum_{l=1}^{\infty}(2l+1)
\sum_{k=1}^{\infty}\sum_{\lambda =0}^{1}
\gamma _{\nu ,k}^{(\lambda )4}T_{\nu }^{AB}(\eta ,
\gamma _{\nu ,k}^{(\lambda )})\psi _\mu (\gamma _{\nu ,
k}^{(\lambda )}/a)f_{\lambda l}^{(q)}(\gamma _{\nu ,k}^{(\lambda )},
\gamma _{\nu ,k}^{(\lambda )}x)
, \quad q=\varepsilon ,p_\perp ,
\label{qab}
\end{equation}
where $x=r/a$, and $\omega a=\gamma _{\nu ,k}^{(\lambda )}$ 
 are solutions to the equations 
(\ref{modesab}) or (\ref{modesabC}).
To sum over $k$ we will use the formula (\ref{cor3form}) with
\begin{equation}
h(z)=z^4\psi _\mu (z/a)f_{\lambda l}^{(q)}(z,zx),
\label{hzab}
\end{equation}
assuming a class of cutoff functions for which (\ref{hzab}) 
satisfies to the conditions (\ref{cond31}) and (\ref{cor3cond1})
uniformly with respect to $\mu $.
The corresponding restrictions on $\psi _\mu $ can be obtained using
 the asymptotic formulae for Bessel functions. From (\ref{qab}) 
by applying to the sum over 
$k$ the formula (\ref{cor3form}) for the EMT components one obtains
\begin{eqnarray}
q & = & \frac{1}{8\pi ^2a^4}\sum_{l=1}^{\infty}(2l+1)
\sum_{\lambda =0}^{1}
\left\{ \int_{0}^{\infty}{z^3\psi _\mu (z/a)\frac{f_{\lambda l}^{(q)}
(z,zx)}{\Omega _{1\lambda l}(z)}dz}+\right. \nonumber \\ 
& & \left.+\frac{1}{x^2}\int_{0}^{\infty}{
\frac{e_l^{(\lambda )}(\eta z)}{e_l^{(\lambda )}(z)} 
\frac{z\chi _\mu (z/a)F_{\lambda l}^{(q)}(z,zx)}{\left[ (\partial /
\partial y)^\lambda G_{\lambda l}(z,y)
\right] _{y=z\eta }}dz}\right\} ,
 \label{sum1kab}
\end{eqnarray}
where we use the notations
\begin{equation}
e_l^{(\lambda )}(y)\equiv \left( \frac{d}{dy}\right) ^\lambda e_l(y),
\quad  s_l^{(\lambda )}(y)\equiv \left( \frac{d}{dy}
\right) ^\lambda s_l(y) \label{RBdernot} 
\end{equation}
for the Riccati-Bessel functions derivatives,
\begin{equation}
\Omega _{1\lambda l}(z)=\left\{ \begin{array}{ll}
j_l^2(z)+n_l^2(z), & \lambda =0 \\
 \left[ zj_l(z)\right] ^{'2}+[zn_l(z)]^{'2}, & \lambda =1
\end{array} \right.,
\label{Omega8}
\end{equation}
and
\begin{eqnarray}
G_{\lambda l}(x,y) & = & 
e_l^{(\lambda )}(x)s_l(y)-e_l(y)s_l^{(\lambda )}(x), \quad \lambda =0,1,
\label{Gl01}\\
F_{\lambda l}^{(\varepsilon )}(z,y) & = & 
\left[ \frac{\partial }{\partial y}G_{\lambda l}(z,y)\right] ^2+
\left[ \frac{l(l+1)}{y^2}-1\right]G^2_{\lambda l}(z,y),
\label{Fepsab} \\
F_{\lambda l}^{(p_\perp )}(z,y) & = &
l(l+1)G^2_{\lambda l}(z,y)/y^2. \label{Fepsabper}
\end{eqnarray}
The function $\chi _\mu $ is determined by (\ref{Fp}). 
To obtain the vev for 
the EMT components outside of a single conducting spherical shell 
with radius $a$ let us consider the limit $b\to \infty$. In this limit the 
second integral on the right of formula (\ref{sum1kab})
tends to zero (for large $\eta =b/a$ the subintegrand is
proportional to $e^{-2\eta z}$), whereas the first one 
does not depend on $b$. Hence one obtains
\begin{equation}
q=\frac{1}{8\pi ^2a^4}\sum_{l=1}^{\infty}(2l+1)\sum_{\lambda =0,1}
\int_{0}^{\infty}{z^3\psi _\mu (z/a)\frac{f_{\lambda l}^{(q)}(z,zx)}
{\Omega _{1\lambda l}(z)}dz}, \quad q=\varepsilon ,p_\perp .
\label{qa1}
\end{equation}
To regularize the expressions (\ref{qa1}) we have to subtract the
Minkowskian part, namely the expression (\ref{EMTMin}). It can be 
easily seen that
\begin{equation}
\frac{f_{\lambda l}^{(q)}(z,zx)}{\Omega _{1\lambda l}(z)}-
D_l^{(q)}(zx)=-\frac{1}{2}\sum_{m=1,2}\Omega ^{(m)}_{\lambda l}(z)
D_l^{(mq)}(zx). \label{rel81}
\end{equation}
Here the functions $D_l^{(mq)}(y)$ are obtained from the relations
(\ref{Dl}) by replacing $j_l\to h_l^{(m)}$, with $h_l^{(m)},\, m=1,2$
being spherical Hankel functions, and
\begin{equation}
\Omega _{\lambda l}^{(m)}(z)=\left\{ \begin{array}{l}
j_l(z)/h_l^{(m)}(z),\quad \lambda =0 \\
\left[ zj_l(z)\right] ^{'}/[zh^{(m)}_l(z)]^{'},\quad \lambda =1
\end{array} \right.
\label{Omegaem}
\end{equation}
The function $h_l^{(1)}(z)k$ ($h_l^{(2)}(z)$) has no zeros for 
$0\leq {\mathrm{arg}} z\leq \pi/2$ 
($-\pi /2\leq {\mathrm{arg}} z\leq 0$) and from this 
it follows that
\begin{equation}
\sum_{m=1,2}\int_{0}^{\infty}{z^3\psi _\mu (z/a)
\Omega ^{(m)}_{\lambda l}(z)D_l^{(mq)}(zx)dz}=
2\int_{0}^{\infty}{z^3{\mathrm{Re}}\left[ \psi _\mu (iz/a)
\Omega ^{(1)}_{\lambda l}(iz)D_l^{(1q)}(izx)\right] dz}.
\label{rel82}
\end{equation}
By introducing Ricatti-Bessel functions (\ref{RicBes}), for the 
regularized components of the vacuum EMT outside the sphere we find
\begin{equation}
q(a,r)=-\frac{1}{8\pi ^2a^4}\sum_{l=1}^{\infty}\int_{0}^{\infty}
{\chi _\mu (z/a)F_l^{(q)}(z,x)dz}, \quad r>a, 
\quad q=\varepsilon ,p_\perp ,p,
\label{qreg8}
\end{equation}
where for $x>1$ the functions $F_l^{(q)}(z,x)$ are defined as
\begin{eqnarray}
F_l^{(\varepsilon )}(z,x) & = & \frac{z}{x^2}
\left[ \frac{s_l(z)}{e_l(z)}+\frac{s'_l(z)}{e'_l(z)}\right] \left[ l
e_{l+1}^2(zx)+(l+1)e_{l-1}^2(zx)-(2l+1)e_{l}^2(zx)\right] 
\label{Foutside8}\\
F_l^{(p_\perp )}(z,x) & = & (2l+1)\frac{l(l+1)}{zx^4}
\left[ \frac{s_l(z)}{e_l(z)}+\frac{s'_l(z)}{e'_l(z)}\right] e_{l}^2(zx),
\quad F_l^{(p)}=F_l^{(\varepsilon )}-2F_l^{(p_\perp )}.
\label{Fpoutside8}
\end{eqnarray}
The exterior mode sum consideration given in this section follows
\cite{Grig1, Grig3}. For the case of exponential cutoff function
the formulae (\ref{qreg8}) and (\ref{Fpoutside8}) can be obtained
also from the results \cite{Brevik1, Brevik2}, where Green's function
formalism was used. Note that the expressions for the exterior
components are obtained from the interior ones replacing
$s_l\to i_l$, $i_l\to s_l$. In particular, the exterior components
can be presented in the form analog to (\ref{regqnew}).

Let us consider the behaviour of the functions $F_l^{(q)}(z,x)$
in various limiting cases. By using the corresponding formulae
for Bessel functions one obtains:

(a) When $l$ is fixed and $z\to 0$
\begin{equation}
F_l^{(\varepsilon )}(z,x)\sim -\pi (l+1/2)x^{-2(l+2)},
\quad F_l^{(p_\perp )}(z,x)\sim -\pi (l+1)x^{-2(l+1)}/2.
\label{asympoutzo}
\end{equation}

(b) When $l$ is fixed and $z$ is large
\begin{equation}
F_l^{(\varepsilon )}\sim 2F_l^{(p_\perp )}\sim -l^2(l+1)^2
(2l+1)\frac{\pi }{2(zx)^4}\exp [-2z(x-1)];
\label{asympzinfout}
\end{equation}

(c) For large $l$ by using the uniform asymptotic expansions for
Bessel functions \cite{abramowiz} one finds
\begin{equation}
F_l^{(q)}(\nu z,x)\sim \Phi _l^{(q)}(z,x)
\exp \left\{ -2\nu [\eta (zx)-\eta (z)]\right\} , \quad \nu =l+1/2
\label{phiasympout}
\end{equation}
with
\begin{eqnarray}
\Phi _l^{(\varepsilon )}(z,x) & = & -\frac{\pi \nu }{x^3}t(zx)t^3(z)
\left\{ 1+\frac{1}{12\nu}\left[ t(z)(t^2(z)+3)+t(zx)
(5t^2(zx)-9)\right] \right\} , \label{phiasympepsn}\\
\Phi _l^{(p)}(z,x) & = & \frac{\pi }{2x^3}t^2(zx)t^3(z),
\label{phiasymppen}
\end{eqnarray}
whith notations (\ref{tezeteta}).

It follows from here that for the values $x>1$ the 
expressions (\ref{qreg8}) are finite and hence cutoff 
may be removed. In this case the independence of the result on 
specific form of cutting function is obvious. 

The expressions (\ref{qreg8}) with $\psi _\mu =1$ diverge at 
sphere surface. The leading terms of these divergences may be 
found using (\ref{phiasympout}) and are as
following
\begin{equation}
\varepsilon \sim 2p_\perp \sim \frac{1}{30\pi ^2a(r-a)^3},
\quad p(r) \sim \frac{-1}{60\pi ^2a^2(r-a)^2},
\label{asymout8}
\end{equation}
and
\begin{equation}
\lim_{r\to a}(\varepsilon /p_\perp )'=
\lim_{r\to a}(p /p_\perp )'=-1.
\label{asympoutder8}
\end{equation}
Comparing (\ref{asymout8}) with (\ref{nearsurface}) we see that
the cancellation of interior and exterior leading divergent terms 
occurs in calculating the total energy and force acting on sphere.
The same cancellations take place for the next subleading
divergent terms as well (see below).
Formulas (\ref{asymout8}) are particular cases of general
asymptotic expansions of the vacuum EMT components for conformally
invariant fields near an arbitrary smooth boundary given in 
\cite{Deutsch}.

For distances far from the sphere one finds 
\begin{equation}
p_\perp \sim \frac{1}{4\pi ^2a^4x^7}\int_{0}^{\infty}
{z^2e_1^2(z)dz}=\frac{5a^3}{16\pi ^2r^7},\quad 
\varepsilon \sim -4p \sim \frac{a^3}{2\pi ^2r^7},
\quad r\gg a.
\label{fardist}
\end{equation}
The results of numerical calculations of the vacuum EMT components 
outside the sphere are given in \cite{Brevik1, Brevik2, Grig2}.
In \cite{Brevik1, Brevik2} calculations are carried out by using
the uniform asymptotic expansions for Riccati-Bessel functions. 
The accuracy of this approximation is estimated in \cite{Grig2},
 where exact relations are used in numerical calculations. 
The simple approximating formulas with the same accuracy as those
for the asymptotic calculations are presented as well.
The energy density and azimuthal pressure are positive, and
radial pressure is negative. The latter means that the 
exterior vacuum forces tend to expand sphere. As we will see
below this dominates the interior  contraction force.

Note that the continuity equation (\ref{conteq}) now may be written
in the following integral form
\begin{equation}
p(r)=\frac{1}{r^3}\int_{\infty}^{r}{\varepsilon (t)t^2dt}=
\frac{2}{r^2}\int_{\infty}^{r}{p_\perp (t)tdt},
\label{intrel8}
\end{equation}
where the integration constant is determined from the asymptotic
relations (\ref{asymout8}).
From (\ref{intrel7}) and (\ref{intrel8}) it follows
that
\begin{equation}
E(a)=\int{\varepsilon (r)dV}=4\pi a^3[p(a-)-p(a+)],
\label{totenergy8}
\end{equation}
where $E(a)$ is the total vacuum energy for a spherical shell with
radius $a$, $p(a\pm )=\lim_{r\to 0}p(a\pm r)$. 
By using the expressions for 
$p(r)$ given above one can obtain the following formula for 
the total energy (the same result can be obtained also by integrating
the energy density)
\begin{eqnarray}
E(a) & = & \frac{-1}{2\pi a}\sum_{l=1}^{\infty}(2l+1)\int_{0}^{\infty}
{dz\chi _\mu (z/a)z\left( \ln \vert s_l(z)e_l(z)\vert \right) '
\left[ 1+\left(\frac{l(l+1)}{z^2}+1\right) \frac{s_l(z)e_l(z)}
{s'_l(z)e'_l(z)}\right]}= \nonumber\\
 & = & \frac{-1}{2\pi a}\sum_{l=1}^{\infty}(2l+1)\int_{0}^{\infty}
 {dz\chi _\mu (z/a)z\frac{d}{dz}\ln \left\{ 1-\left[ s_l(z)e_l(z) 
 \right] ^{'2}\right\} }.
 \label{toten}
\end{eqnarray}
By taking the cutting function $\psi _{\mu }(\omega )=e^{-\mu \omega }$
one obtains the expression for the Casimir energy of the sphere
derived in  \cite{MiltonSph} by Green function method. Note that
in this method the factor $\psi _\mu (iz/a)=e^{-i\omega \mu }$
appears automatically as a result of the point splitting 
procedure. The evaluation of (\ref{toten}) leads to the 
result $E=0.092353/2a$ for the Casimir energy 
of a spherical
conducting shell \cite{Boyer, DaviesSph, Balian, MiltonSph, 
Leseduarte, Nesterenko}. This corresponds
to the repulsive vacuum force on the sphere.
 Here the cancellation
of interior and exterior divergent terms in the energy 
density occurs. The discussion on cancellations of
divergences between interior and exterior modes see 
\cite{MiltonSph, Deutsch, Candelas, Brevik1}.

\section{Electromagnetic vacuum in spherical 
layer between perfectly conducting
surfaces}

\renewcommand{\theequation}{10.\arabic{equation}}

\setcounter{equation}{0}

Electromagnetic vev of the EMT in the region between two cocentric 
perfectly conducting surfaces with radii $a$ and $b$, $a<b$, may 
be obtained from the results of previous section. The corresponding
nonrenormalized components are given by (\ref{sum1kab}). Using this 
formula they can be presented in the form
\begin{equation}
q(a,b,r)=q(a,r)+q^{(ab)}(r),\quad a<r<b, 
\quad q=\varepsilon ,p_\perp ,p,
\quad p=\varepsilon -2p_\perp ,
\label{q9}
\end{equation}
where $q(a,r)$ is given by (\ref{qa1}), and \cite{Sah2shert}
\begin{equation}
q^{(ab)}(r)=\frac{1}{8\pi ^2a^2r^2}\sum_{l=1}^{\infty}(2l+1)
\sum_{\lambda =0,1}
\int_{0}^{\infty}{z\psi _\mu (z/a)\Omega _{\lambda l}(z,\eta )
F_{\lambda l}^{(q)}(z,zx)dz},\quad x=r/a.
\label{qab92}
\end{equation}
Here the functions $F_{\lambda l}^{(q)}$ are defined by relations
(\ref{Fepsab}), (\ref{Fepsabper}), $\eta =b/a$, and 
\begin{equation}
\Omega _{\lambda l}(z,\eta )= \frac{e_l^{(\lambda )}(z\eta )/
e_l^{(\lambda )}(z)}{e_l^{(\lambda )}(z)s_l^{(\lambda )}(z\eta )-
e_l^{(\lambda )}(z\eta )s_l^{(\lambda )}(z)}
\label{Omeganew9}
\end{equation}
(see notation (\ref{RBdernot})). In (\ref{q9}) the 
dependence on $b$ is contained in the summand 
$q^{(ab)}$ only. This quantity is finite for $a\leq r<b$ and the 
regularization of $q(a,b,r)$ is equivalent 
to the renormalization of the 
first summand. This procedure have been done in previous section,
where we have seen that $q(a,r)$ (see expressions (\ref{qreg8}) 
and (\ref{Foutside8})) coincides with the corresponding quantity for 
the exterior region of a single shell with radius $a$. The expressions
(\ref{qab92}) for $a\leq r<b$ are finite when $\mu \to 0$ and hence
for these values the cuttoff function may be removed putting
$\chi _\mu =1$.

It can be seen that the quantities (\ref{q9}) may be written also 
in the form
\begin{equation}
q(a,b,r)=q(b,r)+\tilde q^{(ab)}(r),\quad q=\varepsilon ,p_\perp ,p,
\label{q9in}
\end{equation}
where
\begin{equation}
\tilde q^{(ab)}(r)=\frac{1}{8\pi ^2b^2r^2}\sum_{l=1}^{\infty}(2l+1)
\sum_{\lambda =0,1}
\int_{0}^{\infty}{z\tilde \Omega _{\lambda l}(z,\sigma )
F_{\lambda l}^{(q)}(z,zy)dz},
\label{qab93}
\end{equation}
with $y=r/b$, $\sigma =a/b$, and
\begin{equation}
\tilde \Omega _{\lambda l}(z,\sigma )\equiv 
\frac{s_l^{(\lambda )}(z\sigma )/ s_l^{(\lambda )}(z)}
{e_l^{(\lambda )}(z\sigma )s_l^{(\lambda )}(z)-
e_l^{(\lambda )}(z)s_l^{(\lambda )}(z\sigma )}.
\label{Omega9ba}
\end{equation}
In (\ref{q9in}) $\tilde q^{(ab)}(r)\to 0$ when $a\to 0$ and 
$q(b,r)$ coincides with the corresponding quantities inside 
a single conducting shell with radius $b$ 
(the latter can be seen also by direct evaluation 
of $q(b,r)$). Note that in (\ref{qab93}) the sum 
and integral are convergent for $a<r\leq b$.

As we said above from the expressions $q(a,b,r)$ in limiting cases
$a\to 0$ or $b\to \infty$ may be obtained the vacuum stress inside 
and outside of a single shell. Consider now the another limiting 
case: $h=b-a=const$, $b\to \infty$. For $a/b\to 1$ the main 
contribution in (\ref{qab92}) is due to large $l$. This allows us 
to use asymptotic formulae for Bessel functions. For instance,
in the case of the energy density one has
\begin{equation}
\varepsilon \approx \varepsilon ^{(ab)}\approx -\frac{1}{\pi ^2b^4}
\int_{0}^{\infty}{z^2\frac{\Lambda (z,a/b)}{\sqrt{1+z^2}}dz},
\label{rel9ba}
\end{equation}
where
\begin{equation}
\Lambda =\sum \nu ^3\left\{ e^{2\nu [\eta (z)-\eta (za/b)]}-
1\right\} ^{-1}\approx \frac{b^4}{16h^4(1+z^2)^2}
\int_{0}^{\infty}{\frac{s^3ds}{e^s-1}}=\frac{\pi ^4b^4}
{240h^4(1+z^2)^2}.
\label{rel92b}
\end{equation}
By substituting this into (\ref{rel9ba}) we receive the standard result
for the Casimir parallel plate 
configuration: $\varepsilon =-\pi ^2/720h^4$.

Let us present the quantites $q=\varepsilon ,p,p_\perp $ in the form
\begin{equation}
q=q(a,r)+q(b,r)+\Delta q(a,b,r),\quad a<r<b,
\label{qsum9}
\end{equation}
where "interference" term may be written in two ways
\begin{eqnarray}
\Delta q(a,b,r) & = &  q^{(ab)}(a,b,r)-q(b,r) \label{int91}\\
\Delta q(a,b,r) & = &  \tilde q^{(ab)}(a,b,r)-q(a,r). \label{int92}
\end{eqnarray}
Here $q^{(ab)}$ and $\tilde q^{(ab)}$ are 
defined by relations (\ref{qab92}) and (\ref{qab93}). 
It can be seen that $\Delta q(a,b,r)$ is finite for all
$a\leq r\leq b$, $a<b$. Near the surface $r=a$ it is convenient to use
(\ref{int91}), as for $r\to a$ both summands in this formula are 
finite. For the same reason the formula (\ref{int92}) is convenient
for calculations near the surface $r=b$.

So far we have considered the electromagnetic vacuum in the region
between two perfectly conducting spherical surfaces. Consider now 
a system consisting two cocentric thin spherical shells with radii
$a$ and $b$, $a<b$. In this case the vev for the EMT components
may be written in the form
\begin{equation}
q(a,b,r)=q(a,r)\theta (a-r)+q(b,r)\theta (r-b)+
\left[ q(a,r)+q^{(ab)}(r)\right] \theta (r-a)\theta (b-r),
\label{bettwosph}
\end{equation}
where $\theta (x)$ is the unit step function. By using the continuity
equation (\ref{conteq}) it is easy to see that the total Casimir
energy for the system under consideration can be presentaed in the
form
\begin{equation}
E^{(ab)}=E(a)+E(b)+4\pi \left[ b^3\tilde p^{(ab)}(b)-a^3p^{(ab)}
(a)\right] ,
\label{Casenergytwosph}
\end{equation}
where $E(i)$ is the Casimir energy for a single sphere with radius
$i,\, i=a,b$. As it follows from (\ref{qab92}) 
and (\ref{qab93}) the additional vacuum pressures on the spheres
are equal to \cite{Sah2shert, Sahdis}
\begin{eqnarray}
p^{(ab)}(a) & = & \frac{1}{8\pi ^2a^4}\sum_{l=1}^{\infty}(2l+1)
\int_{0}^{\infty}{dz z\left\{ \left[ \frac{l(l+1)}{z^2}+1
\right] \Omega _{1l}(z,\eta )-\Omega _{0l}(z,\eta ) \right\} }
\label{intforcin9}\\
\tilde p^{(ab)}(b) & = & \frac{1}{8\pi ^2b^4}
\sum_{l=1}^{\infty}(2l+1)\int_{0}^{\infty}{dz z
\left\{ \left[ \frac{l(l+1)}{z^2}+1\right] \tilde \Omega _{1l}
(z,\sigma )-\tilde \Omega _{0l}(z,\sigma ) \right\} } ,
\label{intforcout9}
\end{eqnarray}
where $\Omega _{\lambda l}$ and $\tilde \Omega _{\lambda l}$ 
are defined by relations (\ref{Omeganew9}) and (\ref{Omega9ba}).

The vacuum force per unit area of the inner sphere is equal to
\begin{equation}
F^{(a)}=F_1^{(a)}+\Delta F^{(a)},\quad \Delta F^{(a)}=-p^{(ab)}(a)
\label{forcein9}
\end{equation}
where $F_1^{(a)}$ is the force per unit area of a single sphere with
radius $a$, and $\Delta F^{(a)}$ is due to the existence of the second
sphere ("interaction" force). By similar 
way vacuum force acting on per unit area of outer sphere is
\begin{equation}
F^{(b)}=F_1^{(b)}+\Delta F^{(b)},\quad \Delta F^{(b)}=
\tilde p^{(ab)}(b).
\label{forceout9}
\end{equation}
The results of numerical calculations of quantities $\Delta q(a,b,r)$,
$q=\varepsilon ,p,p_\perp $, as well as those for $\Delta F^{(a,b)}$
are presented in \cite{Sah2shert, Sahdis}. Note that 
as it follows from the results
of these calculations the quantities (\ref{intforcin9}) and
(\ref{intforcout9}) are always negative, and therefore
the interaction forces between two spheres are 
always attractive (as in the parallel plate configuration). 
The total Casimir energy is positive for small values
of $a/b$ and is negative for values close to 1. At $a/b\approx 0.7$ this 
energy is zero.

\section{EMT vev inside a perfectly conducting cylindrical shell}

\renewcommand{\theequation}{11.\arabic{equation}}

\setcounter{equation}{0}

In this and next sections we will consider the case of
perfectly conducting cylindrically symmetric boundaries.
The Casimir effect for a perfectly conducting 
cylindrical shell was considered in \cite{Miltoncyl}
(see also \cite{DeRaadcyl}) and for a dielectric cylinder
in \cite{Brevikcyl} by using the Green function formalism.
Recently the problem is reconsidered in \cite{Miltoncyl1, 
Nestcyldiel} using the mode summation technique and in 
\cite{Lambiase, Romeocyl, Romeocyldiel}, within the 
framework of the zeta-function regularization scheme. In 
these papers global quantities, such as the total enehgy and 
stress on a shell, are investigated. Local characteristics
of the electromagnetic vacuum are considered in 
\cite{Sah2} for the interior and exterior regions of a 
conducting cylindrical shell, and in \cite{Sah3} for 
two coaxial shells. In this papers the mode summation method
is used combined with generalized Abel-Plana formula. Our
consideration below is based on these works (see also 
\cite{Sahdis}).
The vev of the EMT for electromagnetic field inside 
a perfectly conducting cylindrical surface with radius 
$a$ can be found by the way similar to the
spherical case. As an eigenfunctions we use the vector
potentials corresponding to the cylindrical waves of magnetic 
($\lambda =0$) and electric ($\lambda =1$) type:
\begin{equation}
{\mathbf{A}}_{\alpha }=\beta _{\lambda m}\left\{ \begin{array}{ll}
-{\mathbf{e}}_3\times \nabla _t\left\{ J_m(\gamma r)\exp \left[ i
(m\varphi +kz-\omega t)\right] \right\}, & \lambda =0 \\
(1/i\omega )\left[ {\mathbf{e}}_3+(ik/\gamma ^2)
\nabla _t\right] J_m(\gamma r)\exp \left[ i
(m\varphi +kz-\omega t)\right] , & \lambda =1 
\end{array} \right. , 
\label{cylinsmod}
\end{equation}
where the cylindrical coordinates $(r,\varphi ,z)$ are used with
unit vectors ${\mathbf{e}}_i$, $\gamma ^2=\omega ^2-k^2$, $m$ is 
an integer, $\nabla _t$ is the transverse to the $z$ axis part
of the nabla operator. From the standard boundary conditions
we obtain the following equations for the possible values of 
the quantum number $\gamma $:
\begin{eqnarray}
J'_m(\gamma a) & = & 0, \quad \lambda =0, \nonumber\\
J_m(\gamma a) & = & 0, \quad \lambda =1. \label{cylboundin}
\end{eqnarray}
The constants $\beta _{\lambda m}$ are determined from the normalization
condition and are equal to
\begin{equation}
\beta _{\lambda m}^{2}=\frac{\gamma ^{2}}{\pi \omega a^2}\left[ J^{'2}_m(
\gamma a)+(1-m^2/\gamma ^2a^2)J^2_m(\gamma a)\right] ^{-1}=
\frac{\gamma ^3}{\pi \omega a}T_m(\gamma a),
\label{cylnormin}
\end{equation}
where $T_m(z)$ is defined by (\ref{teka}).
As independent quantum numbers we will choose the set 
$\alpha =(mk\gamma \lambda )$. In this case $\omega ^2=\gamma ^2+k^2$ 
and $\gamma $ takes discrete values being solutions to 
(\ref{cylboundin}). By using the standard formula (\ref{emtgform})
with (\ref{cylinsmod}) one obtains
\begin{equation}
\langle 0\vert T^i_k\vert 0\rangle ={\mathrm{diag}}
\left( \varepsilon ,\, -p_1,\, -p_2,\, -p_3 \right) ,
\label{cylinemt}
\end{equation}
where the energy density $\varepsilon $, the pressure $p_i$ in direction
${\mathbf{e}}_i$ can be presented in the form
(below the index $c$ will specify quantities for the 
cylindrical geometry)
\begin{equation}
q_c(a,r)=\frac{1}{8\pi }\sum_{m=-\infty}^{+\infty}
\int_{-\infty}^{+\infty}
{dk\sum_{\lambda ,\gamma}\beta _{\lambda m}^2f_m^{(q)}(\gamma r)},
\quad q_c=\varepsilon, p_1,p_2
\label{cylqin}
\end{equation}
 with
\begin{eqnarray}
f_m^{(\varepsilon )}(y) & = & \left( \frac{2k^2}{\gamma ^2}+1
\right) \left[ J_m^{'2}
(y) +\frac{m^2}{y^2}J_m^2(y)\right] +J_m^2(y) \label{fepspi}\\
f_m^{(p_i)}(y) & = & -(-1)^i\left[ J_m^{'2}
(y) -\left( \frac{m^2}{y^2}+(-1)^i\right) J_m^2(y)\right]  ,
\quad i=1,2, \label{fepspi12}
\end{eqnarray}
and $p_3=\varepsilon -p_1-p_2$. The latter corresponds to the zero trace
of the vacuum EMT. The quantities (\ref{cylqin}) are divergent. To make them 
finite we introduce the cutoff function $\psi _\mu (\gamma )$ and 
consider the finite quantities
\begin{equation}
q_c=\frac{1}{8\pi ^2a^4}\sum_{m=-\infty}^{+\infty}\int_{-\infty}^{+\infty}
{dk\sum_{\lambda =0}^{1}\sum_{n=1}^{\infty }
\frac{j_{m,n}^{(\lambda )3}\psi _\mu (j_{m,n}^{(\lambda )}/a)}
{\sqrt{k^2a^2+j_{m,n}^{(\lambda )2}}}T_m(j_{m,n}^{(\lambda )})
f_m^{(q)}(j_{m,n}^{(\lambda )}x)},
\quad q=\varepsilon, p_1,p_2
\label{cylqin1}
\end{equation}
with $x=r/a$, $\gamma a=j_{m,n}^{(\lambda )}$ are the
roots of the equations (\ref{cylboundin}) for $\lambda =0$
and $\lambda =1$, correspondingly. To calculate the sums over zeros 
of the functions (\ref{cylboundin})
here we use the summation formula obtained in section 2, namely the 
formula (\ref{sumJbranch}). Let us choose as a function $f(z)$ in 
GAPF
\begin{equation}
f(z)=\frac{z^3}{\sqrt{z^2+k^2a^2}}\psi _\mu (z/a)f_m^{(q)}(zx).
 \label{f10ingapsf}
\end{equation}
This function has branch point on the imaginary axis and we have to
use the version (\ref{sumJbranch}) with lower sign. 
Here we will assume a class of cutoff functions for which 
(\ref{f10ingapsf}) satisfies conditions (\ref{condf}) and
(\ref{case21}) uniformly with respect to $\mu $. By using the 
asymptotic formulae for Bessel functions these conditions can
be easily translated in terms of $\psi _\mu $.
We choose $A=0,\, B=1$ in the case $\lambda =0$ and  
 $A=1,\, B=0$ in the case $\lambda =1$
(see (\ref{efnot1})), and $\nu =m$. Using the relation
\begin{equation}
f_m^{(q)}(ye^{-\pi i/2})=e^{2m\pi i}f_m^{(q)}(ye^{\pi i/2})
\label{rel101}
\end{equation}
we see that the subintegrand of the first integral on 
rhs of the formula (\ref{sumJbranch}) is proportianal to 
$\psi _\mu (iz/a)-\psi _\mu (-iz/a)$. Consequently after 
removing the cutoff ($\psi _\mu \to 1$) the contribution of the first
integral will be zero. For this reason we shall write only the
second integral on the right of (\ref{sumJbranch}). 
For the simplicity we will assume also that the cutoff function
has no poles in the right-half plane. In this case the residue 
terms are zero. It can be seen 
that the residue term on the right vanishes as well. Hence by 
applying GAPF to the sums over zeros of Bessel functions in
(\ref{cylqin1}) and omitting the term which will vanish after 
the cutoff removing we obtain
\begin{eqnarray}
q_c & = & \frac{1}{8\pi ^2}\sum_{m=-\infty}^{+\infty}
\int_{-\infty}^{+\infty}
dk\left\{ \int_{0}^{\infty}{\frac{z^3\psi _\mu (z)}
{\sqrt{z^2+k^2}}f_m^{(q)}(zr)dz}\right. +\nonumber\\
 & + & \left. \frac{e^{-m\pi i}}{\pi a^3}
 \int_{|ak|}^{\infty}{\left[ \frac{K_m(z)}
 {I_m(z)}+\frac{K'_m(z)}{I'_m(z)}\right] 
 f_m^{(q)}(zxe^{\pi i/2})\frac{z^3\chi _\mu (z/a)dz}
 {\sqrt{z^2-a^2k^2}}}\right\} ,
\label{cylqin2}
\end{eqnarray}
where the function $\chi _\mu (y)$ is defined in (\ref{Fp}). The second
integral on the right of this formula vanishes in the limit 
$a\to \infty$, whereas the first one does not depend on $a$.
It follows from here that the latter corresponds to the Minkowskian
part without boundaries. This can be seen also directly by explicit 
summation over $m$ using the formula $\sum_{m=-\infty}^{+\infty}
J_{n \pm m}^2(z)=1$. For instance, in the case of the energy density 
one has
\begin{eqnarray}
\varepsilon ^{(0)} & = & \frac{1}{8\pi ^2}\sum_{m=-\infty}^{+\infty}
\int_{-\infty}^{+\infty}{dk\int_{-\infty}^{+\infty}{dz
\frac{z^3\psi _\mu (z)}{\sqrt{z^2+k^2}}f_m^{(\varepsilon )}(zr)}}=
\nonumber\\
& = & \frac{1}{4\pi ^2}\int_{-\infty}^{+\infty}{dk
\int_{0}^{+\infty}{dzz\sqrt{z^2+k^2}\psi _\mu (z)}}=
\frac{1}{2\pi }\int_{0}^{+\infty}{\omega ^3\tilde \psi _\mu (\omega )
d\omega },
\label{cylMink}
\end{eqnarray}
with $\omega ^2=z^2+k^2$. Hence GAPF allows us 
to extract the contribution of unbounded
space without specifying the cutoff function. The remained part is
finite for $x<1$ and $\mu \to 0$, and can be written in the form
\begin{equation}
q_c=\frac{1}{2\pi ^3a^4}\sum_{m=0}^{\infty}{'}e^{-m\pi i}
\int_{0}^{\infty}{dt \int_{0}^{\infty}{dyz^2
\left[ \frac{K_m(z)}{I_m(z)}+\frac{K'_m(z)}{I'_m(z)}
\right] \chi _\mu (z/a)f_m^{(q)}(zxe^{\pi i/2})}},
\label{cylqin2new}
\end{equation}
with $z^2=t^2+y^2$ and $t=ka$. Here we have 
introduced a new integration variable $y$ and
the prime on the summation sign indicates that the 
$m=0$ term is to be halved. For $q=\varepsilon $ from 
(\ref{fepspi}) one has
\begin{equation}
e^{-m\pi i}z^2f_m^{(q)}(zxe^{\pi i/2})=z^2I_m^2(zx)+
(t^2-y^2)\left[ I_m^{'2}(zx)+\frac{m^2}{z^2x^2}I_m^2(zx)\right] 
\label{epscontzero}
\end{equation}
and it can be easily seen that the contribution of the summand 
containing $t^2-y^2$ in (\ref{cylqin2new}) is zero. Introducing
the polar coordinates $(z,\theta )$ on the plane $(t,y)$ for the
EMT components inside a perfectly conducting cylindrical surface
from (\ref{cylqin2new}) one finds \cite{Sah2}
\begin{equation}
q_c(a,r)=\frac{1}{4\pi ^2a^4}\sum_{m=0}^{\infty}{'}\int_{0}^{\infty}
{dz\chi _\mu (z/a)F_m^{(q)}(z,x)}, \quad r<a,
\quad q=\varepsilon ,p_i,
\label{cylqin3}
\end{equation}
where the following notations are introduced
\begin{eqnarray}
F_{cm}^{(q)}(z,x) & = & z^3\left[ \frac{K_m(z)}
 {I_m(z)}+\frac{K'_m(z)}{I'_m(z)}\right] \left\{ \begin{array}{ll}
 I_m^2(zx), & q=\varepsilon \\
 \left( 1+m^2/z^2x^2\right) I_m^2(zx)- I_m^{'2}(zx),
 & q=p_1
 \end{array} \right.\label{cylFqin}\\
F_{cm}^{(p_3)}(z,x) & = & -F_{cm}^{(\varepsilon )},\quad F_{cm}^{(p_2)}=
2F_{cm}^{(\varepsilon )}-F_{cm}^{(p_1)}. \label{cylFqin1} 
\end{eqnarray}
In particular we see that inside the cylinder $\varepsilon =-p_3$. 
This relation is the same as in the case of the Minkowski
vacuum. This is natural, as we have no constraint on $z$
 direction. On cylinder axis ($x=0$) the $m=0$ term contributes 
 only and we have
\begin{equation}
\varepsilon (0)=p_1(0)=p_2(0)=\frac{1}{8\pi ^2a^4}
\int_{0}^{\infty}{dzz^3\left[ \frac{K_0(z)}
 {I_0(z)}+\frac{K'_0(z)}{I'_0(z)}\right]}=-0.0168a^{-4}
 \label{cylaxes}
\end{equation}
with $q'(0)=0$. The vacuum EMT satisfy continuity equation 
which can be written now as
\begin{equation}
\frac{dp_1}{dr}+\frac{2}{r}(p_1-\varepsilon )=0,
\label{cylcont}
\end{equation}
or in the integral form
\begin{equation}
E_c(r)=2\pi \int_{0}^{r}{\varepsilon (t)tdt}=\pi r^2p_1(r),
\quad r<a. \label{cylcontint}
\end{equation}
To determine the integration constant here we have used the
relations (\ref{cylaxes}) between the EMT components on the 
cylinder axis. As we see the 
total energy per unit length inside the cylinder with radius $r$
is equal to the radial pressure on the surface of this cylinder
multiplied by the corresponding volume.

Let us consider the behavior of the functions (\ref{cylFqin1}) 
in two limiting cases:

\bigskip

1) For fixed $m$ and large $zx$ from the asymptotic expansions
of Bessel functions we find
\begin{equation}
F_{cm}^{(\varepsilon )}\sim \frac{1}{2}F_{cm}^{(p_2)}\sim -
\frac{z}{2x}e^{-2z(1-x)},\quad F_{cm}^{(p_1)}\sim -
\frac{1}{2x^2}e^{-2z(1-x)}.
\label{cylinzlarge}
\end{equation}

\bigskip

2) For large $m$ by using the uniform asymptotic expansions
of Bessel functions one obtains
\begin{equation}
F_{cm}^{(q)}(mz,x)\sim \Phi _{cm}^{(q)}(z,x)
\exp \left\{ -2m [\eta (z)-\eta (zx)]\right\} , 
\label{phiasympcyl}
\end{equation}
with
\begin{eqnarray}
\Phi _{cm}^{(\varepsilon )}(z,x) & = & -\frac{m}{2}z^5t(zx)t^3(z)
\left\{ 1-\frac{1}{12m}\left[ t(z)(t^2(z)-3)+t(zx)
(5t^2(zx)-3)\right] \right\} \label{phiasympepscyl}\\
\Phi _{cm}^{(p_1)}(z,x) & = & -\frac{1}{2}z^5t^2(zx)t^3(z),
\label{phiasymppecyl}
\end{eqnarray}
where the standard notations are used.

It follows from here that at cylinder surface, $r\to a$, 
the expressions for the EMT components are divergent and 
near the surface the corresponding quantities are dominated by
large $m$. Hence to obtain the asymptotic behaviour 
we can use the corresponding asymptotic formulae for modified
Bessel functions. Then after the elementary summation over
$m$ we find the following asymptotic behaviour
\begin{equation}
\varepsilon \sim \frac{1}{2}p_2\sim \frac{-1}{60\pi ^2a(a-r)^3},
\quad p_1\sim \frac{-1}{60\pi ^2a^2(a-r)^2}.
\label{cylasymp}
\end{equation}
This formulae are special cases of the general expansions for 
the EMT near a smooth boundary of arbitrary shape \cite{Deutsch}. 
Note that, as it follows from (\ref{cylinzlarge}) now, 
unlike the spherical case, in (\ref{cylqin3}) with 
$\chi _\mu =1$ the integral over $z$ diverges for $x=1$.

The results of the
numerical calculations for vacuum EMT components (\ref{cylqin3})
are presented in \cite{Sah2, Sahdis}. Note that $\varepsilon ,p_i<0$,
 $i=1,2$ everywhere inside the cylinder. The ratio of the energy
 density to the azimuthal pressure 
is a decreasing function on $r$ and $0.5\leq \varepsilon /p_2\leq 1$.

\section{Vacuum EMT outside a perfectly conducting cylinder}

\renewcommand{\theequation}{12.\arabic{equation}}

\setcounter{equation}{0}

First we consider the vev of the electromagnetic EMT 
in the region between two coaxial
cylindrical surfaces with radii $a$ and $b$, $a<b$. The corresponding
eigenfunctions have the form (\ref{cylinsmod}) with replacement
$J_m(\gamma r)\to P_{\lambda m}(\gamma a, \gamma r)$, where
\begin{equation}
P_{\lambda m}(x,y)=\left\{ \begin{array}{ll}
J_m(y)Y_m(x)-Y_m(y)J_m(x), & \lambda =1 \\
J_m(y)Y'_m(x)-Y_m(y)J'_m(x), & \lambda =0 \end{array} \right.
\label{radfout}
\end{equation}
From the boundary conditions on $r=a,b$ one obtains that the eigennumbers 
$\gamma  $ have to be solutions to the following equations
\begin{eqnarray}
P_{1m}(\gamma a,\gamma r)\vert _{r=b} & = & 0,\quad \lambda =1 
\label{cyloutcond}\\
\left[ \frac{\partial }{\partial r}P_{0m}(\gamma a,\gamma r)\right] _
{r=b} & = & 0,\quad \lambda =0 \label{cyloutcond1}
\end{eqnarray}
These equations have infinite number of simple real solutions. Now the 
normalization coefficients $\beta _{\lambda m}$ are in form
\begin{equation}
\beta _{\lambda m}^2=\frac{\pi z^4}{4a^4\omega }\left\{ \begin{array}{ll}
\left[ J_m^2(z)/J_m^2(z\eta )-1\right] ^{-1}, & \lambda =1 \\
\left[ \left( 1-m^2/z^2\eta ^2\right)J_m^{'2}(z)/
J_m^{'2}(z\eta )-1+m^2/z^2\right] ^{-1}, & \lambda =0
\end{array} \right.
\label{cylnormout}
\end{equation}
where $z=\gamma a$, $\eta =b/a$.

From Eq.(\ref{emtgform}) it follows that the vacuum EMT has 
diagonal form (\ref{cylinemt}) with components
\begin{equation}
q_c(a,b,r)=\frac{1}{8\pi }\sum_{m=-\infty}^{+\infty}
\int_{-\infty}^{+\infty}
{dk\sum_{\gamma ,\lambda }}\beta _{\lambda m}^2f_{\lambda m}^{(q)}
(\gamma a, \gamma r),\quad q_c=\varepsilon ,\, p_i,
\label{cylqout}
\end{equation}
where the expressions for the functions $f_{\lambda m}^{(q)}
(\gamma a,y)$ are
obtained from (\ref{fepspi}) and (\ref{fepspi12}) replacing 
$J_m(y)\to P_{\lambda m}(\gamma a,y)$. 

The eigenvalue equations (\ref{cyloutcond}) and (\ref{cyloutcond1}) 
can be written in terms of the function (\ref{bescomb1}) as
\begin{equation}
C_m^{AB}(\eta ,\gamma b)=0,\quad A=\lambda ,\, B=1-\lambda ,
\quad \lambda =0,1
\label{cyloutcondCAB}
\end{equation}
(see the notation (\ref{efnot1})). Note that the normalization 
coefficients can be expressed in terms of the function (\ref{tekaAB}):
\begin{equation}
\beta _{\lambda m}^2=\frac{\pi z^{5-2\lambda }}{4a^4\omega }
T_m^{AB}(\eta ,z).
\label{cylnormoutTAB}
\end{equation}
Using these relations and introducing a cutoff function
$\psi _\mu $ the divergent quantities (\ref{cylqout}) can be 
written in the form of the following finite integrosums
\begin{equation}
q_c=\frac{1}{32a^3}\sum_{m=-\infty}^{+\infty}\int_{-\infty}^{+\infty}
{dk\sum_{\lambda =0}^{1}\sum_{n=1}^{\infty }
\frac{(\gamma _{m,n}^{(\lambda )})^{5-2\lambda }\psi _
\mu (\gamma _{m,n}^{(\lambda )}/a)}
{\sqrt{k^2a^2+\gamma _{m,n}^{(\lambda )2}}}T_m^{AB}
(\eta ,\gamma _{m,n}^{(\lambda )})
f_{\lambda m}^{(q)}(\gamma _{m,n}^{(\lambda )},
\gamma _{m,n}^{(\lambda )}x)},
\label{cylqout1}
\end{equation}
where $q_c=\varepsilon, p_i$ and $\gamma a=\gamma _{m,n}^{(\lambda )}$, 
$n=1,2,\ldots $ are the
solutions to the eigenvalue equations (\ref{cyloutcond}),  
(\ref{cyloutcond1}) or (\ref{cyloutcondCAB}). By choosing 
in the formula (\ref{cor3form})
\begin{equation}
h(z)=\frac{z^{5-2\lambda }}{\sqrt{z^2+k^2a^2}}\psi _\mu (z/a)
f_{\lambda m}^{(q)}(z,zx).
 \label{cyloutfg}
\end{equation}
(as noted above this
formula is valid in the case when the corresponding function 
has branch point on the imaginary axis (see also Remark to 
the Theorem 2))
one obtains
\begin{eqnarray}
\sum_{n=1}^{\infty }
h(\gamma _{m,n}^{(\lambda )})T_m^{AB}(\eta ,
\gamma _{m,n}^{(\lambda )}) & = &
\frac{2}{\pi ^2}\int_{0}^{\infty }{\frac{h(x)dx}{\bar J_m^2(x)+
\bar Y_m^2(x)}}-
\nonumber\\
& & -\frac{1}{2\pi }
\int_{0}^{\infty }{\frac{\bar K_m(\eta x)}
{\bar K_m(x)}\frac{\left[ h(
xe^{\pi i/2})+h(xe^{-\pi i/2})\right] dx}
{\bar K_m(x)\bar I_m(\eta x)-
\bar K_m(\eta x)\bar I_m(x)}} .
\label{reloutcyl1}
\end{eqnarray}
Here in accordance with (\ref{efnot1}) and (\ref{cyloutcondCAB})
\begin{eqnarray}
\bar J_m(z)=J_m(z), \quad \lambda =1 \label{barimast} \\
\bar J_m(z)=zJ'_m(z), \quad \lambda =0,
\end{eqnarray}
and in similar way for other Bessel functions in (\ref{reloutcyl1}).
To obtain the EMT components for the outside region of a perfectly
conducting cylindrical shell we consider the limit $b\to \infty$. 
It can be seen that the second sum on the right of (\ref{reloutcyl1})
is zero in this limit and the first one does not depend on $b$. 
Hence for the outside region of a single cylinder we obtain
\begin{equation}
q_c(a,r)=\frac{1}{16\pi ^2a^4}\sum_{m=-\infty}^{+\infty}
\int_{-\infty}^{+\infty}
{dk\int_{0}^{\infty}{dz\sum_{\lambda =0,1}\frac{z^3\psi_\mu(z/a)}
{\sqrt{k^2+z^2/a^2}}\frac{f_{\lambda m}^{(q)}(z,zx)}
{\bar J_m^{2}+\bar Y_m^{2}}}}\, .
\label{cylqout2}
\end{equation}
To regularize we subtract from these quantities the contribution
of unbounded Minkowski spacetime which can be presented in the form
(see (\ref{cylMink})):
\begin{equation}
q^{(0)}=\frac{1}{8\pi ^2a^4}\sum_{m=-\infty}^{+\infty}\int_{-\infty}^{+\infty}
{dk\int_{0}^{\infty}{dz\frac{z^3\psi_\mu(z/a)}
{\sqrt{k^2+z^2/a^2}}f_{m}^{(q)}(zx)}}
\label{qMinkcyl}
\end{equation}
with the function $f_m^{(q)}$ defined as (\ref{fepspi}), 
(\ref{fepspi12}).
By using the definitions of $f_{\lambda m}^{(q)}$ and $f_{m}^{(q)}$
it is easy to see that
\begin{equation}
\frac{f_{\lambda m}^{(q)}(z,zx)}{\bar J_m^{2}+\bar Y_m^{2}}
-f_{m}^{(q)}(zx)=-\frac{1}{2}\sum_{n=1,2}\Omega _{\lambda m}^{(n)}(z)
f_m^{(nq)}(zx), \label{reloucyl2}
\end{equation}
where by definition the expression for $f_m^{(nq)}(zx)$ is obtained
from that for $f_m^{(q)}(zx)$ replacing $J_m(zx)\to H^{(n)}_m(zx)$,
$n=1,2$, and 
\begin{equation}
\Omega _{\lambda m}^{(n)}(z)=\left\{ \begin{array}{ll}
J_m(z)/H^{(n)}_m(z), & \lambda =1 \\
J'_m(z)/H^{(n)'}_m(z), & \lambda =0 
\end{array} \right.
\label{reloutcyl3}
\end{equation}
Hence
\begin{equation}
{\mathrm{reg}}\, q_c(a,r)=-\frac{1}{16\pi ^2a^4}
\sum_{m=-\infty}^{+\infty}
\int_{-\infty}^{+\infty}{dk\int_{0}^{\infty}{dz
\sum_{\lambda ,n}\Omega _{\lambda m}^{(n)}(z)f_{m}^{(nq)}(zx)}}.
\label{reloutcyl4}
\end{equation}
By rotating the integration contour for $z$ by angle $\pi /2$
for $n=1$ and by angle $-\pi /2$ for $n=2$ (note that the function
$H^{(1)}_m(z)$ ($H^{(2)}_m(z)$) has no zeros for $0\leq 
{\mathrm{arg}}z\leq \pi /2$ ($-\pi /2\leq {\mathrm{arg}}z\leq 0$))
and introducing Bessel modified functions for the regularized 
components we obtain (the ${\mathrm{reg}}$ sign is suppressed)
\begin{eqnarray}
q_c & = & \frac{1}{16\pi ^3a^4}\sum_{m=-\infty}^{+\infty}
\int_{-\infty}^{+\infty}dk\left\{ i\int_{0}^{|ak|}{dz
\left[ \psi _\mu \left( \frac{iz}{a}\right) -
\psi _\mu \left( -\frac{iz}
{a}\right) \right] \frac{F_{cm}^{(q)}(z,x)}{\sqrt{k^2-z^2/a^2}}}+\right.
\nonumber \\
& & \left.+2\int_{|ak|}^{\infty}{dz
\chi _\mu \left( \frac{z}{a}\right) 
\frac{F_{cm}^{(q)}(z,x)}{\sqrt{z^2/a^2-k^2}}}\right\}  
\label{reloutcyl5}
\end{eqnarray}
(the definition of the functions $F_{cm}^{(q)}$ see below). 
In (\ref{reloutcyl5}) the 
integrals are convergent for $x>1$ and $\mu =0$ and hence the 
cutoff can be removed. In this limit the first integral is 
zero and for regularized components of EMT after
transformations, similar to the interior case, one obtains
\begin{equation}
q_c(a,r)=\frac{1}{4\pi ^2a^4}\sum_{m=0}^{\infty}{'}
\int_{0}^{\infty}{dz\chi _\mu (z/a)F_{cm}^{(q)}(z,x)},
\quad r>a, \quad q=\varepsilon ,\, p_i,
\label{cylqout3}
\end{equation}
where for $x>1$ the functions $F_{cm}^{(q)}(z,x)$ are defined as
\begin{eqnarray}
F_{cm}^{(q)}(z,x) & = & z^3\left[ \frac{I_m(z)}{K_m(z)}+
\frac{I'_m(z)}{K'_m(z)}
\right] \left\{ \begin{array}{ll}
K^2_m(zx), & q=\varepsilon \\
\left( 1+m^2/z^2x^2\right) K^2_m(zx)-K^{'2}_m(zx), & q=p_1
\end{array} \right.
\label{cyloutF} \\
F_{cm}^{(p_3)}(z,x) & = & -F_{cm}^{(\varepsilon )},
\quad F_{cm}^{(p_2)}=
2F_{cm}^{(\varepsilon )}-F_{cm}^{(p_1)}. \label{cyloutF1}
\end{eqnarray}
It follows from here that $\varepsilon =-p_3$.

The asymptotic expressions for the functions (\ref{cyloutF1})
are as follows:

\bigskip

1) For fixed $m$ and large $z$:
\begin{equation}
F_{cm}^{(\varepsilon )}\sim \frac{1}{2}F_{cm}^{(p_2)}\sim 
 \frac{z}{2x}e^{2z(1-x)},\quad F_{cm}^{(p_1)}\sim -
\frac{1}{2x^2}e^{2z(1-x)}.
\label{cyloutzlarge}
\end{equation}

\bigskip

2) For large $m$ from the uniform asymptotic expansions
of Bessel functions one obtains
\begin{eqnarray}
F_{cm}^{(\varepsilon )}(mz,x) & \sim  & \frac{m}{2}z^5t(zx)t^3(z)
\left\{ 1+\frac{1}{12m}\left[ t(z)(t^2(z)-3)+t(zx)
(5t^2(zx)-3)\right] \right\} \times \nonumber\\
 & & \times \exp \left\{ 2m[\eta (z)-
\eta (zx)]\right\} , \label{phiasympepscylout}\\
F_{cm}^{(p_1)}(mz,x) & \sim  & -\frac{1}{2}z^5t^2(zx)t^3(z)
\exp \left\{ 2m [\eta (z)-\eta (zx)]\right\} ,
\label{phiasymppecylout}
\end{eqnarray}

 As in the case of the interior components of the vacuum EMT 
is divergent when $r\to a$ with asymptotic behaviour
\begin{equation}
\varepsilon \sim \frac{1}{2}p_2\sim \frac{1}{60\pi ^2a(r-a)^3},
\quad p_1\sim \frac{-1}{60\pi ^2a^2(r-a)^2}
\label{cylasympout}
\end{equation}
Comparing with (\ref{cylasymp}) we see that in calculating 
the total energy for the infinitely thin cylindrical shell 
the leading divergences cancel.

The asymptotic expressions for the vev at large distances from
the cylinder axis, $r\gg a$, can be found from (\ref{cylqout3})
introducing new integration variable $y=zx$ and expanding the 
integrands over $1/x$. In this limit the main contribution
comes from the lowest order mode with $m=0$ and one obtaines
\begin{equation}
q_c(a,r)\sim \frac{c^{(q)}}{8\pi ^2r^4\ln (r/a)},\quad 
c^{(\varepsilon )}=-c^{(p_1)}=\frac{1}{3},\,\, c^{(p_2)}=1,
\quad r\gg a \label{asympfaraxis}
\end{equation}
Here compared to the spherical case the corresponding 
quantities tend to zero more slowly.

From the continuity equation for the vacuum EMT one has the following
integral relation
\begin{equation}
p_1(r)=\frac{2}{r^2}\int_{\infty }^{r}{\varepsilon (t)tdt}=
-\frac{E^{out}_c(r)}{\pi r^2},
\label{cylcontintout}
\end{equation}
where $E^{out}_c(r)$ is the total energy (per unit length) outside 
cylinder with radius $r$. Combining this relation with 
(\ref{cylcontint}) for the total vacuum energy of the cylindrical
shell per unit length we obtain
\begin{equation}
E_c=E_c^{in}(a)+E_c^{out}(a)=\pi a^2\left[ p_1(a-)-p_1(a+)\right] .
\label{cyltotenergy}
\end{equation}
By taking into account the corresponding expressions for
the radial pressure this yields
\begin{eqnarray}
E_c & = & -\frac{1}{4\pi a^2}\sum_{m=0}^{\infty }{'}\int_{0}^{\infty }
{dz\chi _\mu (z/a)\big( \ln [I_m(z)K_m(z)]\big) {'}
\left[ z^2+\left( z^2+m^2\right) \frac{I_m(z)K_m(z)}
{I'_m(z)K'_m(z)}\right] }= \nonumber\\
& = & -\frac{1}{4\pi a^2}\sum_{m=0}^{\infty }{'}\int_{0}^{\infty }
{dz\chi _\mu (z/a)z^2\frac{d}{dz}\ln \left[ 1-z^2
\big( I_m(z)K_m(z)\big) '^{2}\right] }.
\label{cyltotenergy1}
\end{eqnarray}
In the last expression integrating by part and omitting the
boundary term we obtain the Casimir energy in the form used
in numerical calculations. The corresponding results are
presented in \cite{Miltoncyl, Miltoncyl1, Romeocyl}.
Note that in the evaluation of the Casimir energy for a
perfeclty conducting cylindrical shell by Green function 
method to perform the complex frequency rotation procedure
an additional cutoff function have to be introduced
 (see \cite{Miltoncyl}). This is 
related to the abovmentioned divergency of the integrals
over $z$ for $x=1$. 

The results of the numerical evaluations for 
the energy density and pressures distributions 
(formula (\ref{cylqout3}))
are presented in \cite{Sah2, Sahdis}. The energy density 
and azimuthal pressure in the exterior region are always
positive, and radial pressure is negative. The ratio
of the energy density to the azimuthal pressure is 
decreasing function on $r$, and 
$1/3\leq \varepsilon /p_2\leq 0.5$. Note that this ratio
is continous function for all $r$ and monotonically 
decreases from 1 at the cylinder axis to 1/3 at infinity.

\section{Vacuum EMT between two coaxial cylindrical shells}

\renewcommand{\theequation}{13.\arabic{equation}}

\setcounter{equation}{0}

By using the results from previous section the vev of the 
electromagnetic EMT in the region between two coaxial 
conducting cylindrical surfaces may be presented in 
the form (\ref{cylinemt}) with components
\begin{equation}
q_c(a,b,r)=q_c(a,r)+q_c^{(ab)}(r),\quad a<r<b, 
\quad q_c=\varepsilon ,\, p_i,
\label{cylab12}
\end{equation}
where $q_c(a,r)$ is given by (\ref{cylqout2}), and
\begin{equation}
q_c^{(ab)}(r) = -\frac{1}{16\pi a^3}\sum_{m=0}^{\infty}{'}
\int_{0}^{\infty}
{dk\sum_{\lambda =0}^{1}\int_{0}^{\infty}{ 
\frac{\bar K_m(z\eta )}{\bar K_m(z)}
\frac{\left[ h(ze^{\pi i/2})+h(ze^{-\pi i/2})\right] dz}
{\bar K_m(z)\bar I_m(z\eta )-\bar K_m(z\eta )
\bar I_m(z)}}}. \label{cylrel121}
\end{equation}
Here the function $h(z)$ is 
defined according to (\ref{cyloutfg}). As we
have shown the first summand on the right of (\ref{cylab12}) presents 
the corresponding quantity for the vacuum outside a single perfectly
conducting cylindrical shell with radius $a$. As we shall see later 
$q^{(ab)}(r)$ is finite for $a\leq r<b$ at $\mu =0$, and hence 
regularization is necessary for $q_c(a,r)$ only. This have been 
done in previous section. We have shown that result does not depend on
specific form of the cutoff function and can be presented in the form
(\ref{cylqout3}) and (\ref{cyloutF1}). 

In (\ref{cylrel121}) the integral over $z$ can be presented as 
a sum of two integrals along segments $(0,\vert ak\vert )$ and 
$(\vert ak\vert ,\infty )$. By using the relation (\ref{rel6})
and the explicit form of $h(z)$ it is easy to see that the first
integral will contain the cutoff function in the form 
$\psi _\mu (iz/a)-\psi _\mu (-iz/a)$ and hence vanishes after 
the cutoff removing. For this reason below we will consider
the second integral only. After the transformations similar to 
those we used to obtain (\ref{cylqin3}), the quantities 
$q^{(ab)}$ can be written in the form
\begin{equation}
q_c^{(ab)}(r)=\frac{1}{4\pi ^2a^4}\sum_{m=0}^{\infty}{'}
\int_{0}^{\infty}{dz\sum_{\lambda =0}^{1} 
z^3\Omega _{\lambda m}^c(\eta ,z)F_{\lambda m}^{(q)}(z,zx)},
\quad x=r/a,
\label{cylqab122}
\end{equation}
where 
\begin{eqnarray}
\Omega _{1m}^c(\eta ,z) & = & \frac{K_m(z\eta )/K_m(z)}
{K_m(z)I_m(z\eta )-K_m(z\eta )I_m(z)}\, , \label{Omegacm}\\
\Omega _{0m}^c(\eta ,z) & = & \frac{K'_m(z\eta )/K'_m(z)}
{K'_m(z)I'_m(z\eta )-K'_m(z\eta )I'_m(z)}\, ,
\label{Omegacm1}
\end{eqnarray}
and
\begin{eqnarray}
F_{c\lambda m}^{(\varepsilon )}(z,y) & = & Q_{\lambda m}^2(z,y),
\quad F_{c\lambda m}^{(p_3)}=F_{c\lambda m}^{(\varepsilon )}-
F_{c\lambda m}^{(p_1)}-F_{c\lambda m}^{(p_2)}\label{cylabF123}\\
F_{c\lambda m}^{(p_i)}(z,y) & = & \left( 1-(-1)^i\frac{m^2}{y^2}
\right) Q_{\lambda m}^2(z,y)+(-1)^i\left[ \frac{\partial }
{\partial y}Q_{\lambda m}(z,y)\right] ^2, \quad i=1,2
\nonumber
\end{eqnarray}
Here we have introduced the notation
\begin{eqnarray}
Q_{1m}(z,y) & = & K_m(z)I_m(y)-I_m(z)K_m(y) \label{cylabOm124}\\
Q_{0m}(z,y) & = & K'_m(z)I_m(y)-I'_m(z)K_m(y). \nonumber
\end{eqnarray}
The quantities (\ref{cylab12}) with (\ref{cylqout3}) and
(\ref{cylqab122}) present the regularized vev of the EMT
components in the region between two coaxial conducting
cylindrical surfaces.
Let us consider the limiting cases of the term 
(\ref{cylqab122}). First 
let $a/r,\, a/b \ll 1$. After replacing $z\to z\eta $ and expanding
the subintegrand over $a/r$ and $a/b$ it can be seen that
\begin{equation}
q_c^{(ab)}(r)\approx q_c(b,r), \quad a/r,\, a/b \ll 1,\, r<b,
\label{cylablim1}
\end{equation}
where $q_c(b,r)$ are the components for the vacuum EMT inside a single 
cylindric shell with radius $b$ (see (\ref{cylqin3}), (\ref{cylFqin})).

When $a\to b$ the sum over $m$ in (\ref{cylqab122}) diverges. Consequently
for $b-a\ll b$ the main contribution to $q_c^{(ab)}$ is due to large $m$.
By using the uniform asymptotic expansions for Bessel functions 
in this limit one obtaines
\begin{equation}
\varepsilon \approx -\frac{1}{2\pi ^2a^4}\sum_{m=0}^{\infty}{'}
\int_{0}^{\infty}{\frac{m^3z^3dz}{e^{2m[\eta (zb/a)-\eta (z)]}-1}}
\approx -\frac{\pi ^2}{720(b-a)^4}, \label{cyltoplate}
\end{equation}
which coincides with the corresponding quantity for the 
Casimir parallel plate configuration.

From (\ref{cylab12}) and (\ref{cylqab122}) it can be seen that the 
vev of EMT components can be written also in the form
\begin{equation}
q_c(a,b,r)=q_c(b,r)+\tilde q_c^{(ab)}(r), \quad a<r<b,
\quad q_c=\varepsilon ,p_i.
\label{cylinb12}
\end{equation}
Here $q_c(b,r)$ are vev inside a single cylindrical surface with radius 
$b$ (see (\ref{cylqin3}), (\ref{cylFqin}) with replacement $a\to b$), and
\begin{equation}
\tilde q_c^{(ab)}(r)=\frac{1}{4\pi ^2b^4}\sum_{m=0}^{\infty}{'}
\int_{0}^{\infty}{dz\sum_{\lambda =0}^{1}
z^3\tilde \Omega _{\lambda m}^c(\sigma ,z)F_{\lambda m}^{(q)}(z,zy)}
 \label{cylabqin12}
\end{equation}
where $y=r/b$, $\sigma =a/b$, and
\begin{eqnarray}
\tilde \Omega _{1m}^c(\sigma ,z) & = & \frac{I_m(z\sigma )/I_m(z)}
{I_m(z)K_m(z\sigma )-I_m(z\sigma )K_m(z)}\, , \label{Omegacmtilde} \\
\tilde \Omega _{0m}^c(\sigma ,z) & = &\frac{I'_m(z\sigma )/I'_m(z)}
{I'_m(z)K'_m(z\sigma )-I'_m(z\sigma )K'_m(z)}\, .
\label{Omegacmtilde1}
\end{eqnarray}
The quantites (\ref{cylabqin12})are finite for all 
$a<r\leq b$ and diverge on surface $r=a$.

From the above it follows that if we present the vacuum EMT 
components between cylindrical surfaces in the form 
\begin{equation}
q_c=q_c(a,r)+q_c(b,r)+\Delta q_c(a,b,r)
\label{cyldelq}
\end{equation}
then the quantities 
\begin{equation}
\Delta q_c(a,b,r)=q_c^{(ab)}(r)-q_c(b,r)= 
\tilde q_c^{(ab)}(r)-q_c(a,r) \label{cyldelq1}
\end{equation}
are finite for all $r$ from $a\leq r\leq b$. In (\ref{cyldelq1})
the first presentation is convenient near the surface $r=a$, as in
this case both summands are finite. Similarly the second presentation
is convenient near $r=b$.

Let us consider a system of two coaxial thin cylindrical shells with radii
$a$ and $b$, $a<b$. The vacuum EMT components
may be written in the form
\begin{equation}
q_c=q_c(a,r)\theta (a-r)+q_c(b,r)\theta (r-b)+
\left[ q_c(a,r)+q_c^{ab}(r)\right] \theta (r-a)\theta (b-r).
\label{bettwocyl}
\end{equation}
 Similar to the spherical case using the continuity
equation (\ref{cylcont}) the total Casimir
energy for this system may be written as
\begin{equation}
E_c^{(ab)}=E_c(a)+E_c(b)+\pi b^2\tilde p_{c1}^{(ab)}(b)-
\pi a^2p_{c1}^{(ab)}(a),
\label{Casenergytwocyl}
\end{equation}
where $E_c(i)$ is the Casimir energy for a single cylindrical shell 
with radius $i,\, i=a,b$. For the additional vacuum pressures on the 
cylindrical surfaces from (\ref{cylqab122}) and (\ref{cylabqin12}) 
one has:
\begin{eqnarray}
p_{c1}^{(ab)}(a) & = & \frac{1}{4\pi ^2a^4}\sum_{m=0}^{\infty}{'}
\int_{0}^{\infty}{zdz\left[ \left( \frac{m^2}{z^2}+1
\right) \Omega ^c_{0m}(\eta ,z)-
\Omega ^c_{1m}(\eta ,z)\right] },
\label{forceincyl1} \\
\tilde p_{c1}^{(ab)}(b) & = & \frac{1}{4\pi ^2b^4}\sum_{m=0}^{\infty}{'}
\int_{0}^{\infty}{zdz\left[ \left( \frac{m^2}{z^2}+1
\right) \tilde \Omega ^c_{0m}(\sigma ,z)-
\tilde \Omega ^c_{1m}(\sigma ,z)\right] },
\label{forceoutcyl1} 
\end{eqnarray}
where $\Omega ^c_{\lambda m}(\eta ,z)$ and 
$\tilde \Omega ^c_{1m}(\sigma ,z)$ are defined in (\ref{Omegacm}),
(\ref{Omegacm1}), (\ref{Omegacmtilde})
and (\ref{Omegacmtilde1}). In (\ref{forceincyl1}) and 
(\ref{forceoutcyl1}) the first summands in 
braces come from the magnetic waves contribution, and second ones 
from the electric type waves.

Let us now consider the interaction forces between cylindrical surfaces.
The force acting per unit area of the inner surface can be presented in
the form 
\begin{equation}
F_c^{(a)}=F_{c1}^{(a)}+\Delta F_c^{(a)},\quad \Delta F_c^{(a)}
=-p_c^{(ab)}(a),
\label{forceincyl}
\end{equation}
where $F_{c1}^{(a)}$ is the force acting on a single 
cylindrical surface with radius $a$, and $\Delta F_c^{(a)}$ is 
additional force due to the existence of the 
outer surface and is determined from (\ref{forceincyl1}). The latter is 
finite without additional subtractions. 

By similar way the force acting per unit area of the outer cylinder
\begin{equation}
F_c^{(b)}=F_{c1}^{(b)}+\Delta F_c^{(b)},
\quad \Delta F_c^{(b)}=\tilde p_c^{(ab)}(b)
\label{forceoutcyl}
\end{equation}
where additional term $\Delta F_c^{(b)}$ is 
due to the existence of the inner
cylinder and is defined by (\ref{forceoutcyl1}).

The results of the numerical calculations for quantities
$\Delta q(a,b,r)$ are given in \cite{Sah3}.
 Note that the sign of $\Delta \varepsilon $ and $\Delta p_{c1}$
 is the same as in the case of interior of the parallel plate
 configuration. In particular 
the additional forces $\Delta F_c^{(a,b)}$
always have attractive nature.

\section{Summary}

\renewcommand{\theequation}{14.\arabic{equation}}

\setcounter{equation}{0}

In the present paper we considered a possible way for generalization
of Abel-Plana summation formula, proposed in \cite{Sah1}. 
The generalized version contains two meromorphic functions $f(z)$ 
and $g(z)$ and is formulated in the form of Theorem 1. 
The special choice $g(z)=-if(z)\cot \pi z$
with $f(z)$ being an analytic function in the right half-plane
gives APF with additional residue terms coming from the poles of
$f(z)$. Another consequence from GAPF is the summation
formula (\ref{apsf3}) over the points with integer
values of an analytic function. An application of this 
formula to the Casimir effect is given in \cite{Sah}.

Further we consider the applications to the series
and integrals involving Bessel functions.
First of all, in section 3 choosing the function $g(z)$ in the 
form (\ref{gebessel}) we derive two types of summation 
formulae for the series $\sum_{k}^{}T_\nu (
\lambda _{\nu ,k})f(\lambda _{\nu ,k})$ (the definition
$T_\nu (z)$ see (\ref{teka})) with $\lambda _{\nu ,k}$
being the zeros of the function $\bar J_\nu (z)=
AJ_\nu (z)+BzJ'_\nu (z)$. Such a type of series arises in a number
of problems of mathematical physics with spherical and 
cylindrical symmetry.
As a special case they include
Fourier-Bessel and Dini series (see \cite{Watson}).
Using the formula (\ref{sumJ1}) the difference
between the sum over zeros of $\bar J_\nu (z)$ and 
corresponding integral can be presented
in terms of an integral involving Bessel modified functions plus 
residue terms. For a large class of functions the last integral
converges exponentially fast and is useful for numerical
calculations. The mode summation method for calculating 
the vev of the EMT inside perfectly conducting spherical 
and cylindrical shells used in \cite{Grig1, Grig2, Sah2} is based
on this formula. In this method the independence of the regularized
EMT components on specific form of the cutoff function 
becomes obvious. APF is a special case of (\ref{sumJ1}) with
$\nu =1/2$, $A=1$, $B=0$ and an analytic function $f(z)$.
Choosing  $\nu =1/2$, $A=1$, $B=2$ we obtain APF in the form
(\ref{apsf2half}) useful for fermionic field calculations.
Note that the formula (\ref{sumJ1}) may be used also for some
functions having poles and branch points on the imagianary axis.

The second type of summation formulae, formula (\ref{sumJ2}), 
considered in subsection 3.2 (Theorem 3), is valid for functions
satisfying condition (\ref{caseb}) and presents the difference 
between the sum over zeros of $\bar J_\nu (z)$ and corresponding
integral in terms of residues over poles for $f(z)$ in the right
half-plane (including purely imaginary ones). It may be used to 
summarize a large class of series of this type in finite terms. 
In particular, the examples we found in literature, when the 
corresponding sum may be presented in closed form, are special
cases of this formula. A number of new series summable by this 
formula and some classes of functions to which it can be 
applied is presented.

In Section 4 we consider applications to the series
of type $\sum_{k}^{}T^{AB}_\nu (\lambda ,\gamma _{\nu ,k})
h(\gamma _{\nu ,k})$ (with $T^{AB}_\nu (\lambda ,z)$ defined as
(\ref{tekaAB})), where $\gamma _{\nu ,k}$ are zeros of the
function $\bar J_\nu (z)\bar Y_\nu (\lambda z)-
\bar J_\nu (\lambda z)\bar Y_\nu (z)$. The corresponding 
results are formulated in the form of Corollary 2 and 
Corollary 3. Using the formula (\ref{cor3form})
the difference between the sum and corresponding integral
can be expressed as an integral containing Bessel modified functions
plus residue terms. For the large class of functions $h(z)$ this
integral converges exponentially fast. The formula of the 
second type, (\ref{cor2form}), allows to find in closed form
the sums of some types of the series over $\gamma _{\nu ,k}$. 
To evaluate the corresponding integral the formula can be used derived 
in section 7. This yields to the another summation formula,  
(\ref{th4form}), containing residue terms only.
The similar formulae can be obtained for the series over zeros
of the function $J'_\nu (z)Y_\nu (\lambda z)-
J_\nu (\lambda z)Y'_\nu (z)$ as well.
Note that the several examples we found in literature when
the corresponding sum was evaluated in closed form are special cases 
of the formulae considered here. We present new examples and some 
classes of functions satisfying the corresponding conditions.
The possibilities are endless.

The results from GAPF for the integrals of type
${\mathrm{{\mathrm{p.v.}}}}\int_{0}^{\infty }{F(x)\bar J_\nu (x)dx}$ 
(see notation (\ref{efnot1})) and
${\mathrm{{\mathrm{p.v.}}}}\int_{0}^{\infty }{F(x)[J_\nu (x)\cos \delta +
Y_\nu (x)\sin \delta ]dx}$ are considered in section 5. 
The corresponding formulae have the form (\ref{intJform41}), 
(\ref{intJform42}) and (\ref{intJYform43}). In particaular
the formula (\ref{intJform41}) is useful to express the
integrals containing Bessel functions with oscillating
subintegrand through the integrals of modified Bessel
functions with exponentially fast convergence. The results 
obtained in \cite{Schwartz} are special cases of these
formulae. The illustrating examples of applications of the 
formulae for integrals are given in section 6 (see 
(\ref{examp51})-(\ref{examp54ad}) and (\ref{examp56})-
(\ref{examp59ad})). Looking the standard books (see, e.g., 
\cite{Erdelyi}, \cite{Watson}-\cite{Oberhettinger}) one will find many 
particular cases which follow from these formulae. 
Many new integrals can be evaluated as well. We consider
also two examples of functions having purely imaginary
poles, (\ref{examp510}) and (\ref{exampforpole}), with
corresponding formulae (\ref{intJ5sum}) and 
(\ref{exampforpoleform}) (two special cases of these formulae
see \cite{Watson}).

By choice of the functions $f(z)$ and $g(z)$ in accord with 
(\ref{fg61}) formulae (\ref{intJYth65}) and (\ref{th6form})
for integrals of type
\[
{\mathrm{p.v.}}\int_{0}^{\infty }{\frac{J_\nu (x)Y_\mu (\lambda x)-
J_\mu (\lambda x)Y_\nu (x)}{J_\nu ^2(x)+Y_\nu ^2(x)}F(x)dx}
\]
can be derived from GAPF. The corresponding results are 
formulated in the form of Theorem 5 and Theorem 6 in section 7.
The several examples for the integrals of this type we have been 
able to find in literature  are particular cases of the formula 
(\ref{intJYth65}). New examples when the integral is evaluated
in finite terms are presented. Some classes of functions are 
distinguished to which the corresponding formulae may be 
applied.

In the following sections, based on 
\cite{Grig1}-\cite{Sah3}, the physical applications
of the summation formulae obtained from GAPF are reviewed.  
We consider the vacuum expectation values of the energy-momentum
tensor for the electromagnetic field inside and outside the 
perfectly conducting spherical and cylindrical shells, as well
as between two conducting cocentric spherical and coaxial 
cylindrical surfaces.The corresponding mode sums contain the series over
zeros of Bessel functions and their combinations. The application 
of the summation formulae from sections 3 and 4 allows 
(i) to extract from corresponding divergent quantities the 
contribution of the unbounded space in explicitly cutoff 
independent way, and (ii) to obtain for the 
regularized values strongly convergent integrals. To compare
note that in the Green function method after the subtraction of 
the Minkowskian part the additional complex frequency rotation
is used. In the regularization scheme based on the 
summation formulae of APF type the complex frequency
rotation is made automatically. The corresponding global
quantities such as total Casimir energy or forces acting
on the surfaces can be obtained from the EMT components.
It is shown that in the geometries with two surfaces the 
additional vacuum forces due to the existence of the second surface
always have attractive nature. In the limiting case of the large
radii the corresponding results for the Casimir parallel plate
configuration are obtained.
 
Of course the applications of the summation formulae obtained
from GAPF are not restricted by the Casimir effect only. Similar
types of series will arise in considerations of various
physical phenomenon near the boundaries with spherical and
cylindrical symmetries, for example in calculations of the electron
self-energy and the electron anomalous magnetic moment
(for the similar problems in the plane boundary case see, e.g.,
\cite{Fischbach, Svozil} and references therein). 
The dependence of these 
quantities on boundaries originates from the modification of the
photon propagator due to the boundary conditions imposed by the 
walls of the cavity.

\section*{Acknowledgements}

 I am indebted to Prof. G. Sahakyan, Prof. E. Chubaryan
and Prof. A. Mkrtchyan for general encouragement, valuable
comments and suggestions. I am grateful to
Levon Grigoryan for fruitful colaboration and stimulating
discussions. A part of this
paper was written during my stay in Tehran, when I given 
lectures on the Casimir effect at Sharif University of Technology. 
I acknowledge the Physics Department and Prof. Reza Mansouri
for the hospiality. I would also like acknowledge the hospitality of 
the Abdus Salam International Centre for Theoretical Physics, 
Trieste, Italy.


\begin{thebibliography}{99}

\bibitem{Casimir} H. B. G. Casimir, Proc. Kon. Nederl. Akad. 
Wet. {\bf 51}, 793 (1948).

\bibitem{Mostepanenko} V. M. Mostepanenko and N. N. Trunov, {\it The 
Casimir effect and its applications}. Clarendon Press, Oxford, 1977.

\bibitem{Plunien} G. Plunien, B. Muller and W.Greiner, Phys. Rep.
{\bf 134}, 87 (1986).

\bibitem{Bordag} M. Bordag (Ed.), {\it The Casimir Effect. 50 years
later}. World Scientific, 1999.

\bibitem{Lamor} S. K. Lamoreaux, Am. J. Phys. {\bf 67}, 850 (1999).

\bibitem{Hardy} G. H. Hardy, {\it Divergent series}. Chelsea Publishing
Company, New York, 1991.

\bibitem{Erdelyi} A. Erd$\acute{{\mathrm{e}}}$lyi {\it et al},
{\it Higher transcendental functions}.
Vol.1,2. McGraw Hill, New York, 1953.

\bibitem{Evgrafov} M. A. Evgrafov, {\it Analytic functions}. Nauka, 
Moscow, 1968 (in Russian).

\bibitem{Mamaev} S. G. Mamaev, V. M. Mostepanenko and A. A. 
Starobinsky, Zhurnal Eksperimentalnoi i Teoreticheskoi Fiziki,
{\bf 70}, 1577 (1976). [Sov. Phys.-JETP, {\bf 43}, 823 (1976)]

\bibitem{Sah} A. A. Saharian, Izv. AN Arm. SSR. Fizika,
{\bf 21}, 262 (1986) (Sov. J. Contemp. Phys., 
{\bf 21}, 32 (1986)).

\bibitem{Sah1} A. A. Saharian, Izv. AN Arm. SSR. Matematika,
{\bf 22}, 166 (1987) (Sov. J. Contemp. Math. Analysis, 
{\bf 22}, 70 (1987)).

\bibitem{Sahdis} A. A. Saharian, PhD thesis, Yerevan, 1987 (in Russian).

\bibitem{Grig1} L. Sh Grigoryan and A. A. Saharian, Dokladi AN
Arm. SSR, {\bf 83}, 28 (1986) (Reports NAS RA, in Russian).

\bibitem{Grig2} L. Sh Grigoryan and A. A. Saharian, Uchenie
Zapiski EGU, No.3, 56 (1986) (Scientific Transactions of the 
Yerevan State University, in Russian).

\bibitem{Grig3} L. Sh Grigoryan and A. A. Saharian, Uchenie
Zapiski EGU, No.1, 61 (1987) (Scientific Transactions of the 
Yerevan State University, in Russian).

\bibitem{Sah2shert} L. Sh Grigoryan and A. A. Saharian, Izv. AN Arm. 
SSR. Fizika, {\bf 22}, 3 (1987) (Sov. J. Contemp. Phys., 
{\bf 22}, 1 (1987)).

\bibitem{Sah2} A. A. Saharian, Izv. AN Arm. SSR. Fizika,
{\bf 23}, 130 (1988) (Sov. J. Contemp. Phys., 
{\bf 23}, 14 (1988)).

\bibitem{Sah3} A. A. Saharian, Dokladi AN
Arm. SSR, {\bf 86}, 112 (1988) (Reports NAS RA, in Russian).

\bibitem{Mamaev2} S. G. Mamaev and N. N. Trunov, Izvestiya Vuzov.
Seriya Fzika, {\bf 17}, 88 (1979). (Sov. Phys. J., {\bf 22}, 
966 (1979).)

\bibitem{Watson} G. N. Watson, {\it A treatise on the theory of 
Bessel function}. Cambridge University Press, 1995.

\bibitem{abramowiz} M. Abramowitz and I. A. Stegun, {\it Handbook
of Mathematical functions}. National Bureau of Standards,  
Washington D.C.,1964; reprinted by Dover, New York, 1972.

\bibitem{Magnus} W. Magnus and F. Oberhettinger, {\it Formulas and 
theorems for the functions of mathematical physics}. Chelsea,
New York, 1962.

\bibitem{Prudnikov} A. P. Prudnikov, Yu. A. Brychkov and O. I. Marichev,
{\it Integrals and series. v.2. Special functions}. 1986.

\bibitem{Schwartz} C. Schwartz, J. Math. Phys. {\bf 23}, 2266 (1982).

\bibitem{Erdelyi2} A. Erd$\acute{{\mathrm{e}}}$lyi et al., 
{\it Tables of integral transforms}. Vol.2. McGrew-Hill, New York, 1954.

\bibitem{Luke} Y. L. Luke, {\it Integrals of Bessel functions}. 
McGrew-Hill, 1962.

\bibitem{Gradshteyn} I. S. Gradshteyn and I. M. Ryzhik, {\it Tables of
integrals, series and products}. Academic Press, New York, 1994.

\bibitem{Wheelon} A. D. Wheelon, {\it Tables of summable series and 
integrals involving Bessel functions}. Holden-Day, New York, 1968.

\bibitem{Oberhettinger} F. Oberhettinger, {\it Tables of Bessel 
transforms}. Springer, Berlin, 1972.

\bibitem{Boyer} T. H. Boyer, Phys. Rev. {\bf 174}, 1764 (1968).

\bibitem{DaviesSph} B. Davies, J. Math. Phys. {\bf 13}, 1324 (1972).

\bibitem{Balian} R. Balian and B. Duplantier, Ann. Phys. (N.Y.)
{\bf 112}, 165 (1978).

\bibitem{MiltonSph} K. A. Milton, L. L. DeRaad, Jr., and 
J. Schwinger, Ann. Phys. (N. Y.) {\bf 115}, 388 (1978).

\bibitem{Olaussen1} K. Olaussen and F. Ravndal, Nucl. Phys. 
{\bf B192}, 237 (1981).

\bibitem{Olaussen2} K. Olaussen and F. Ravndal, Phys. Lett.
 {\bf 100B}, 497 (1981).
 
\bibitem{Brevik1} I. Brevik and H. Kolbenstvedt, Ann. Phys. (N.Y.)
{\bf 149}, 237 (1983).

\bibitem{Brevik2} I. Brevik and H. Kolbenstvedt, Can. J. Phys. 
{\bf 62}, 805 (1984).

\bibitem{Nesterenko} V. V. Nesterenko and I. G. Pirozhenko, Phys. Rev. 
{\bf D57}, 1284 (1998).

\bibitem{Bowers} M. E. Bowers and C. R. Hagen, Phys. Rev. {\bf 59},
025007 (1999).

\bibitem{Leseduarte} S. Leseduarte and A. Romeo, Ann. Phys. (N.Y.)
{\bf 250}, 448 (1996).

\bibitem{Cognola} E. Cognola, E. Elizalde and K. Kirsten,
Casimir energies for spherically symmetric cavities, 
hep-th/9906228.

\bibitem{Lambiase} G. Lambiase, V. V. Nesterenko and M. Bordag, 
J. Math. Phys. {\bf 40}, 6254 (1999).

\bibitem{Birrel} N. D. Birrel and P. C. W. Davis, {\it Quantum
fields in curved space}. Cambridge University Press, 1982.

\bibitem{Jackson} J. D. Jackson, {\it Classical electrodynamics}.
John Wiley and Sons, New York, 1975.

\bibitem{Deutsch} D. Deutsch and P. Candelas, Phys. Rev. D,
{\bf 20}, 3063 (1979).

\bibitem{Kennedy} G. Kennedy, R. Critchley and J. S. Dowker,
Ann. Phys. (N.Y.) {\bf 125}, 346 (1980).

\bibitem{Barton1} G. Barton, J. Phys. A: Math. Gen. {\bf 14},
1009, 1981; ibid {\bf 15}, 323 (1982).

\bibitem{Candelas} P. Candelas, Ann. Phys. (N.Y.) {\bf 143}, 241 (1982).

\bibitem{Miltoncyl} L. L. DeRaad Jr., and K. A. Milton,
Ann. Phys. (N.Y.) {\bf 136}, 229 (1981).

\bibitem{DeRaadcyl} L. L. DeRaad, Jr., Fortschr. Phys. {\bf 33},
117 (1985).

\bibitem{Brevikcyl} I. Brevik and G. H. Nyland, Ann. phys. (N.Y.),
{\bf 230}, 321 (1994).

\bibitem{Miltoncyl1} K. A. Milton, A. V. Nesterenko and
V. V. Nesterenko, Phys. Rev. {\bf D59}, 105009 (1999).

\bibitem{Nestcyldiel} V. V. Nesterenko and I. G. Pirozhenko,
Phys. Rev. {\bf D60}, 125007 (1999).

\bibitem{Romeocyl} P. Gosdzinsky and A. Romeo, Phys. Lett. {\bf 441},
265 (1998).

\bibitem{Romeocyldiel} I. Klich and A. Romeo, Regularized 
Casimir energy for an infinite dielectric cylinder subject to
light-velocity conservation, hep-th/9912223.

\bibitem{Fischbach} E. Fischbach and N. Nakagawa, Phys. Rev. {\bf D30},
2356 (1984).

\bibitem{Svozil} M. Kreuzer and K. Svozil, Phys. Rev. {\bf D34}, 
1429 (1986).

\end{thebibliography}
\end{document}